\documentclass[twocolumn]{aastex631}

\usepackage{graphicx}
\usepackage{subfig}
\usepackage{subfig}
\usepackage{rotating}
\usepackage{booktabs}
\usepackage{float}
\usepackage{color}
\usepackage{ulem}
\usepackage{amsmath}
\usepackage{graphicx}
\usepackage{soul}

\newcommand{\OIII}{{\ion{O}{3}}}

\newcommand{\NII}{{\ion{N}{2}}}

\newcommand{\SII}{{\ion{S}{2}}}

\newcommand{\FeII}{{\ion{Fe}{2}} }

\newcommand{\kms}{km s$^{-1}$}

\newcommand{\Hb}{{H$\beta$}}

\newcommand{\ergs}{erg s$^{-1}$}

\newcommand{\snu}{\affil{Astronomy Program, Department of Physics and Astronomy, Seoul National University, 1 Gwanak-ro, Gwanak-gu, Seoul 08826, Republic of Korea}}
\newcommand{\yonsei}{\affil{Department of Astronomy, Yonsei University, 50 Yonsei-ro, Seodaemun-gu, Seoul 03722, Republic of Korea}}

\shorttitle{Strong Outflow Type 1 AGNs by GMOS-IFU}
\shortauthors{Kim et al.}

\begin{document}

\title{Unraveling the Complex Structure of AGN-driven Outflows. VI.

Strong Ionized Outflows in Type 1 AGNs and the Outflow Size--Luminosity Relation}

\correspondingauthor{Jong-Hak Woo}
\email{woo@astro.snu.ac.kr}

\author[0000-0002-2156-4994]{Changseok Kim}\snu
\email{kcs1996kcs@snu.ac.kr}
\author[0000-0002-8055-5465]{Jong-Hak Woo}\snu
\affil{SNU Astronomy Research Center, Seoul National University, 1 Gwanak-ro, Gwanak-gu, Seoul 08826, Republic of Korea}
\author[0000-0003-4509-7822]{Rongxin Luo}\snu\affil{Department of Physics and Astronomy, University of Alabama in Huntsville, Huntsville, AL 35899, USA}
\author[0000-0003-1440-8552]{Aeree Chung}\yonsei
\author[0000-0002-3744-6714]{Junhyun Baek}\yonsei
\author[0000-0003-1270-9802]{Huynh Anh N. Le}
\affil{CAS Key Laboratory for Research in Galaxies and Cosmology, Department of Astronomy, University of Science and Technology of China, Hefei 230026, China}
\author[0000-0002-4704-3230]{Donghoon Son}\snu

\begin{abstract}
We present spatially resolved gas kinematics, ionization, and energetics of 11 type 1 and 5 type 2 active galactic nuclei (AGNs) with strong ionized gas outflows at z $<0.3$ using Gemini Multi-Object Spectrograph Integral Field Unit (GMOS-IFU) data. We find a strongly blueshifted region in [\OIII] velocity maps, representing an approaching cone in biconical outflows, and blueshifted and redshifted regions in H$\alpha$ velocity maps, which show gravitationally rotating kinematics. AGN photoionization is dominant in the central region of most targets, and some of them also show ring-like structures of LINER or composite that surround the AGN-dominated center. Following our previous studies, we kinematically determine outflow sizes by the ratio between [\OIII] and stellar velocity dispersion. Outflow sizes of type 1 AGNs follow the same kinematic outflow size--[\OIII] luminosity relation obtained from the type 2 IFU sample in Kang \& Woo and Luo (updated slope $0.29\pm0.04$), while they are limited to the central kpc scales, indicating the lack of global impact of outflows on the interstellar medium. Small mass outflow rates and large star formation rates of the combined sample support that there is no evidence of rapid star formation quenching by outflows, which is consistent with the delayed AGN feedback.
\end{abstract}


\section{Introduction} \label{sec:intro}
Over the last two decades, the role of active galactic nuclei (AGNs) in galaxy evolution has been one of the main topics for understanding the connection between black holes and host galaxies \citep{Ferrarese00, Kormendy13}. AGNs are thought to regulate star formation via so-called AGN feedback \citep[e.g.,][]{Silk98, Shimizu15, Wylezalek16}. A majority of cosmological simulations implement AGN feedback to prevent overgrowth of massive galaxies and reproduce observed properties \citep[e.g.,][]{Choi15, Weinberger17, Dave19, Wellons22}. Observationally, however, recent studies have reported that AGN luminosity correlates with the star formation rate (SFR), suggesting no sign of instantaneous AGN feedback \citep[e.g.,][]{Netzer09, Woo17, Woo20, Kim22} or perhaps positive AGN feedback \citep{Silk13, Zhuang20, Zhuang21}, while others have suggested no significant trend between AGN and star formation activity \citep{Smirnova-Pinchukova22}. Overall, there is no general consensus on how AGN feedback suppresses (or triggers) star formation due to the absence of direct observational evidence except for a small number of galaxies \citep{Harrison17}.

Various energetic phenomena from AGNs, such as radiation, relativistic jets, and outflows, are potential channels of feedback for ejecting or stabilizing the interstellar medium (ISM). Among them, outflows driven by AGNs are frequently suggested as a promising channel due to its prevalence \citep{Boroson05, Mullaney13, Woo16, Rakshit18} and impact across the host galaxy \citep{Cano-Diaz12, Zubovas12, Cresci15, Rupke17, Baron18, Husemann19}. 

To investigate the physical properties of AGN-driven outflows, spatially resolved spectroscopy data are intensively utilized to map the spatial distribution of warm ionized gas outflows, using, e.g.,  [\OIII] $\lambda 5007$ line emission. Numerous studies mainly targeted relatively luminous AGNs with clear signs of strong ionized outflows \citep[e.g., ][]{Harrison14, Husemann16, Villar-Martin16, Kang18} or the nearest AGNs \citep[e.g., ][]{Mingozzi19, Revalski21, Ruschel-Dutra21}. Large optical IFU surveys such as Mapping Nearby Galaxies at APO (MaNGA), Sydney-AAO Multi-object Integral-field spectroscopy (SAMI), or Calar Alto Legacy Integral Field Area (CALIFA) have enabled statistical studies on spatial properties of the ionized gas kinematics using a large sample of local galaxies \citep{Wylezalek20, Deconto-Machado22, Oh22}. Other studies focused on resolving detailed geometries of the outflows in individual nearby AGNs with high spatial resolution and large field-of-view (FoV) IFU instruments such as Multi Unit Spectroscopic Explorer on the Very Large Telescope \citep[VLT/MUSE, ][]{Mingozzi19, Shin21, Juneau22}. Recently, high-resolution near-infrared (NIR) IFU data using adaptive optics or James Webb Space Telescope (JWST) revealed extreme ionized outflows in luminous quasars at cosmic noon \citep[][]{Kakkad20, Vayner21a, Wylezalek22, Veilleux23, Vayner23, Kakkad23, Cresci23}. 

In addition, submillimeter arrays, e.g., the Atacama Large Millimeter/submillimeter Array (ALMA), have been used to probe various cold molecular outflows in AGNs or ultraluminous infrared galaxies \citep[][]{Cicone14, Lamperti22}, while NIR IFU studies have found warm molecular outflows in local AGNs \citep[e.g., ][]{Ramos-Almeida19, Riffel23}. Molecular outflows have been considered as another crucial ingredient in AGN feedback since the mass outflow rates are considerably larger than those of the ionized outflows \citep[e.g., ][]{Fiore17, Fluetsch19}.

These studies have been performed to reveal the connection with star formation \citep{Cano-Diaz12, Harrison14, Carniani15, Carniani16, Maiolino17, Shin19, Scholtz20, Scholtz21, Bessiere22}, multiphase outflows \citep{Fiore17, Husemann19, Fluetsch19, Bianchin22, Riffel23} and the scaling relations of the size and energetics \citep[][]{Villar-Martin16, Fiore17, Kang18, Baron19, Fluetsch19, Wylezalek20, Ruschel-Dutra21, Kakkad22, Molina22}. Now we have a more complex picture of AGN-driven outflows, which contains diverse sizes, energetics, multiphase, and both positive and negative feedback effects on star formation. Thus, to test the feedback by the AGN-driven outflows, it is essential to apply consistent measurements of the outflow size and energetics to a well-constrained AGN sample.

In a series of papers, we have investigated the spatially resolved properties of AGN-driven outflows in local type 2 AGNs ($z\lesssim0.1$) with strong (or weak) outflows using optical IFU data, including the Gemini Multi-Object Spectrograph Integral Field Unit (GMOS-IFU) \citep{Karouzos16a, Karouzos16b, Bae17, Kang18, Luo19, Luo21}. The main results from these studies are summarized as follows. First, we commonly found blueshifted [\OIII] emission lines at the central region, manifesting the approaching cone in the biconical outflow model. Second, the AGN photoionization region is dominant at the central region and frequently surrounded by the low-ionization nuclear emission-line region (LINER), composite and star formation ionization regions. Third, we developed the methodology of determining outflow size based on kinematics, finding that this kinematically-determined outflow size correlates with [\OIII] luminosity. Last, outflow sizes are relatively small (i.e., a few to several kpc) in general, compared to large galaxy scales, suggesting no global impact of the outflows. While our sample is limited to type 2 AGNs and lacks very luminous ($\mathrm{L}_{[\mathrm{OIII}]}\gtrsim10^{42}\;\mathrm{erg\;s}^{-1}$) and more extreme outflows, our conclusion is based on relatively luminous local AGNs.

In this paper, we enlarge our GMOS-IFU sample to type 1 AGNs for a comprehensive view in terms of AGN unification and present our overall understanding of the nature and role of AGN-driven outflow in galaxy evolution based on the combined sample of type 1 and type 2 AGNs. This type 1 sample also represents luminous and strong outflow targets compared to the type 2 sample in previous studies. In Section \ref{sec:samp_obs}, we summarize the sample properties, IFU and submillimeter observations. Section \ref{sec:Analysis} describes the multicomponent spectral modeling routine and spectral energy distribution (SED) fitting method for SFR estimation. In Section \ref{sec:result}, we show spatially resolved properties. We discuss our implications for AGN feedback in Section \ref{sec:discussion} by focusing on kinematic outflow sizes, narrow line region (NLR) sizes, mass outflow rates and SFRs. Finally, we summarize our key results in Section \ref{sec:summary}.
In this work, all calculations are performed based on the assumption of a flat, $\Lambda$ cold dark matter ($\Lambda$CDM) cosmology with $H_0=70\;\mathrm{kms}^{-1}\mathrm{Mpc}^{-1}$, $\Omega_m=0.3$ and $\Omega_\Lambda=0.7$.

\section{Sample and Observation} \label{sec:samp_obs}
\subsection{Sample Selection}\label{subsec:sample}

\begin{figure}[t]
    \includegraphics[width=\columnwidth]{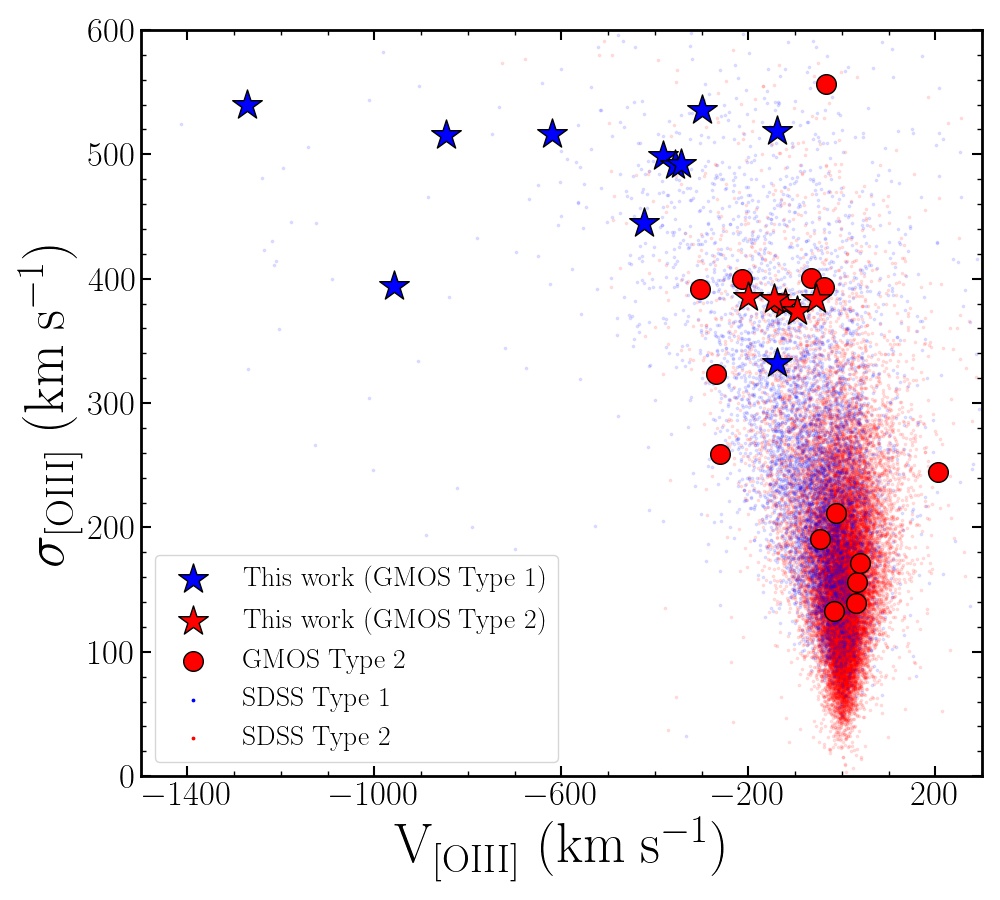}
    \caption{[\OIII] Velocity--velocity dispersion (VVD) diagram of the GMOS sample, which is composed of type 1 (blue) and type 2 AGNs (red stars) in this study and type 2 AGNs (red circles) in our previous studies \citep{Karouzos16a, Karouzos16b, Kang18, Luo19}. Blue and red dots represent the parent SDSS samples of type 1 AGNs \citep{Rakshit18} and type 2 AGNs \citep{Woo16}, respectively.}
    \label{F1_VVD}
\end{figure}

We selected a total of 11 type 1 AGNs from the parent sample of \citet{Rakshit18}, which consists of $\sim$ 5,000 Sloan Digital Sky Survey (SDSS) type 1 AGNs at z $<0.3$. Based on SDSS spectra, \citet{Rakshit18} statistically analyzed [\OIII] kinematics and compared them with bolometric AGN luminosity and the Eddington ratio. To investigate the spatially resolved properties of energetic type 1 AGNs, we first selected five AGNs with high 5100$\mathrm{\AA}$ luminosity ($\mathrm{L}_{5100} > 10^{43.5}\;\mathrm{erg\;s}^{-1}$) and strong outflow signatures ($\mathrm{log}(\sigma_{[\mathrm{OIII}]}/\mathrm{M}_*^{1/4})>-0.2$) for a pilot study. Note that $\sigma_{[\mathrm{OIII}]}/\mathrm{M}_*^{1/4}$ is a proxy of outflow strength, which is the [\OIII] velocity dispersion normalized by the gravitational potential of the total stellar mass. Among the five AGNs, two type 1 AGNs at z $\sim$ 0.1 were observed by GMOS-N IFU in the 2019A semester (ID: GN-2019A-Q-201, PI: Woo).

As a follow-up observation, we selected nine more type 1 AGNs with strong outflows at higher redshifts ($0.2<z<0.3$). We applied a higher 5100$\mathrm{\AA}$ luminosity criterion by $10^{44.5} \mathrm{erg\;s}^{-1}$ and the same outflow strength cut by $\mathrm{log}(\sigma_{[\mathrm{OIII}]}/M_*^{1/4})>-0.2$. All nine targets were observed during the 2019B semester by three Gemini science programs (ID: GN-2019B-Q-137, GN-2019B-Q-236, GN-2019B-Q-327, PI: Woo).

It is worth noting that there are five narrow-line Seyfert 1 galaxies (NLS1) and one post-starburst galaxy (PSB) with AGN classified by published catalogs \citep[See Table \ref{T1_info}, ][]{Cracco16, Rakshit17, Wei18}. NLS1s are thought to be a rapidly growing SMBH stage with generally small black hole masses and high Eddington ratios. PSB is an abrupt transition phase from star-forming galaxy to quiescent galaxy within recent 0.1--1 Gyr.

In addition, we also included five type 2 AGNs with strong ionized gas outflows in this study, which were not dealt with in our series of type 2 IFU studies. These AGNs were selected from the parent sample of \citet{Bae14}, which has $\sim23,000$ type 2 AGNs at $z<0.1$ based on SDSS spectra. We applied [\OIII] kinematics cut by velocity offset $>$ 250 km s$^{-1}$ or [\OIII] velocity dispersion $>$ 400 km s$^{-1}$ and dust-corrected [\OIII] luminosity cut by $L_\mathrm{[OIII], cor}>10^{42} \;\rm{erg\;s}^{-1}$ and constructed a sample of 29 type 2 AGNs. Previously, nine of them were observed in the 2015A and 2015B semesters and analyzed in \citet{Karouzos16a, Karouzos16b} and \citet{Kang18}. Five type 2 AGNs were observed by both Gemini-North and -South in the 2016B semesters (ID: GN-2016B-Q-30, GS-2016B-Q-30, PI: Woo).

\begin{figure*}[]
    \centering
    \scalebox{0.95}{\includegraphics[width=0.99\textwidth]{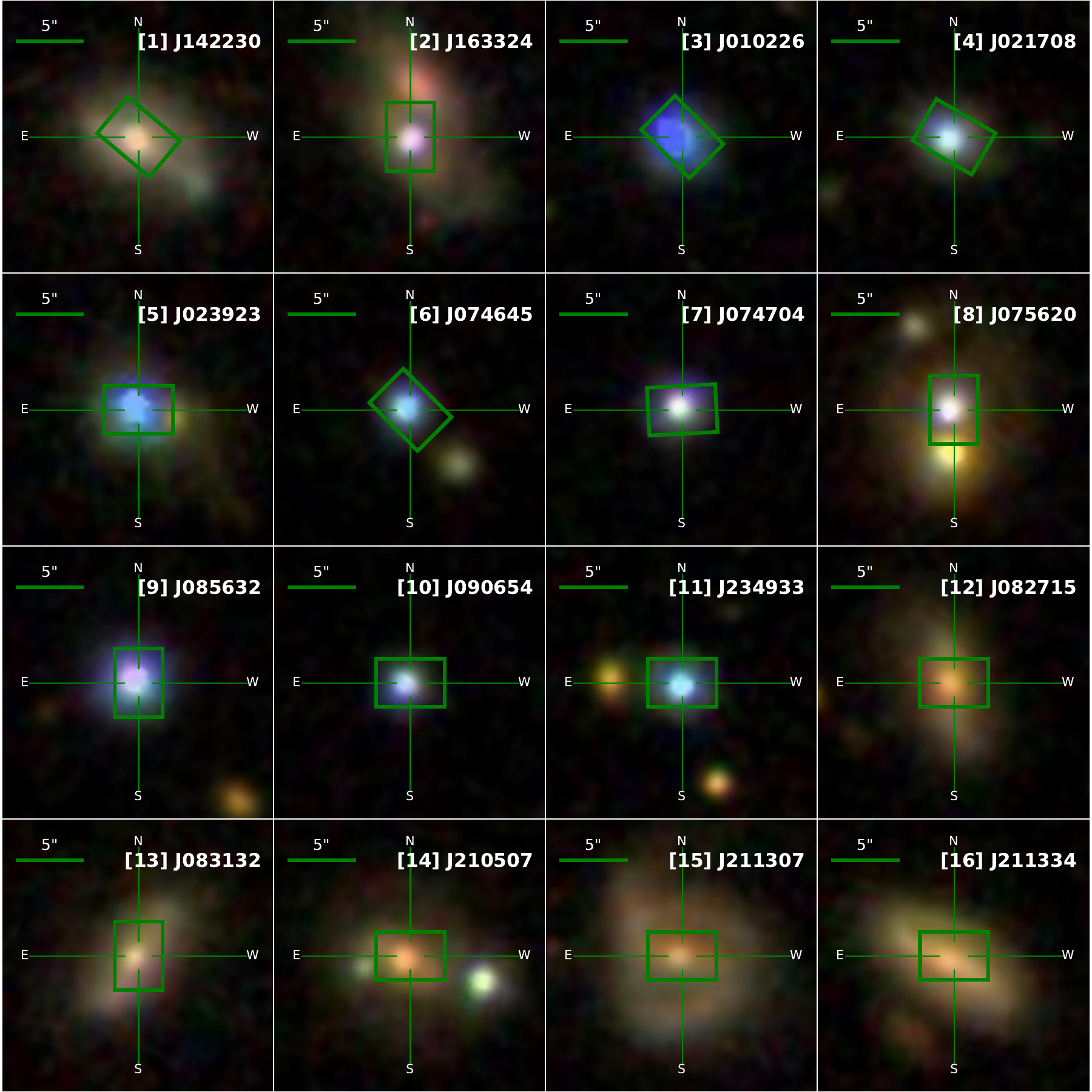}}
    \caption{SDSS Optical $gri$-composite images of the GMOS-IFU sample in this work. Each panel shows a 20$\arcsec \times 20\arcsec$ scale, while the FOV of GMOS-IFU (3\farcs5$\times$5\farcs0) is denoted with a green box.}
    \label{F2_SDSS_images}
\end{figure*}

In Figure \ref{F1_VVD}, we plot the [\OIII] velocity--velocity dispersion diagram of our sample overlaid with the parent samples from \citet{Woo16} and \citet{Rakshit18}. Our sample (star markers) is clearly located at the extreme end of [\OIII] kinematic properties among the parent samples, especially type 1 AGNs (blue stars). Moreover, some of our targets show interacting pairs or disturbed morphology in SDSS images (see Figure \ref{F2_SDSS_images}). We will discuss these host galaxy properties with respect to their kinematic results in terms of AGN feedback through the evolutionary track of galaxies.

\begin{deluxetable*}{cccccccccc}
    \tablecolumns{9}
    \tablecaption{Target Information}\label{T1_info}
    \tablehead{
       & Name & SDSS ID  & z &  Type & obs.  & Tel &  T$_\mathrm{exp}$ & Seeing size  & Remark \\
       &      &          &   &       &       &     &    [sec]   & [\arcsec]    &  \\
       {[}1{]} & {[}2{]} & {[}3{]} & {[}4{]} & {[}5{]} & {[}6{]} & {[}7{]} & {[}8{]} & {[}9{]} & {[}10{]}
       }
        \startdata
        1  & J142230 &J142230.34+295224.2 & 0.114 & Type 1 & 19A & Gemini-North & 5520 & 0.4 & Post-starburst\tablenotemark{a}       \\
        2  & J163324 &J163323.58+471858.9 & 0.116 & Type 1 & 19A & Gemini-North & 5754 & 0.6 & NLS1\tablenotemark{c} \;\& Interacting\\
        3  & J010226 &J010226.31-003904.5 & 0.291 & Type 1 & 19B & Gemini-North & 2000 & 0.8 & NLS1\tablenotemark{b}   \\
        4  & J021708 &J021707.87-084743.5 & 0.291 & Type 1 & 19B & Gemini-North & 4800 & 0.7 & -                       \\
        5  & J023923 &J023922.87-000119.6 & 0.262 & Type 1 & 19B & Gemini-North & 2340 & 1.0 & Interacting             \\
        6  & J074645 &J074644.79+294059.0 & 0.293 & Type 1 & 19B & Gemini-North & 7290 & 0.7 & NLS1\tablenotemark{b,c} \\
        7  & J074704 &J074704.43+483833.3 & 0.221 & Type 1 & 19B & Gemini-North & 7488 & 0.7 & -                       \\
        8  & J075620 &J075620.07+304535.4 & 0.236 & Type 1 & 19B & Gemini-North & 3600 & 0.6 & NLS1\tablenotemark{b} \;\& Interacting  \\
        9  & J085632 &J085632.40+504114.0 & 0.235 & Type 1 & 19B & Gemini-North & 3600 & 0.7 & -                       \\
        10 & J090654 &J090654.48+391455.3 & 0.241 & Type 1 & 19B & Gemini-North & 9000 & 0.7 & NLS1\tablenotemark{c}   \\
        11 & J234933 &J234932.77-003645.8 & 0.279 & Type 1 & 19B & Gemini-North & 2700 & 0.9 & - \\
        12 & J082715 &J082714.96+263618.2 & 0.094 & Type 2 & 16B & Gemini-North & 8820 & 0.7 & - \\
        13 & J083132 &J083132.28+160143.3 & 0.065 & Type 2 & 16B & Gemini-North & 8820 & 0.5 & - \\
        14 & J210507 &J210506.94+094118.8 & 0.096 & Type 2 & 16B & Gemini-South & 5820 & 0.7 & Interacting \\
        15 & J211307 &J211307.21+005108.4 & 0.071 & Type 2 & 16B & Gemini-South & 8820 & 0.7 & - \\
        16 & J211334 &J211333.79-000950.4 & 0.073 & Type 2 & 16B & Gemini-South & 2700 & 0.8 & - \\
        \enddata
    \tablecomments{(1) Target number.; (2) Name.; (3) SDSS ID.; (4) redshift from SDSS DR 14.; (5) AGN Type.; (6) Observed semester.; (7) Telescope.; (8) Science exposure time.; (9) Seeing size.; (10) Target classification.}
    \tablenotetext{a}{Post-starburst quasar \citep{Wei18}}
    \tablenotetext{b}{Narrow-line Seyfert 1 \citep{Cracco16}}
    \tablenotetext{c}{Narrow-line Seyfert 1 \citep{Rakshit17}}
\end{deluxetable*}

\subsection{GMOS-IFU Observation and Data Reduction}\label{subsec:obs}
In this section, we briefly introduce our GMOS-IFU observations of 11 type 1 AGNs in 2019A and 2019B and 5 type 2 AGNs in 2016B. We carried out Gemini-North and -South/GMOS-IFU one-slit mode observations, which cover a 3\farcs5$\times$5\farcs0$\;$ FoV, to secure a wide wavelength range including both [\OIII] and H$\alpha$. The FoV of GMOS-IFU corresponds to physical scales of $\sim$ 10--22 kpc for type 1 and $\sim$ 6--9 kpc for type 2. We adopted the B600 grating at an optimal central wavelength depending on the target's redshift. The B600 grating has a spectral resolution of R $\sim$ 1400 in IFU mode since we used 2 pixel spectral binning to achieve a higher signal-to-noise ratio (S/N). This is converted to $\sigma_\mathrm{inst} \sim 90$ km s$^{-1}$ on the velocity scale, which is enough to resolve narrow emission lines from AGNs. Observations were conducted under the condition of a typical seeing size of 0\farcs7--0\farcs8 FWHM.

Note that most targets were observed by Gemini-North except for three type 2 AGNs (see Table \ref{T1_info}). Due to technical issues in CCD, Gemini-South data show several saturated columns that affect the data reliability at certain wavelengths. While [14]J210507 is not affected by the issue in any important emission lines, [15]J211307 and [16]J211334 are affected in [\NII] $\lambda$6548 and H$\alpha$ $\lambda$6563, respectively. We masked these saturated columns for these two targets in emission line fitting.

We used the Gemini \textit{IRAF} package for the data reduction of five type 2 AGNs observed in 2016, as we did in our previous studies \citep{Karouzos16a, Kang18, Luo19}. For 11 type 1 AGNs observed in 2019, we adopted the newly modified Py3D pipeline, which was originally designed to reduce the data from CALIFA survey \citep{Husemann13} but was also adapted for GMOS-IFU data. We followed a standard GMOS-IFU reduction routine, including bias subtraction, fiber extraction, twilight flat fielding, cosmic-ray removal with Pycosmic \citep{Pycosmic}, wavelength calibration with Cu-Ar lamp, and sky subtraction. Flux calibration was performed using the standard stars, which were observed together with science targets. Extracted spectra from fibers were finally reconstructed into a 3D data cube based on the drizzle algorithm. We applied consistent spatial sampling (0\farcs1) for type 2 AGNs with our previous studies. For type 1 AGNs, however, we increased the spaxel size to 0\farcs2 to achieve a higher S/N of emission lines.

\subsection{SCUBA-2 Observation and Data Reduction}\label{subsec:jcmtobs}
To assist SED modeling, we also carried out submillimeter continuum observations at 450 and 850 $\mu$m using the Submillimeter Common User Bolometer Array 2 (SCUBA-2) in the James Clerk Maxwell Telescope (JCMT). Since we aim to measure dust luminosity based on SED analysis, these fluxes (or 3$\sigma$ upper limits if there is no detection) efficiently provide constraints at the Rayleigh-Jeans tail of dust SED. Here, we present a short summary of our submillimeter observations and data reduction steps \citep[for details, see ][]{Kim22}.

Among the 11 type 1 AGNs, one ([1]J142230) target was already observed in the 2019A semester (M19AP057; PI: Woo), and we additionally observed six more type 1 AGNs in the 2021A semester (E21AK001; PI: Kim). All type 2 AGNs were observed in the 2018B semester (M18BP070; PI: Woo). The minimum and maximum integration times are 0.5 and 4 hrs, respectively.

Since our target galaxies have relatively small angular sizes, all observations were conducted in Daisy mapping mode, which is optimal for compact objects due to its high sensitivity at the center. Data were reduced by the SCUBA-2 data pipeline \texttt{ORAC-DR} \citep{ORACDR15} in the \texttt{Starlink} software \citep{Starlink14}, with the \texttt{REDUCE} \texttt{SCAN} \texttt{FAINT} \texttt{POINT} \texttt{SOURCES} recipe. This recipe automatically reduces the data, including the matched filter and point-spread function (PSF) convolution steps. After the reduction, we cropped the images by a 120\arcsec$\;$ radius circle and made overlay images with optical data. Finally, we checked whether there is a $>3\sigma$ clump cospatial with the optical emission in the overlaid images and determined the detection or upper limit. 
Since there were no detected sources in the sample, we did not perform clump-finding or flux extraction. Instead, we utilized 3$\sigma$ values as upper limits in SED analysis.

\section{Analysis} \label{sec:Analysis}
\begin{figure*}[]
    \includegraphics[width=0.49\textwidth]{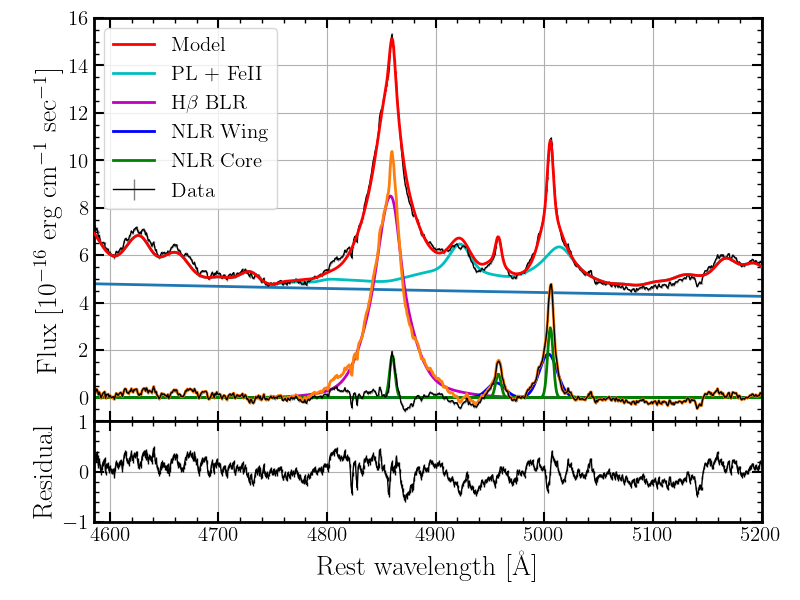}
    \includegraphics[width=0.49\textwidth]{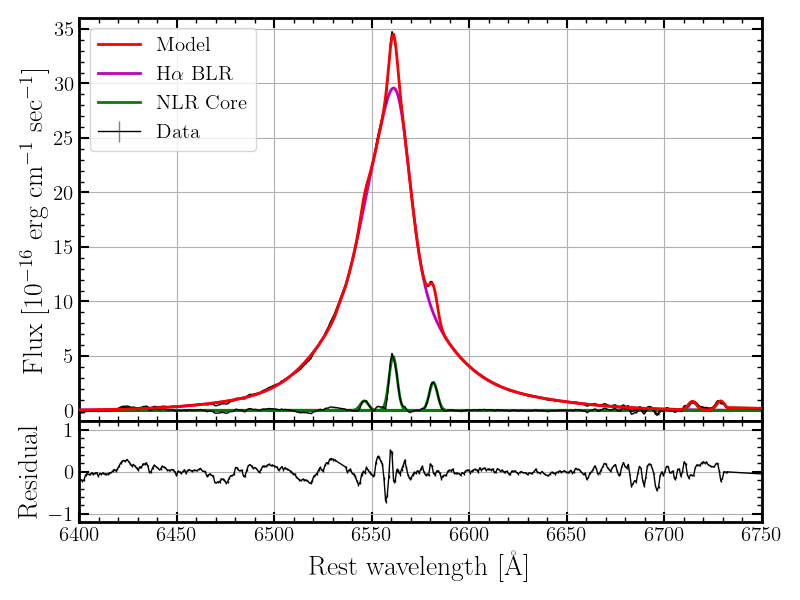}
 \caption{Spectral fitting examples of type 1 AGN ([7]J074704). Left: H$\beta$ \& [\OIII] region. The best-fit model (red) is compared with the observed spectrum (black). NLR core (green), NLR wing (blue), BLR (magenta), and the continuum emission (power-law + \FeII, cyan) are fitted simultaneously. Right: H$\alpha$, [\NII] \& [\SII] region. The power-law continuum is subtracted before the emission line fitting. The best-fit model (red) is compared with the continuum-subtracted residual spectrum (black). NLR core (green) and BLR (magenta) are included in the model for this target.}
    \label{F3_Type1_Fitting_example}
\end{figure*}

\begin{figure*}[]
    \includegraphics[width=0.49\textwidth]{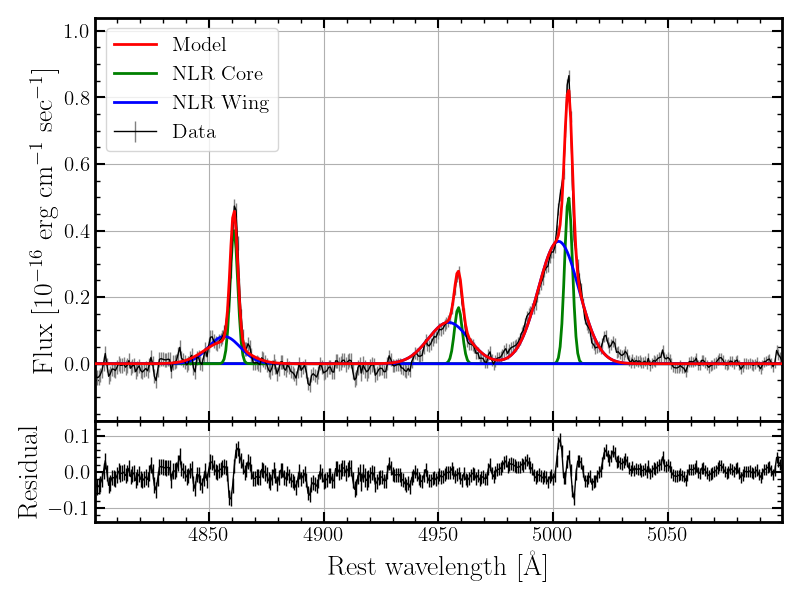}
    \includegraphics[width=0.49\textwidth]{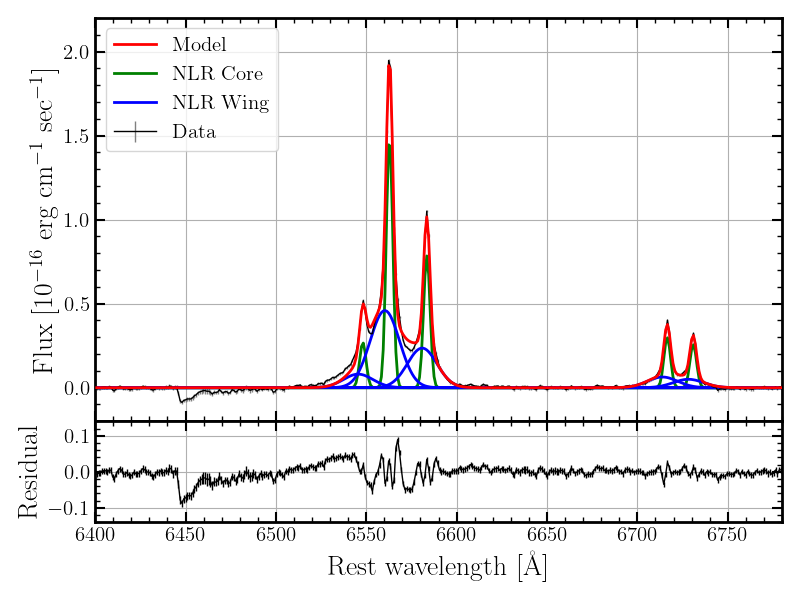}
\caption{Spectral fitting examples of type 2 AGN ([13]J083132). The stellar continuum is subtracted before the emission line fitting. The best-fit models (red) are compared with the continuum-subtracted residual spectra (black) for H$\beta$ \& [\OIII] region (left panel) and H$\alpha$, [\NII] \& [\SII] region (right panel). NLR core (green) and wing (blue) components are included in the model.}
    \label{F3_Type2_Fitting_example}
\end{figure*}

Here, we describe our spectral modeling routine for GMOS-IFU data cubes (see Figure \ref{F3_Type1_Fitting_example}, \ref{F3_Type2_Fitting_example} for fitting examples). All spectral fitting was performed using the Markov Chain Monte Carlo (MCMC) python package \texttt{emcee} \citep{emcee}. We also provide a short summary of the SED fitting method of photometric data in Section \ref{subsec:SEDfitting}.

\subsection{Emission Line Fitting}\label{subsec:IFUfitting}

We decomposed the type 1 AGN spectra into three components: narrow emission lines, broad emission lines and continuum. All narrow and broad emission lines were modeled with a multiple Gaussian profile. For the continuum, we applied a summation of a single power-law model for the accretion disk and a Gaussian-velocity-convolved \FeII template \citep{BG92_FeII} for the pseudocontinuum of broad \FeII emission. Note that the \FeII template is added only to H$\beta$-[\OIII] region fitting since the \FeII contribution is negligible in the H$\alpha$-[\NII]-[\SII] region. The flux ratio of [\OIII] ($\lambda$4959 \& $\lambda$5007) and [\NII] ($\lambda$6548 \& $\lambda$6583) doublets are fixed as 2.993 \& 2.94, respectively, to reduce the number of free parameters. We did not include the stellar continuum from the host galaxy since the accretion disk and \FeII emission dominate the continuum spectrum.

Since broad line regions (BLRs) are spatially unresolved, we assume that the velocity and velocity dispersion of broad emission lines do not change between spaxels. Thus, we first fitted the integrated spectrum, which was extracted from the central nine pixels, to obtain the velocity and velocity dispersion of each broad emission line (i.e., H$\alpha$, H$\beta$, and \FeII). Then, we fitted the spectra of individual pixels to determine the velocity and velocity dispersion of narrow emission lines, while we used the fixed value of the velocity and velocity dispersion of broad emission lines from the central spaxels and only adjusted the flux levels.
In the case of narrow emission lines, we first fitted them with a double Gaussian profile, consisting of a broad wing component and a narrow core component. If one of the Gaussian components showed an amplitude-to-noise ratio (A/N) less than three, we rejected the fitting result and refitted the line with a single Gaussian model.

For type 2 AGNs, we fitted narrow emission lines and stellar continuum from the host galaxy, following the same procedure as described in our previous IFU studies \citep{Karouzos16a, Kang18, Luo19, Luo21}. First, we modeled the continuum emission from stars using the python code pPXF \citep{ppxf17} and E-MILES simple-stellar population (SSP) templates \citep{EMILES16}. We used 47 template spectra with different ages from 0.06 to 12.6 Gyr at solar metallicity. After masking emission lines, we determine the best-fit stellar continuum and subtract it from each spectrum in each spaxel. Second, we fitted narrow emission lines with a single or double Gaussian model as adopted for the type 1 AGNs.

Based on the best-fit parameters, the velocity offset and velocity dispersion of [\OIII] and H$\alpha$ lines were calculated using the definition of the first and second moments of the total line profile as:

\begin{equation}
    \lambda_0 = \frac{\int \lambda f_\lambda d\lambda}{\int f_\lambda d\lambda}
\end{equation}

\begin{equation}
    \Delta\lambda^2 = \frac{\int \lambda^2 f_\lambda d\lambda}{\int f_\lambda d\lambda}-\lambda^2_0
\end{equation}
We calculated velocity offsets with respect to the systematic velocity, which was determined based on the flux-weighted stellar velocity or the [\OIII] narrow component of the integrated spectrum of central pixels. To correct for the instrumental resolution, we subtracted $\sigma_{inst} \sim 93.5$ km s$^{-1}$ from the measured velocity dispersion in quadrature.

\subsection{SED Fitting}\label{subsec:SEDfitting}
To determine global SFRs, we performed multiwavelength SED fitting using a python Code Investigating GALaxy Emission \citep[CIGALE\footnote{http://www.oamp.fr/cigale}][]{CIGALE09,CIGALE19}. We previously performed SED fitting and measured the global SFRs based on the dust luminosity of 52 local AGNs \citep{Kim22}, including six AGNs from our new sample (see Table \ref{T2_outflowinfo}) and nine type 2 AGNs from our previous studies \citep{Karouzos16a, Karouzos16b, Kang18}. Thus, we newly carried out SED fittings of the remaining 10 targets with the same method. In addition, we performed SED fittings of 14 AGNs from \citet{Harrison14}, of which outflow sizes were analyzed by \citet{Kang18}. In total, we obtained the dust luminosity-based SFRs of 39 AGNs with strong outflows for this study.

As \citet{Kim22} described the SED fitting procedure in detail, here we briefly summarize the fitting method. We first constructed a multiwavelength photometric dataset from the UV to radio range, including new submillimeter observations with the JCMT/SCUBA-2. We collected archival data from the NASA/IPAC Extragalactic Database \footnote{https://ned.ipac.caltech.edu/}, which provides UV (far- and near- UV) data from the Galaxy Evolution Explororer (GALEX), optical ($ugriz$) data from SDSS, near-IR (J, H, K) data from Two Micron All Sky Survey Extended Source Catalog (2MASS), Wide-field Infrared Survey Explorer data from the ALLWISE Source Catalog, the Infrared Astronomical Satellite Faint Source Catalog, and 1.4 GHz from the NRAO VLA Sky Survey \citep{VLASS98}. We also found far-infrared fluxes provided by the AKARI/Far-infrared Surveyor all-sky bright-source catalog \citep{AKARI10} and the Herschel/PACS \& SPIRE point-source catalog \citep{Herschel-PACS17, Herschel-SPIRE17}.

Using multiwavelength photometry, we performed SED fittings with the same modules and parameter space as listed in Table 3 in \citet{Kim22}. The SED model contains multiple emission components, including stellar emission, AGN, dust emission, radio and nebular emission. In the fitting, CIGALE first builds all SED models from the input parameter space ($\sim$ 1.5 million cases), and then it calculates the likelihoods of all calculated models. Using the likelihoods, CIGALE estimates physical parameters in the Bayesian framework, exp$^{-\chi^2/2}$-weighted mean and error. We determined the SFRs of our sample based on dust luminosity from SED modeling using the following equation from \citet{Kennicutt98} as revised for the Chabrier initial mass function (IMF):
$$\mathrm{log}\;\mathrm{SFR}\;(\mathrm{M}_\odot \mathrm{yr}^{-1})=\mathrm{log}\;\mathrm{L}_{\mathrm{IR}}\;(\mathrm{erg\;s}^{-1}) - 43.591$$

\section{Result} \label{sec:result}
\subsection{Flux Distribution}\label{subsec:flux}
\begin{figure*}[]
\centering
\includegraphics[width=\textwidth]{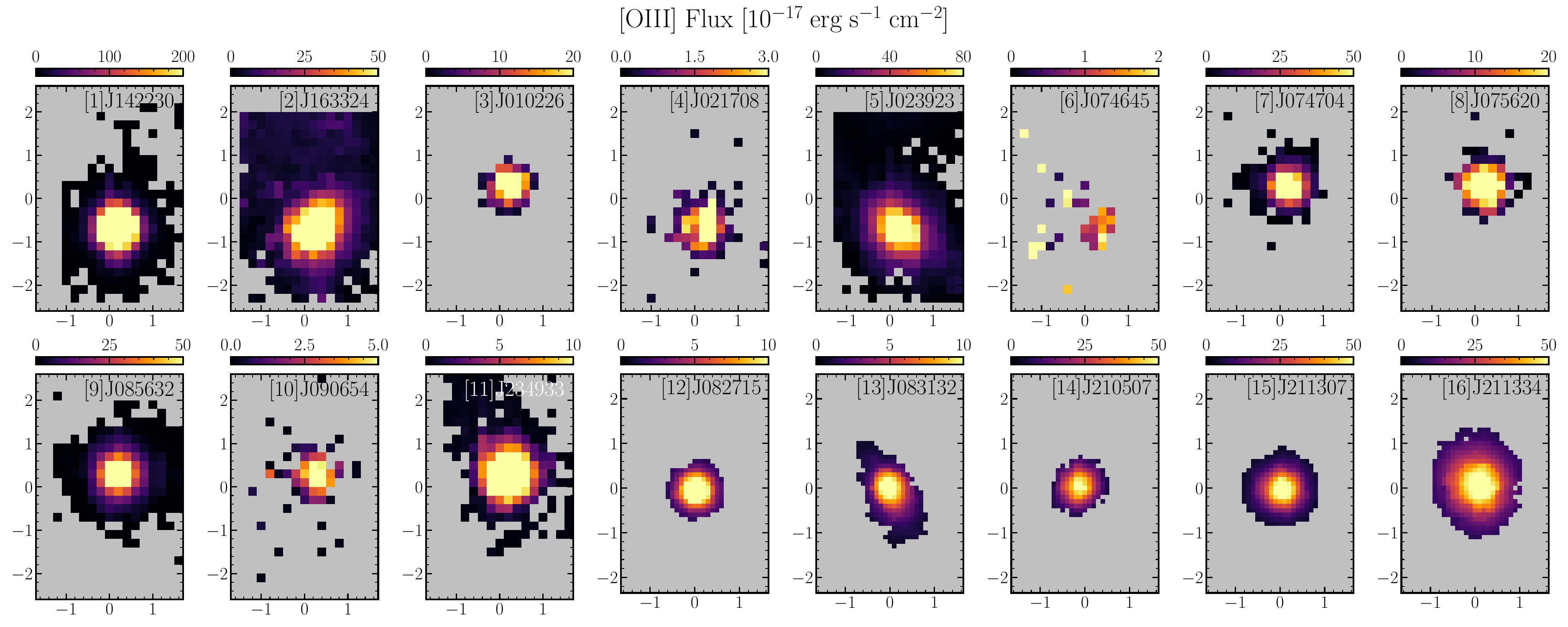}\\
\includegraphics[width=\textwidth]{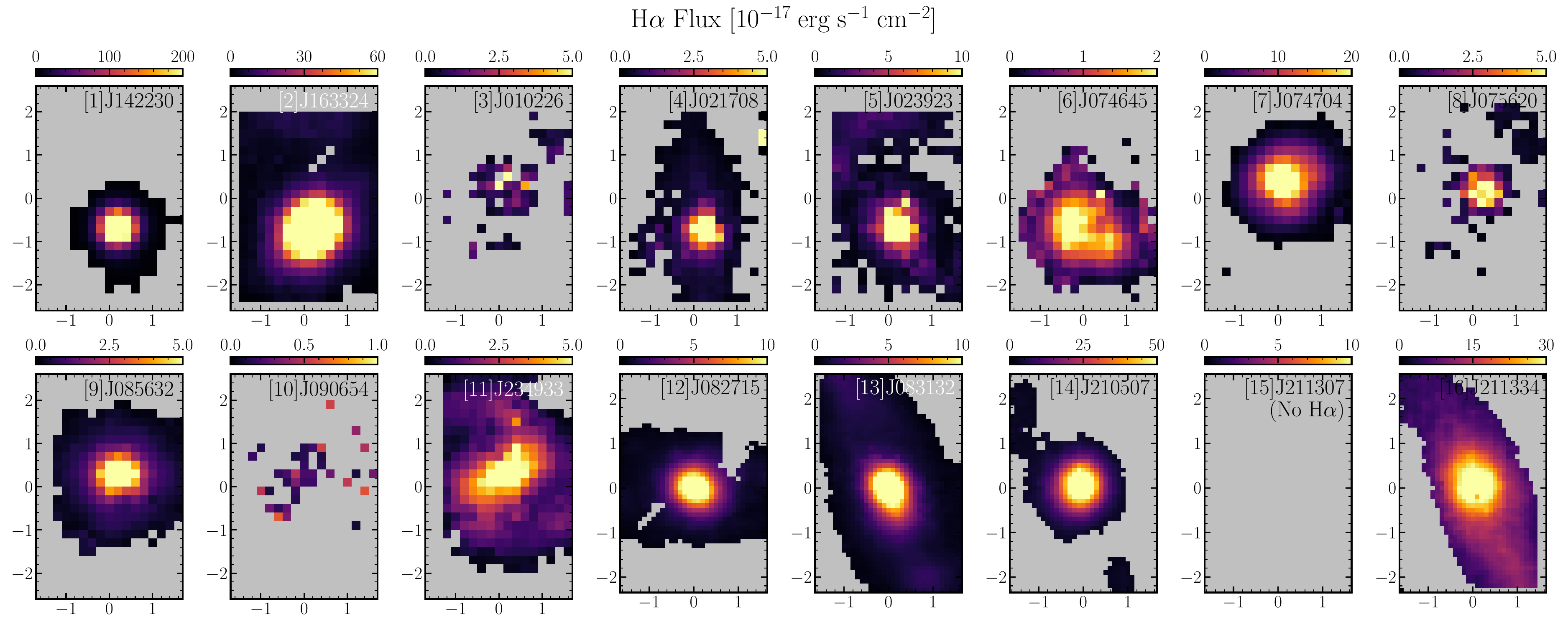}
\caption{Flux maps of [\OIII] (top) and H$\alpha$ (bottom) emission lines. The gray area represents the spaxels where the emission line is not detected (i.e., A/N $<$ 3). Note that H$\alpha$ flux map of [15]J211307 is not shown here due to the saturated columns in the CCD.}
\label{F3_Fluxmaps}
\end{figure*}

We present the total flux distribution of the [\OIII] and H$\alpha$ emissions from the NLR in Figure \ref{F3_Fluxmaps}, excluding pixels with an A/N ratio less than three. The H$\alpha$ flux map of [15]J211307 is excluded due to the CCD saturation issue in the observed H$\alpha$ wavelength.

In general, emission line regions are extended to several kpc scales, while the H$\alpha$ tends to be more extended than the [\OIII]. Interacting galaxies, such as [2]J163324 and [5]J023923, show additional emission line sources at the upper region in the GMOS FOV, indicating the emission from the interacting pair. Barred galaxies (e.g., [13]J083132 and [16]J211334) show elongated features in the H$\alpha$ flux maps, probably due to star-forming regions along the galactic bar.

\subsection{Ionized Gas Kinematics}\label{subsec:kinematics}
\begin{figure*}[]
\centering
\includegraphics[width=\textwidth]{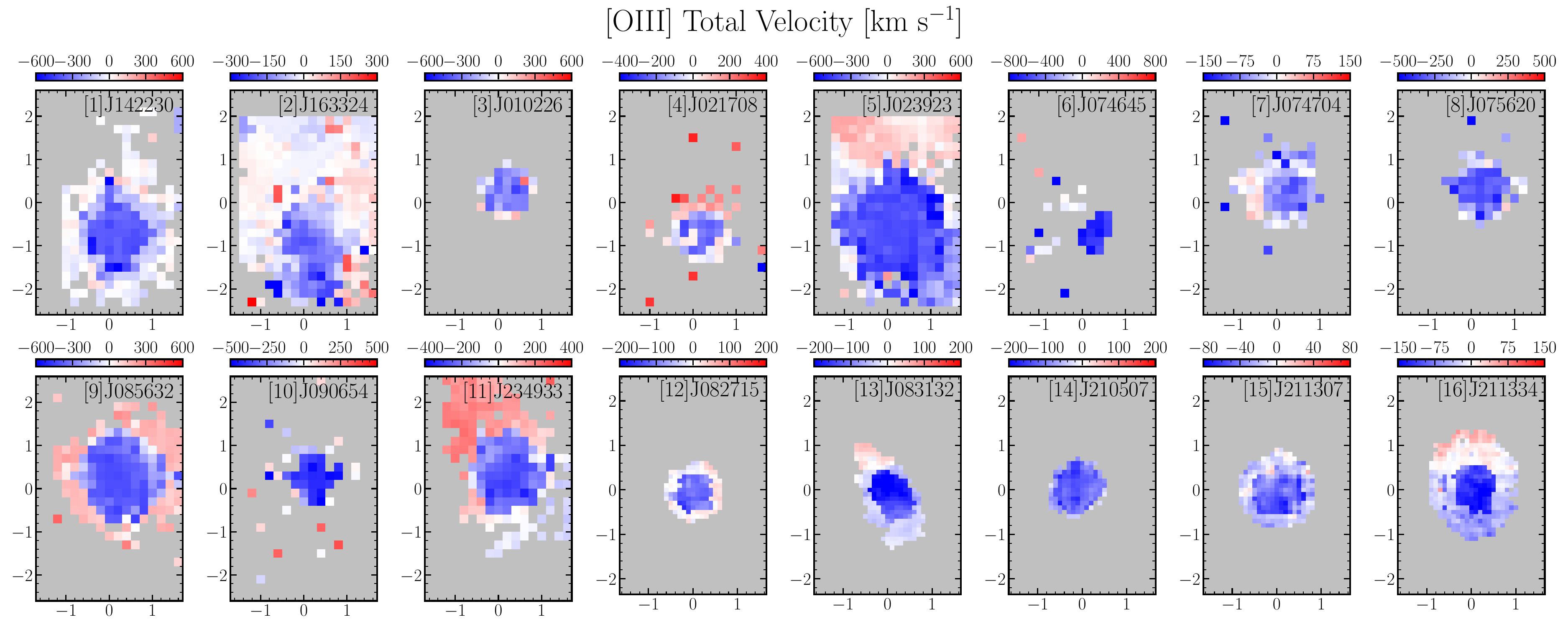}\\
\includegraphics[width=\textwidth]{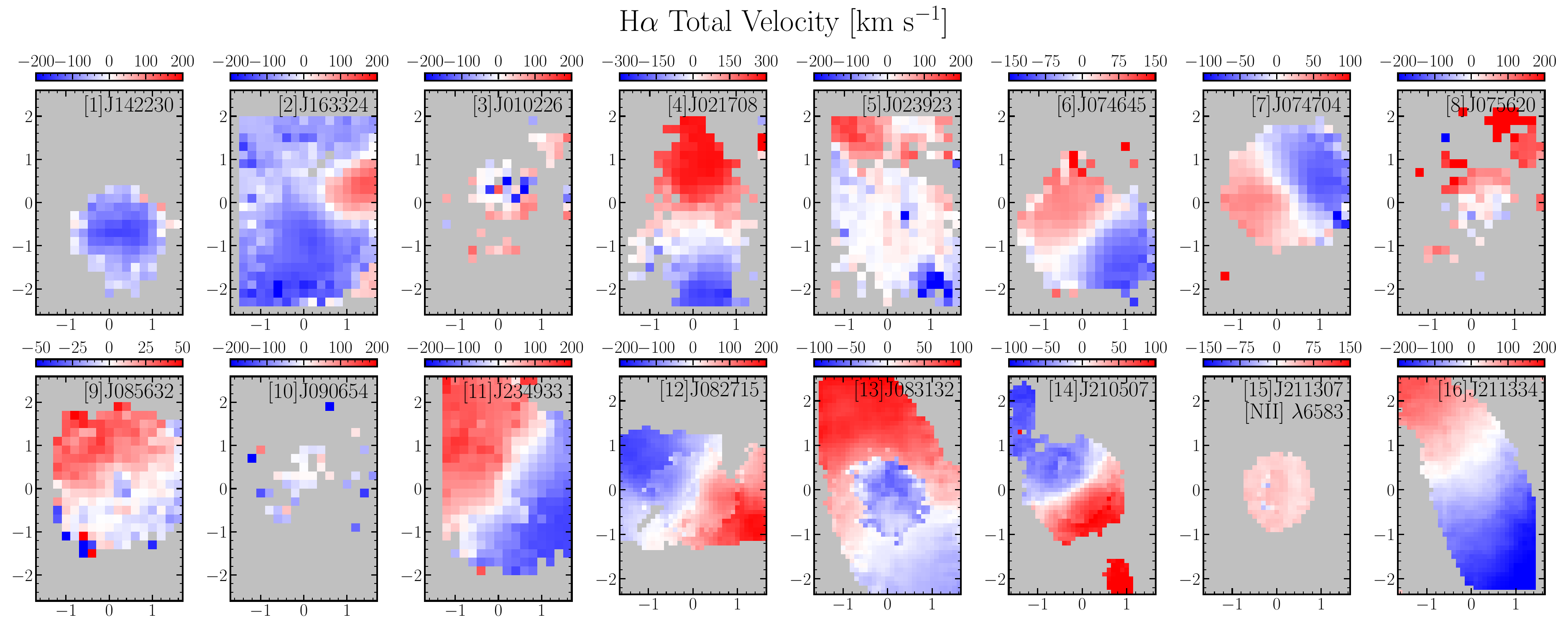}
\caption{Velocity maps of [\OIII] (top) and H$\alpha$ (bottom) emission lines. The blue color indicates a blueshift, i.e., approaching kinematics, while the red color indicates a redshift, i.e., receding kinematics. Note that H$\alpha$ velocity map of [15]J211307 is replaced with [\NII] due to the saturated columns in the CCD.}
\label{F4_Velocitymaps}
\end{figure*}

\begin{figure*}[]
\centering
\includegraphics[width=\textwidth]{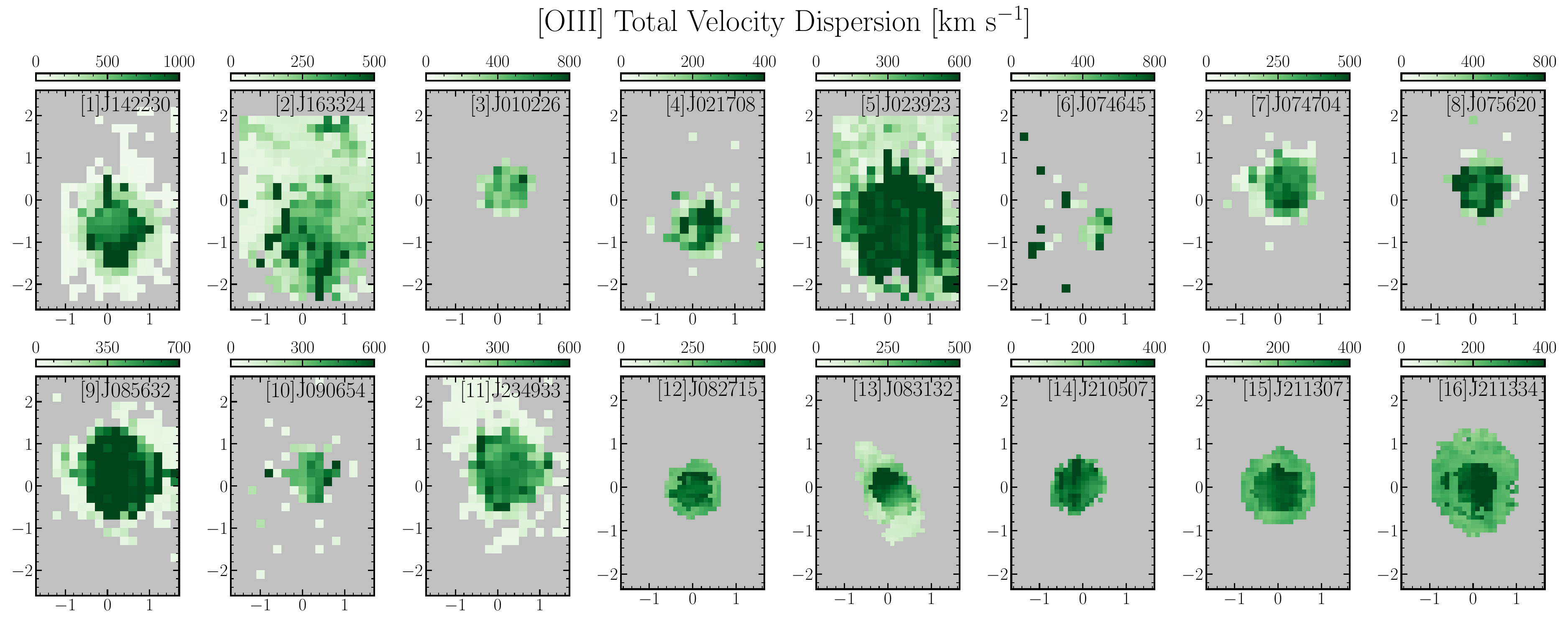}\\
\includegraphics[width=\textwidth]{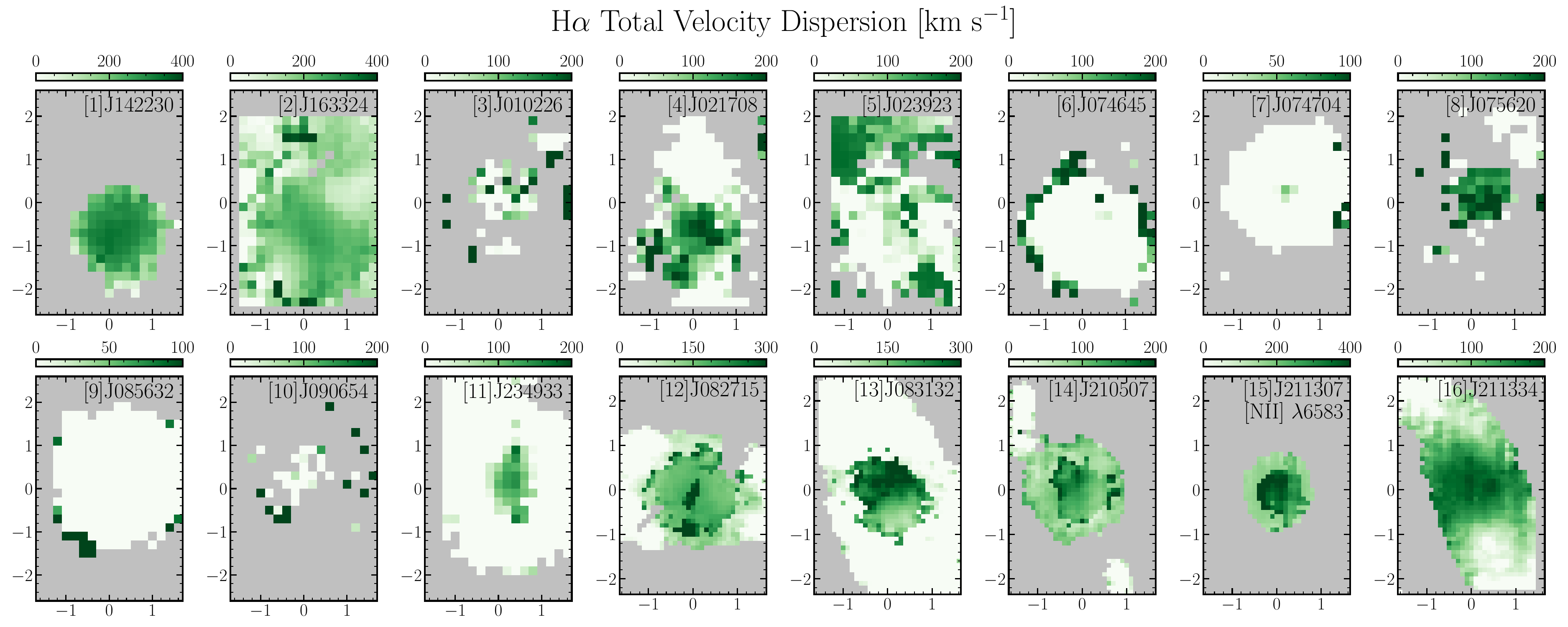}
\caption{Velocity dispersion maps (i.e., the second moment maps) of [\OIII] (top) and H$\alpha$ (bottom) emission lines. Note that H$\alpha$ velocity dispersion map of [15]J211307 is replaced with [\NII] due to the saturated columns in the CCD.}
\label{F5_Vdispmaps}
\end{figure*}

In Figures \ref{F4_Velocitymaps} and \ref{F5_Vdispmaps}, we present the total velocity and velocity dispersion (first and second moments of the line profile) maps of [\OIII] and H$\alpha$, respectively. Note again that H$\alpha$ of [15]J211307 is not available, so we replaced H$\alpha$ with [\NII] kinematics as an alternative since we fixed the kinematics of two lines in the spectral fitting.

We find that all [\OIII] velocity maps are dominated by the blueshifted regions at the center. These blueshifted regions are consistent with an approaching cone in the biconical outflow model. In the bicone outflow scenario, a receding cone is dimmed due to extinction by galactic dust plane. Thus, the flux from an approaching cone dominates in the [\OIII] profile, resulting in blueshift of the line profile. The biconical geometry of ionized outflows has been observed in various studies. Especially, high-resolution observations of nearby active galaxies have revealed that ionized outflows frequently show the biconical structure \citep{Shin19, Mingozzi19, Revalski21, Juneau22}. Using Monte Carlo simulation of biconical outflow models, \citet{Bae16} showed that a simple biconical geometry combined with dust extinction can reproduce the distribution of [\OIII] kinematics in the VVD diagram. Thus, although our results cannot constrain detailed structures of the outflows, we interpret the outflow kinematics using the biconical outflow scenario.

While the blueshifted region originates from the wing component of the [\OIII] line profile, the core component mostly shows blueshifted and redshifted regions due to the rotational motion of the host galaxy (see appendix for velocity maps of core \& wing components, respectively). Similarly, most H$\alpha$ velocity maps follow rotation by gravity except for several targets, such as [1]J142230, [2]J163324 and [13]J083132. Note that the wing component of H$\alpha$ was not detected in the majority of the type 1 AGNs except [1]J142230 and [2]J163324. Nevertheless, we cannot rule out the possibility of outflows traced by H$\alpha$ in type 1 AGNs because the contamination of broad H$\alpha$ emission affects the detectability of the H$\alpha$ wing component.

In the case of velocity dispersion maps, [\OIII] and H$\alpha$ show similar characteristics to the velocity maps, depending on the influence of the wing component. The [\OIII] velocity dispersion maps show enhanced velocity dispersion, ranging from several hundred km s$^{-1}$ to 1000 km s$^{-1}$ at the central region, while H$\alpha$ generally has a $\sim$ 100 km s$^{-1}$ level of velocity dispersion. The H$\alpha$ line of most type 1 AGNs is too narrow to be resolved even at the central pixels (plotted as 0 km s$^{-1}$), probably due to a weak gravitational potential.

Interestingly, targets classified as NLS1 and PSB often show extreme velocity in [\OIII] kinematics. For example, [1]J142230, the only PSB in our sample, shows that the [\OIII] profile has three Gaussian components. Moreover, the most blueshifted Gaussian component among the three has 1046 km s$^{-1}$ of velocity dispersion. This is consistent with a quenching scenario by the energetic AGN-driven winds \citep[e.g.,][]{Baron18}. On the other hand, NLS1s mostly show highly blueshifted [\OIII] kinematics in the relatively confined region at the center. In particular, for one object ([6]J074645), the [\OIII] core component was not detected, while the line profile is dominated by a broad wing component with $\sim$ 800 km s$^{-1}$ blueshifted at the very central region. Considering that NLS1 is thought to be a young AGN with high Eddington ratio \citep{Komossa08}, this high velocity outflow in a small region can be interpreted as a sign of an early launching stage  or an inner part of ionized gas outflows \citep{Cracco16, Schmidt18}. More statistical analysis of spatially resolved [\OIII] kinematics with a larger NLS1 sample is necessary.

\subsection{Ionization Property}\label{subsec:BPT}
\begin{figure*}[]
    \includegraphics[width=0.98\textwidth]{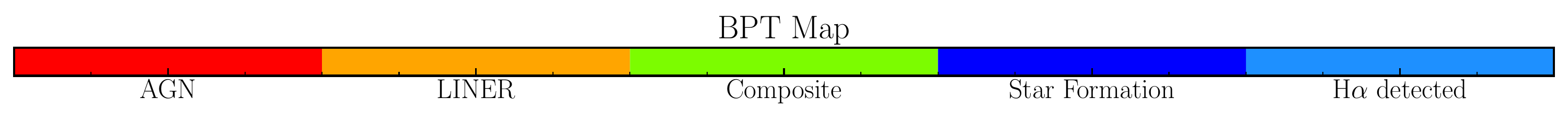}
    \includegraphics[width=0.98\textwidth]{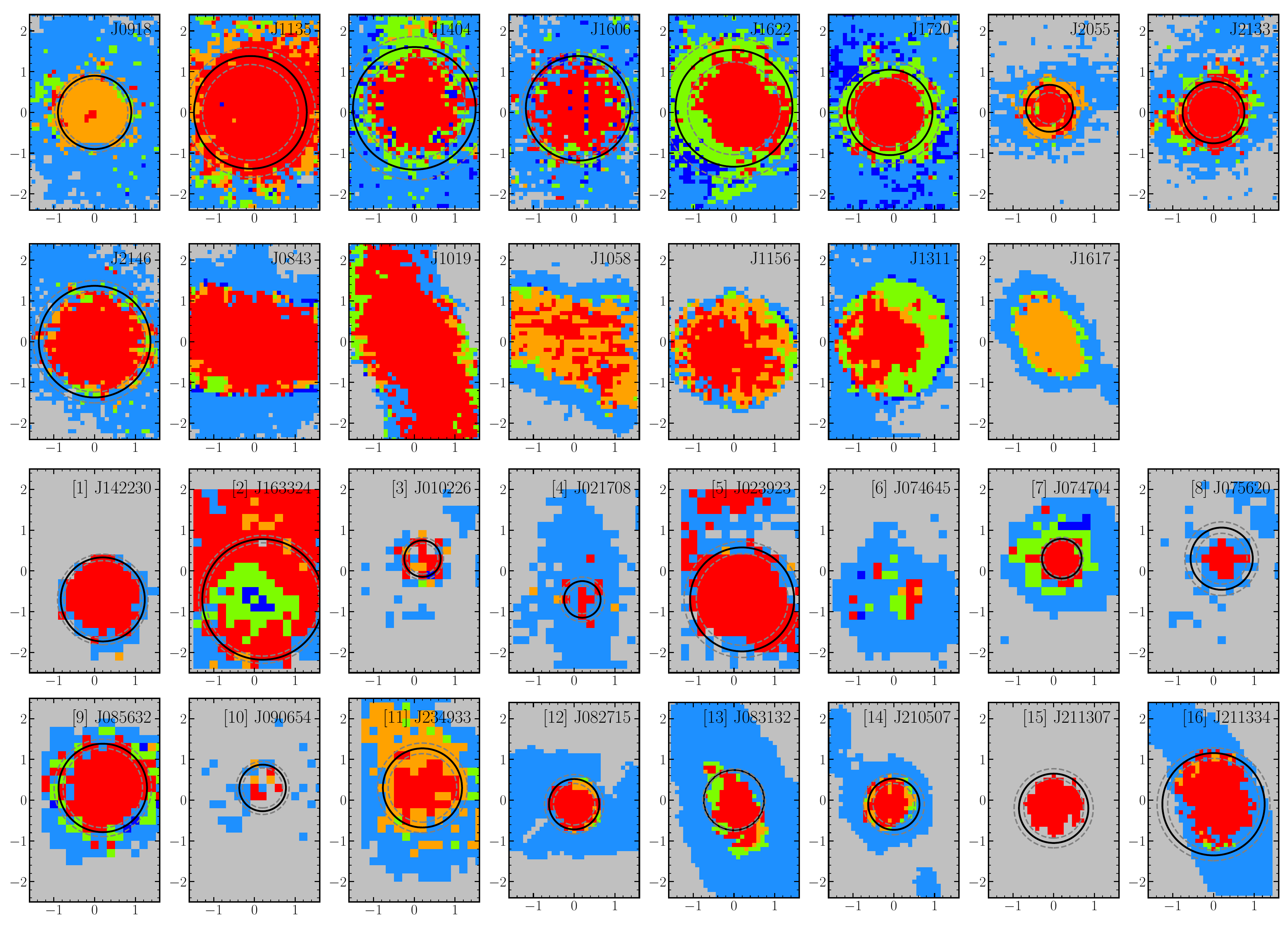}
    \caption{BPT maps of all GMOS-IFU targets. The first and second rows are 15 type 2 AGNs from \citet{Karouzos16a, Karouzos16b, Kang18, Luo19}, while the third and fourth rows are type 1 and type 2 AGNs in this study. The color bar demonstrates the ionization source corresponding to each color. The H$\alpha$ detected region (light blue) is where H$\alpha$ is detected but one or more of the other BPT lines are not detected, which is a proxy of the entire emission line region, while the gray area shows where H$\alpha$ is also not detected. The kinematic outflow size and its uncertainty are shown as black solid and gray dashed circles, respectively. Note that the last six AGNs in the second row are from \citet{Luo19} sample, which has no measured outflow sizes due to no or weak outflow components.}
    \label{F6_BPT}
\end{figure*}

We classified the ionization sources of each pixel into AGN, LINER, composite and star formation based on the Baldwin-Phillips-Terlevich (BPT) diagram using the flux ratios [\OIII]/H$\beta$ and [\NII]/H$\alpha$ \citep{Kewley01, Kauffmann03, CidFernandes10}. For [15]J211307, we only classified AGN photoionization by $\mathrm{log}$([\OIII]/H$\beta$) $>$ 0.6 cut due to the unavailable H$\alpha$ flux. Figure \ref{F6_BPT} shows the spatial distributions of ionization sources by different colors. The outflow size and its uncertainty are drawn as black solid and gray dashed circles, respectively (see Section \ref{subsec:outflowsize}). Along with BPT classification, we also mark in light blue color the region where H$\alpha$ is detected but one or more other BPT diagnostic lines (\Hb, [\OIII] and [\NII]) are not detected. Note that we check the possibility of ``fake AGN" based on WHAN diagnostics \citep{CidFernandes11}, using four type 2 AGNs with available H$\alpha$ emssion. In general, most spaxels show H$\alpha$ equivalent width larger than 10 \AA, indicating a clear sign of AGN emission. Since our sample is selected by high [\OIII] luminosity and strong outflow conditions, we consider that the rest of the targets in the sample are in a similar ionization condition as in the strong AGN class.

It is not surprising that AGN photoionization dominates at the central regions in all targets except [2]J163324 since all of our targets host luminous AGNs at the center. For [7]J021708, [9]J085632, [11]J234933 and [13]J083132, there are radial trends of BPT classification, from AGN to composite and star formation (or LINER). These AGN-dominated regions are surrounded by composite and star formation, as similarly found in the type 2 AGNs in our previous studies \citep{Karouzos16b, Kang18, Luo19}, as shown in the first and second rows in Figure \ref{F6_BPT}. For the type 1 AGN, the AGN-photoionized region is more extended than that for the type 2 AGN, possibly due to the higher bolometric luminosity of the sample than that of the type 2 sample. The only exception, [2]J163324, shows an opposite trend in which the star formation is at the center and the AGN is at the outer part. \citet{Castello-Mor12} reported that most of the missing AGNs in optical emission line diagnostics are NLS1s. \citet{Scharwachter17, Winkel22} also found a similar ionization structure to [2]J163324 in nearby NLS1s. Especially, \citet{Winkel22} detected a prominent signature of circumnuclear star formation in Mrk 1044. Since [2]J163324 is also one of the NLS1s and interacting systems, a large amount of circumnuclear star formation may dominantly ionize the ISM at the central region. (SFR$_{\mathrm{IR, SED}} \sim 50 \mathrm{M}_\odot\;\mathrm{yr^{-1}})$.

In most cases, the size of the outflow is comparable to the size of the AGN photoionization region but much smaller than the entire emission line region. Furthermore, H$\alpha$ is detected far beyond the outflow region, implying that star formation is ongoing at the large disk scale. We discuss the size of the outflow and NLR in detail in Section \ref{subsec:NLRsize}.

\section{Discussion} \label{sec:discussion}

\subsection{Kinematic Outflow Size}\label{subsec:outflowsize}

\begin{figure*}[]
    \includegraphics[width=0.98\textwidth]{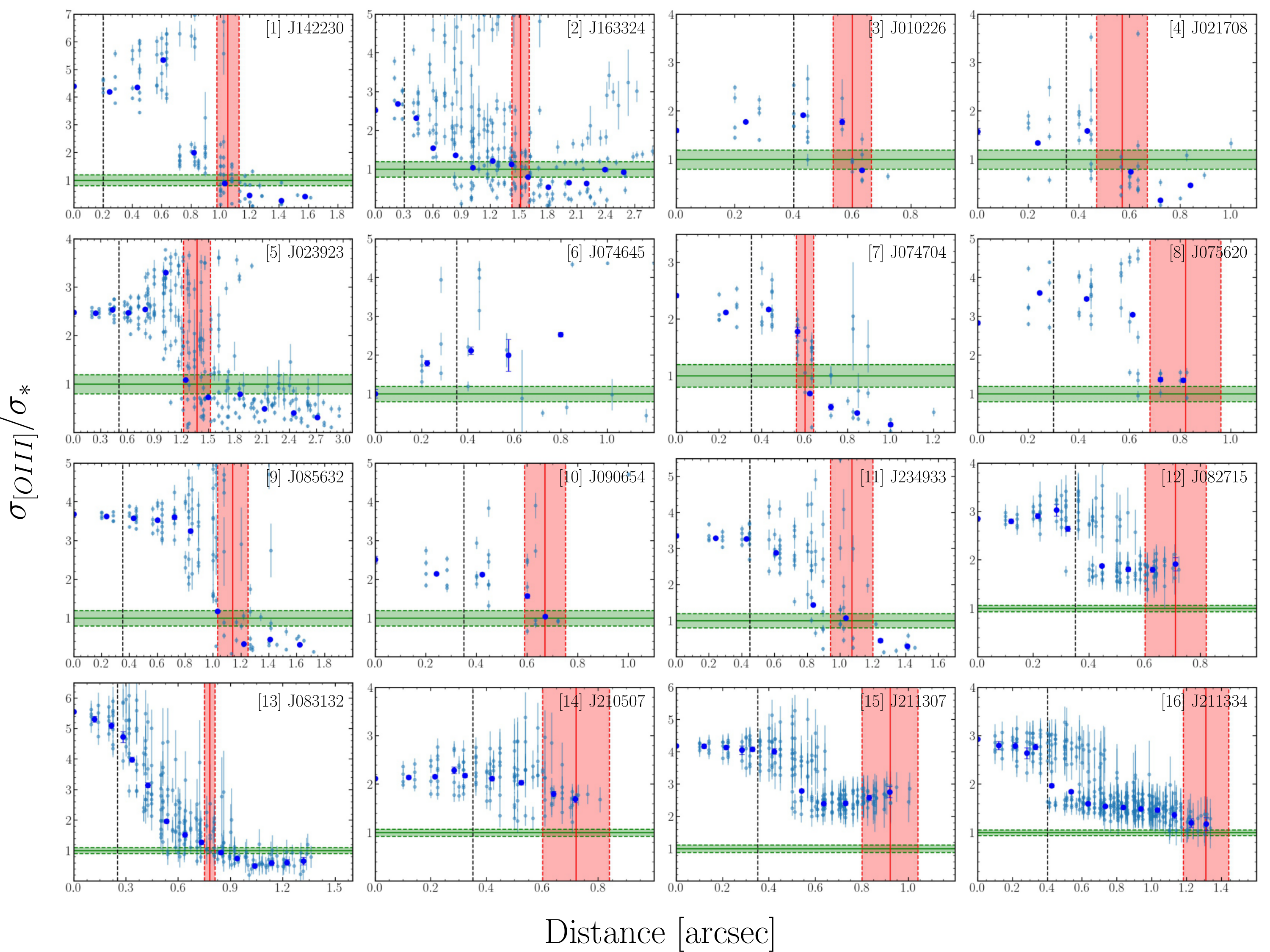}
    \caption{Radial profiles of the ratio between [\OIII] and stellar velocity dispersion. Light blue points show individual spaxels, while large blue points are error-weighted means of each radial bin. The green horizontal line and shaded region show where the velocity dispersion ratio becomes unity and 1 $\sigma$ error. Kinematic outflow size and its error are plotted as a red vertical line and shaded region. Black vertical lines represent the HWHM of the seeing size. We obtained kinematic outflow sizes of 15 among 16 targets except [6]J074645, which lacks [\OIII] detected spaxels.}
    \label{F7_sigma_ratios}
\end{figure*}

\begin{deluxetable*}{cccccccccc}
    \tablecolumns{10}
    \tablecaption{Target Outflow Properties}\label{T2_outflowinfo}
    \tablehead{
       & Name  & $v_{out}$ & $\sigma_{out}$ & $\sigma_{*}$  & log $\mathrm{R}_{\mathrm{out}}$ & log $\mathrm{L}_{[\mathrm{OIII}]}$  & log $\dot{\mathrm{M}}_{\mathrm{out}}$ & log SFR & Ref of SFR    \\
       &       & [\kms]    & [\kms]         & [\kms]        & [kpc]                   & [\ergs]                              & [$\mathrm{M}_{\odot}\;yr^{-1}$]          & [$\mathrm{M}_{\odot}\;yr^{-1}$] \\
       {[}1{]} & {[}2{]}    & {[}3{]}        & {[}4{]}      & {[}5{]}                & {[}6{]}                              & {[}7{]}                                & {[}8{]}  & {[}9{]} & {[}10{]}}
        \startdata
        1  & J142230 & -374.73$\pm$2.49  & 768.96$\pm$3.02 & 168.86 &  0.33$\pm$0.03 & 42.59 &  0.41$\pm$0.03 & 1.79$\pm$0.02 & \citet{Kim22} \\
        2  & J163324 & -106.16$\pm$2.57  & 331.27$\pm$2.04 & 119.83 &  0.49$\pm$0.03 & 42.12 & -0.62$\pm$0.03 & 1.70$\pm$0.02 & This work  \\
        3  & J010226 & -281.05$\pm$13.18 & 474.58$\pm$4.77 & 226.25 &  0.29$\pm$0.05 & 42.01 & -0.33$\pm$0.05 & 2.11$\pm$0.12 & This work  \\
        4  & J021708 & -178.28$\pm$10.30 & 372.65$\pm$6.30 & 203.25 &  0.29$\pm$0.08 & 41.21 & -1.25$\pm$0.08 & 1.96$\pm$0.04 & This work  \\
        5  & J023923 & -415.98$\pm$1.64  & 597.67$\pm$1.39 & 230.79 &  0.71$\pm$0.05 & 42.86 &  0.22$\pm$0.05 & 1.85$\pm$0.11 & This work  \\
        6  & J074645 &  ...    &  ...    & 179.21 &   ...          &  ...  & ...   & 1.39$\pm$0.17 & This work  \\
        7  & J074704 &  -74.63$\pm$2.40  & 407.46$\pm$2.68 & 174.03 &  0.24$\pm$0.03 & 42.12 & -0.29$\pm$0.03 & 1.59$\pm$0.08 & This work  \\
        8  & J075620 & -308.57$\pm$6.62  & 718.57$\pm$4.21 & 202.13 &  0.46$\pm$0.07 & 42.03 & -0.33$\pm$0.07 & 2.40$\pm$0.05 & This work  \\
        9  & J085632 & -381.62$\pm$2.20  & 720.37$\pm$1.83 & 198.01 &  0.61$\pm$0.04 & 42.36 & -0.12$\pm$0.04 & 1.85$\pm$0.15 & This work  \\
        10 & J090654 & -408.30$\pm$7.05  & 419.26$\pm$4.26 & 172.88 &  0.34$\pm$0.05 & 41.18 & -1.18$\pm$0.05 & 1.25$\pm$0.11 & This work  \\
        11 & J234933 & -248.25$\pm$3.84  & 449.13$\pm$2.19 & 188.13 &  0.61$\pm$0.05 & 42.21 & -0.48$\pm$0.05 & 1.90$\pm$0.12 & This work  \\
        12 & J082715 &  -81.13$\pm$3.53  & 398.72$\pm$3.55 & 147.51 &  0.03$\pm$0.07 & 41.24 & -0.98$\pm$0.07 & 0.95$\pm$0.16 & \citet{Kim22} \\
        13 & J083132 & -150.90$\pm$3.09  & 384.40$\pm$3.59 & 91.05  & -0.04$\pm$0.02 & 40.84 & -1.30$\pm$0.02 & 0.09$\pm$0.10 & \citet{Kim22} \\
        14 & J210507 & -122.28$\pm$3.70  & 343.46$\pm$3.05 & 144.14 &  0.05$\pm$0.07 & 41.79 & -0.49$\pm$0.07 & 1.05$\pm$0.07 & \citet{Kim22} \\
        15 & J211307 &  -38.88$\pm$1.67  & 328.97$\pm$2.37 & 84.24  &  0.06$\pm$0.06 & 41.69 & -0.64$\pm$0.06 & 0.39$\pm$0.19 & \citet{Kim22} \\
        16 & J211334 &  -67.50$\pm$2.80  & 307.31$\pm$2.16 & 135.89 &  0.24$\pm$0.04 & 42.04 & -0.49$\pm$0.04 & 0.93$\pm$0.07 & \citet{Kim22} \\
        \enddata
    \tablecomments{(1) Target number; (2) name; (3) flux-weighted mean [\OIII] outflow velocity within outflow size; (4) flux-weighted mean [\OIII] outflow velocity dispersion within outflow size; (5) stellar velocity dispersion; (6) outflow size; (7) [\OIII] luminosity within outflow size; (8) mass outflow rate; (9) Star formation rate based on dust luminosity; (10) reference of star formation rate.}
\end{deluxetable*}

We determine how far the outflow extends using the spatially resolved gas kinematics as introduced by \citet{Kang18} and \citet{Luo21}. The edge of outflow ($\mathrm{R}_\mathrm{out}$) is defined as where the [\OIII] velocity dispersion becomes comparable to stellar velocity dispersion, which is flux-weighted within the central region, representing the gravitational potential of host galaxies. For type 1 AGNs, we estimate stellar velocity dispersion using the $\mathrm{M}_{BH}$--$\sigma_*$ relation \citep{Woo15} due to the lack of stellar absorption line detection, while we calculate $\mathrm{M}_{BH}$ using Equation 2 of \citet{Woo15} with the measured $L_{5100}$ and width of the broad H$\beta$ line. As presented in Figure \ref{F7_sigma_ratios}, the radial profiles of the ratio between [\OIII] and stellar velocity dispersion clearly show a decreasing trend. 
Even in the case of no gas outflows, gas velocity dispersion would show a radial decrease as typically observed in spatially resolved gas and stellar kinematics in nearby galaxies. However, we find that gas velocity dispersion (or normalized gas velocity dispersion) decreases rather strongly as presented in Figure \ref{F7_sigma_ratios}, while we expect generally milder decrease without gas outflows \citep[e.g., Appendix D in][]{Martinsson13}. Since gas kinematics is dynamically colder than stellar kinematics in non-AGN galaxies, gas velocity dispersion is typically smaller than stellar velocity dispersion \citep{Oh22}. In contrast, our targets show much higher gas velocity dispersion than stellar velocity dispersion, and this also supports that the radial decrease is mainly due to outflow kinematics.

We mark the outflow size with a red vertical line, where the ratio becomes unity. If the ratio does not converge to unity, we alternatively choose the last radial point as a lower limit of the outflow size. Among 16 targets, the outflow size of [6]J074645 cannot be determined due to the lack of usable data points in the radial distribution. After determining the angular outflow sizes, we correct for the seeing effect by subtracting the half width half maximum (HWHM) of the seeing size in quadrature and convert the angular sizes to the physical outflow size.

It is possible that we may miss weak outflows located in the outer region because we decide to detect outflows by comparing gas velocity dispersion (i.e., the 2nd moment of the total line profile) with stellar velocity dispersion. For example, if outflow is weak and gas kinematics are dominated by the disk kinematics in the outer region, our method does not detect the presence of outflows. Also, gas velocity dispersion can be smaller than stellar velocity dispersion in disk kinematics owing to the thinner gas disk than stellar disk. Thus, we could miss outflows comparable to stellar kinematics but stronger than gas kinematics. Nevertheless, we take a conservative approach to detect prominent outflow signatures. Although we may miss weak outflows on a larger scale, we expect that the impact of those weak outflows on the ISM is not significant.

We find that the outflow size of the sample ranges from a few to $\sim$ 5 kpc, similar to the range of the type 2 AGNs of \citet{Kang18} (see Table \ref{T2_outflowinfo}). This result indicates that the impact of the outflow on the ISM is generally limited to the central kpc scales. The fact that H$\alpha$ emissions are detected in a much larger region beyond the edge of the outflow also supports the lack of global impact by outflow (see Figure \ref{F6_BPT}) since there is still ongoing star formation outside of outflows.

Previous IFU studies have reported the size of ionized gas outflow based on various definitions. One of the frequently used parameters to determine the presence of outflow is $W_{80}$, which is a non-parametric line width containing 80\% of the [\OIII] total flux. For example, \citet{Harrison14} found that the outflow region ($W_{80}$ $>$ 600 km/s) is more extended than 10 kpc in luminous type 2 AGNs. Since our sample includes a majority of their targets \citep[from][]{Kang18}, we note that this $W_{80}$ method tends to yield a much larger outflow size than our kinematical approach with gas velocity dispersion (i.e., the 2nd moment of the total line profile). \citet{Wylezalek20} also estimated the typical outflow size of the low and high [\OIII] luminosity AGNs, respectively, in the MaNGA sample. By comparing the mean radial profiles of $W_{80}$ with those of non-AGN targets and by defining the edge of outflows where AGN and non-AGN targets have similar $W_{80}$ velocities, they reported outflow sizes of 8 and 15 kpc, respectively for low and high [\OIII] luminosity AGNs, which are also much larger than the size measurements based on our method. Using $W_{80}$, \citet{Ruschel-Dutra21} suggested that outflow regions in nearby AGNs are $\gtrsim$ 0.1--1 kpc scales.

While there were various observational reports of large galactic scale AGN-driven ionized gas outflows \citep[$\gtrsim$ 10 kpc, e.g.,][]{Cresci15, Baron18}, our results are more consistent with those of \citet{Carniani15, Carniani16, Villar-Martin16, Fischer18, Kakkad20, Singha22, Deconto-Machado22}, who found several kpc scales or smaller outflow sizes. Most of these methods can be classified as spectro-astrometry, which utilizes spatial offset of the flux centroid of the [\OIII] wing component. In the case of \citet{Deconto-Machado22}, the size of the kinematically disturbed region (KDR) was defined based on the velocity dispersion residuals ($\sigma_{[OIII]}-\sigma_*$) of AGN and control samples. They found that the KDR region sizes vary from 0.2 to 2.3 kpc (with a mean of 0.96 kpc). Considering the lower [\OIII] luminosity of their MaNGA AGN sample (i.e., $10^{39}\;-\; 10^{42}\;\mathrm{erg\;s}^{-1}$), their results are broadly consistent with our measurements. On the other hand, \citet{Fischer18} defined the outflow size as the largest radius requiring multiple components in spectral fitting or centroid velocities higher than 300 km/s. Their outflow sizes are mostly less than 1 kpc scale (with a mean of $\sim$ 0.6 kpc), despite their similar [\OIII] luminosity range to that of our sample (i.e., $10^{42}\;-\; 10^{43}\;\mathrm{erg\;s}^{-1}$). \citet{Fischer18} also reported the size of the disturbed rotation region (R$_{dist}$) with FWHM larger than 250 km s$^{-1}$ in a single component in emission lines, and those values are slightly larger (mean $\sim$ 1.1 kpc).

As \citet{Husemann16} pointed out, the size of outflow region can be overestimated due to the seeing effect. Based on their PSF deblending technique for type 1 AGNs, they found that the [\OIII] line width decreased in general after correcting for the seeing effect. Recently, \citet{Singha22} investigated whether the [\OIII] wing component is spatially resolved, using a sample of type 1 AGNs. Their method utilizes the spatial distribution of broad H$\beta$ emission from the BLR as a PSF template and compares with the spatial profile of the [\OIII] wing component. We adopt the method of \citet{Singha22} to verify spatially resolved outflows. Note that our combined sample mainly consists of type 2 AGNs, for which no BLR emission is available as a proxy of the PSF. For the majority of type 1 AGNs in our sample, the number of spaxels with the detection of the [\OIII] wing component is too small to perform this test. Thus, we are able to do this test for a few type 1 AGNs , and the results indicate that the outflow region is clearly resolved for these objects (i.e., [1]J142230 and [5]J023923). It is possible that the outflow region may not be spatially resolved for some of our targets and our measurements are upper limits in these cases. 
Nevertheless, even if we assume that a significant fraction of our targets has unresolved [\OIII] outflows, it leads to smaller outflow sizes. In fact, their size estimation, which is defined as an offset of the [\OIII] wing component from the center of H$\beta$ BLR flux, is much smaller than our kinematic outflow size ($\lesssim 100$ pc). Therefore, this suggests that our interpretation, i.e., lack of global impact by ionized outflows, is still valid.

In summary, we obtain the kinematic outflow sizes of 15 among 16 new targets. Similar to our previous measurements, the outflow sizes are generally a few kpc scale, and do not exceed 5 kpc for a majority of our targets. The range of the outflow size implies that AGN feedback by strong outflows may affect star formation in the central several kpc but the overall impact on a large disk scale is rather limited.

\subsection{Outflow Size--Luminosity Relation}\label{subsec:RLrelation}

\begin{figure}[]
    \includegraphics[width=0.49\textwidth]{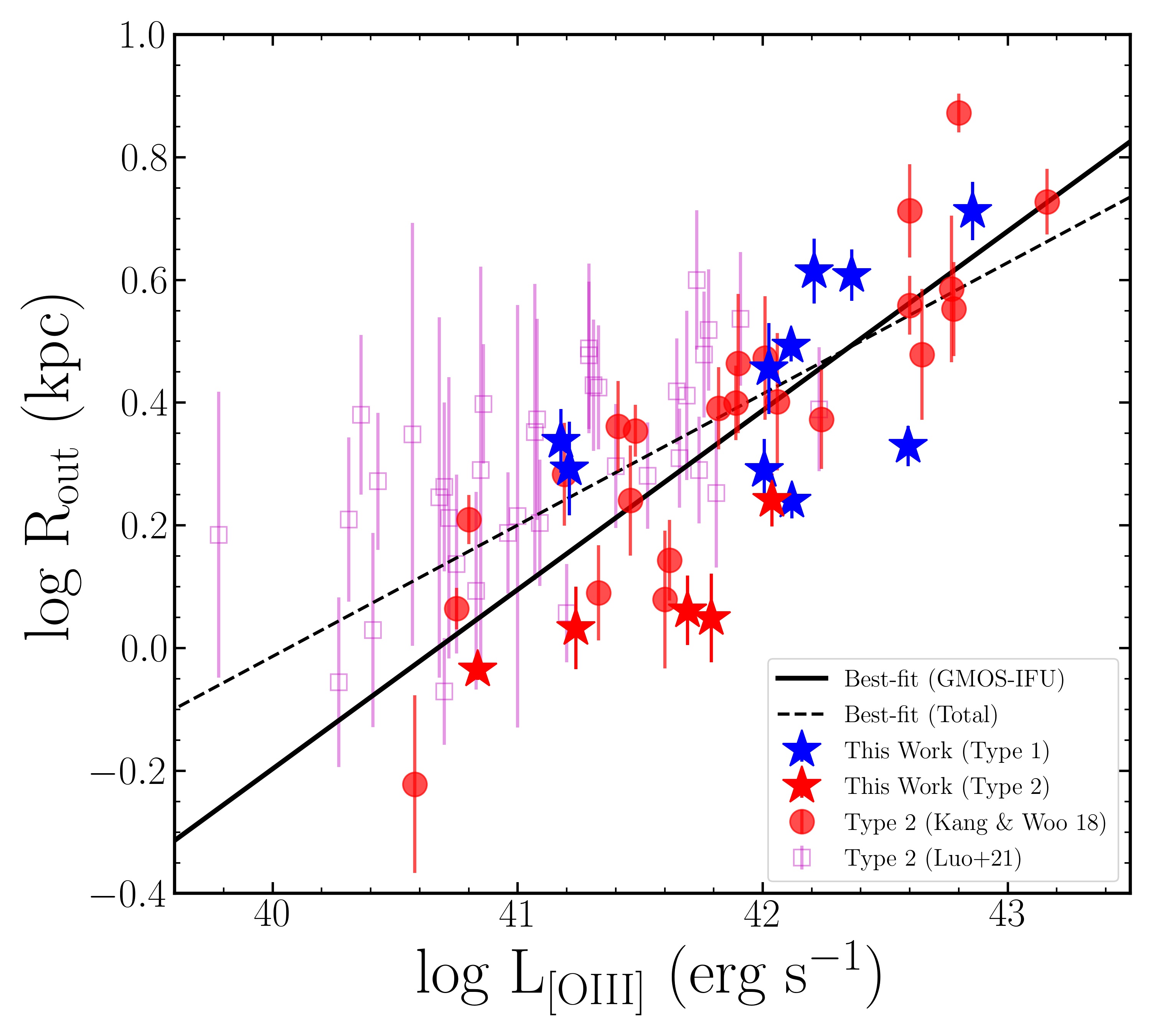}
    \caption{Outflow size--[\OIII] luminosity relation ($\mathrm{R}_{\mathrm{out}}$-- $\mathrm{L}_{[\mathrm{OIII}]}$ relation) plot of the total sample. Blue and red markers indicate GMOS-IFU type 1 and type 2 samples, and star markers are the targets from this study. Magenta squares are SNIFS type 2 AGN samples from \citet{Luo21}. The dashed line is the best-fit relation from the total sample (75 AGNs, slope $\sim$ 0.22), and the solid line is the best-fit using only GMOS-IFU targets (38 AGNs, slope $\sim$ 0.29).}
    \label{F8_Rout_Loiii}
\end{figure}

By comparing the kinematically determined outflow size with AGN luminosity, we extend the outflow size--luminosity relation using both type 1 and type 2 AGNs. \citet{Kang18} reported a clear correlation of the kinematic outflow size with the [\OIII] luminosity based on a combined GMOS-IFU sample of 23 type 2 AGNs, including six AGNs from \citet{Karouzos16a, Karouzos16b} and 14 AGNs from \citet{Harrison14}. We also presented the outflow size--luminosity relation using 37 type 2 AGNs observed by the SuperNova Integral Field Spectrograph (SNIFS) in \citet{Luo21}. Here, we combine the new outflow size measurements of 15 AGNs with our previous outflow size measurements. The total sample consists of 75 AGNs, and 38 of them are based on GMOS-IFU data and consistent analysis, including both type 1 and type 2 AGNs.

We compare the kinematic outflow size with the extinction uncorrected [\OIII] luminosity, which is measured within the outflow size in Figure \ref{F8_Rout_Loiii}.
Our new targets (star markers) follow the same positive trend of \citet{Kang18} and \citet{Luo21}. Considering the orientation effect, we expect smaller outflow sizes in type 1 AGNs since the launching direction of outflows is closer to the line-of-sight than that of type 2 AGNs.
If the orientation effect is dominant, the type 1 AGNs are likely to deviate from the trend with systematically smaller sizes than type 2 AGNs.
However, we do not see a clear difference between the two types, implying that the orientation effect is not significant due to a relatively large opening angle of biconical outflows, which would reduce the systematic difference due to the orientation.

We perform a linear regression using the combined sample of GMOS-IFU and SNIFS and obtain the best-fit relation (dashed line) as follows:
\begin{equation}\label{eq_Rout_Loiii}
    \begin{split}
        \mathrm{log}\;\left(\frac{\mathrm{R}_\mathrm{out}}{\mathrm{kpc}}\right) =  & (0.22\pm0.03) \\
        & \times\mathrm{log}\;\left(\frac{\mathrm{L}_\mathrm{[OIII]}}{10^{42}\;\mathrm{erg\;s}^{-1}}\right)+(0.42\pm0.02)
    \end{split}
\end{equation}
Considering the fact that the outflow size is more uncertain for the SNIFS sample \citep{Luo21}, especially for lower luminosity targets, due to the limited seeing conditions (0\farcs9--2\farcs1, median 1\farcs2) and the potential systematic difference between GMOS-IFU and SNIFS data, we perform a linear regression using only GMOS-IFU targets and obtain the best-fit relation (solid line) as follows:
\begin{equation}\label{eq_Rout_Loiii_GMOS}
    \begin{split}
    \mathrm{log}\;\left(\frac{\mathrm{R}_\mathrm{out}}{\mathrm{kpc}}\right) =  & (0.29\pm0.04) \\
    & \times\mathrm{log}\;\left(\frac{\mathrm{L}_\mathrm{[OIII]}}{10^{42}\;\mathrm{erg\;s}^{-1}}\right)+(0.39\pm0.02)
    \end{split}
\end{equation}

We find that the best-fit slope of 0.29$\pm$0.04 based on the combined type 1 and type 2 AGNs is consistent with that of \citet{Kang18}. In general, more luminous and energetic AGNs deliver ionized outflows farther out and impact the ISM on larger scales of host galaxies. However, we emphasize again that the outflow size is confined to the central several-kpc scales. For instance, if we consider an AGN with M$_{\rm BH}$ = $10^8\;\mathrm{M}_\odot$ and Eddington luminosity, the expected outflow size is $\sim$ 3.5 kpc. Thus, except for the most luminous AGNs, the global impact of AGN-driven outflows on star formation is unlikely to be significant on large global scales.

\subsection{Outflow Size versus Photoionization Size}\label{subsec:NLRsize}
\begin{figure*}[]
    \includegraphics[width=0.49\textwidth]{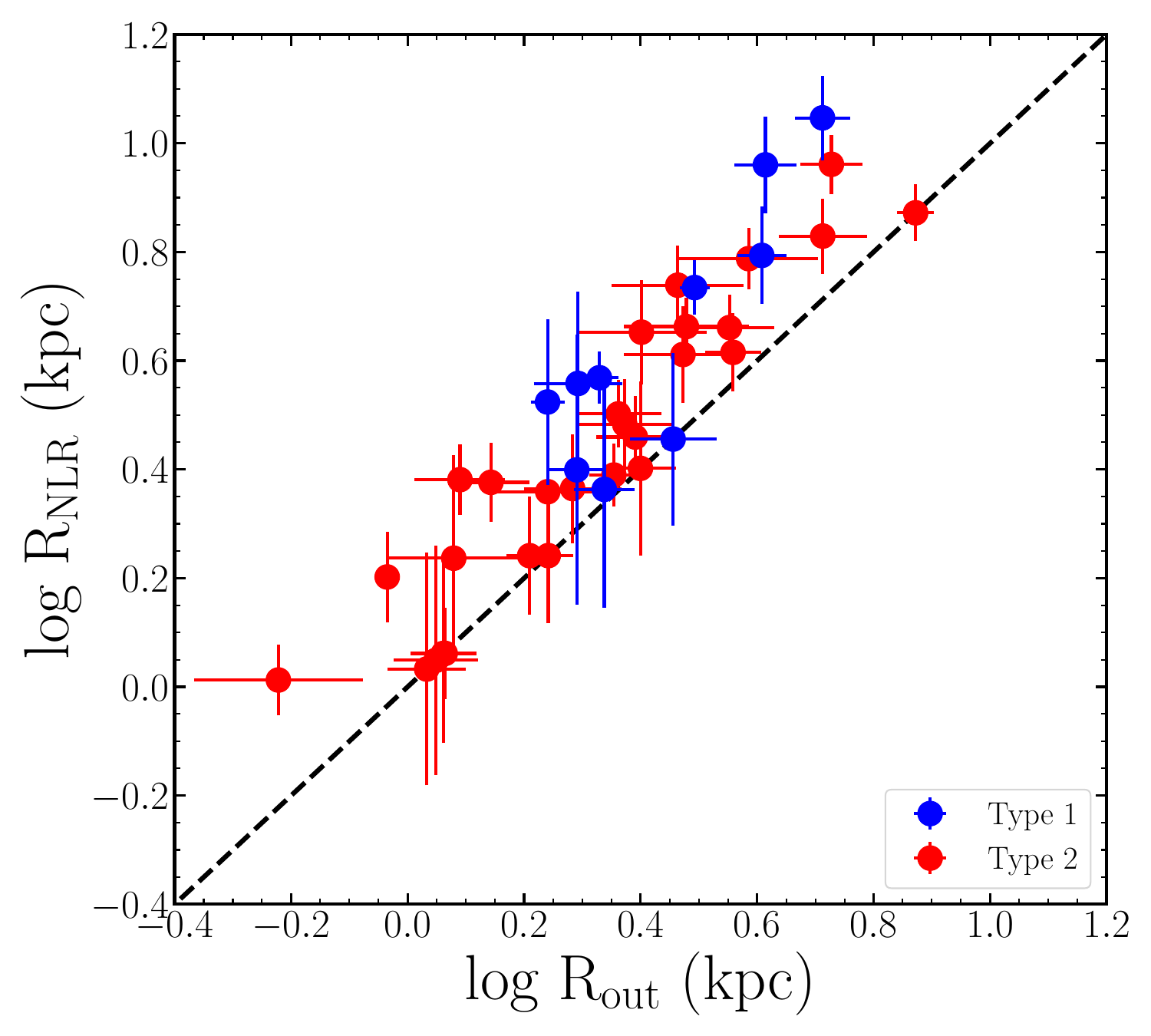}
    \includegraphics[width=0.49\textwidth]{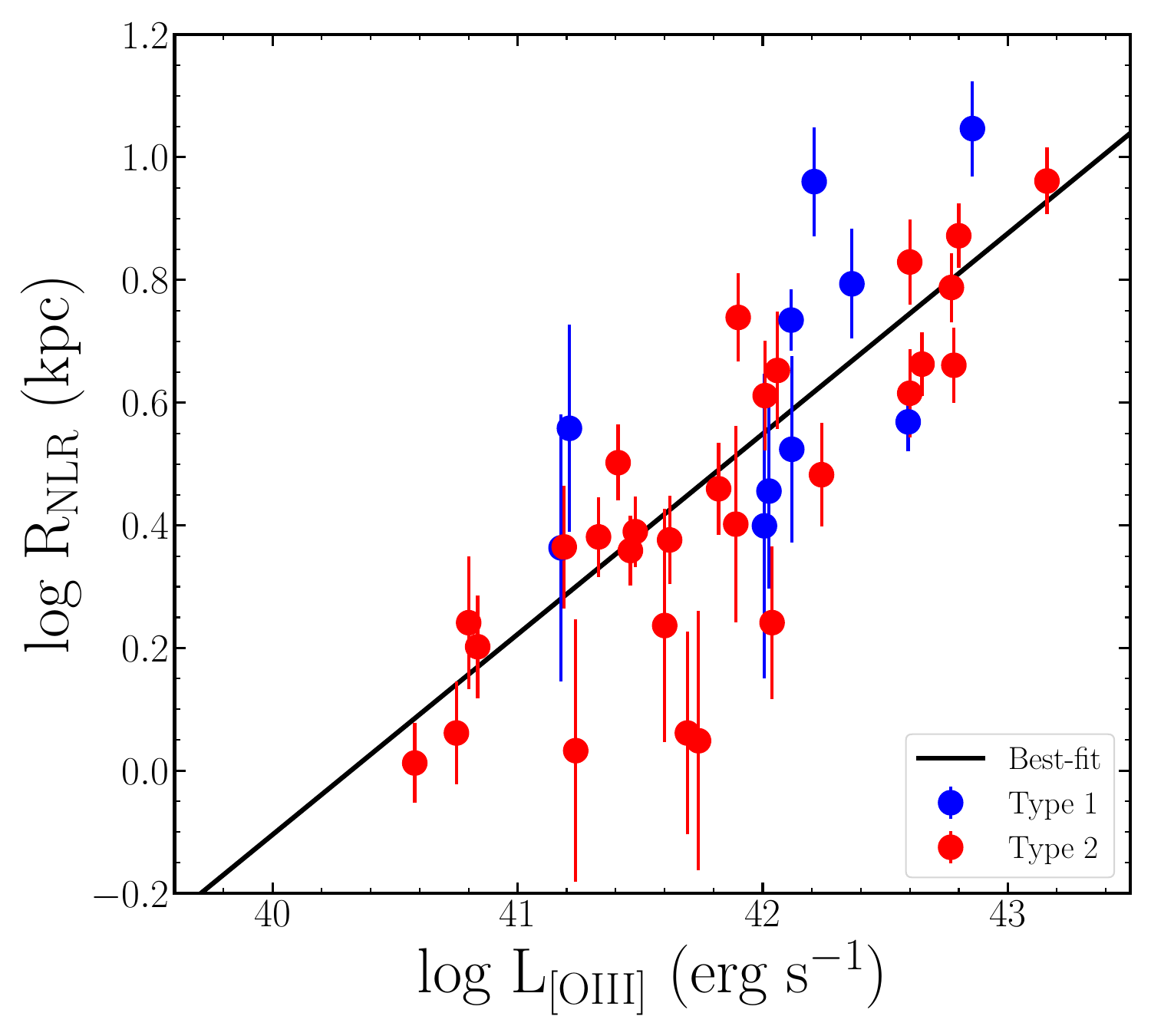}
    \caption{Left: $\mathrm{R}_{\mathrm{out}}$--$\mathrm{R}_{\mathrm{NLR}}$ comparison plot. The black dashed line shows the one-to-one relation. Right: The photoionization size--[\OIII] luminosity plot for the GMOS-IFU targets. The black solid line shows the best-fit relation (slope $\sim$ 0.33).}
    \label{F9_RNLR}
\end{figure*}

To examine the different spatial scales between outflows and photoionization, we estimate the size of the photoionization (i.e., NLR size) and compare it with the outflow size. The photoionization size is defined by the largest radial point of [\OIII] detection in Figure \ref{F7_sigma_ratios}.
By directly comparing the photoionization size with the outflow size in Figure \ref{F9_RNLR} (left panel), we find that the photoionization size is larger than the outflow size by 0.14 dex (i.e., $\sim$ 38\%) on average, indicating that outflow propagates less efficiently than ionizing photons. This result is presumably due to several obstacles in the spreading of outflows, i.e., contact with the ISM and the gravitational effect of host galaxies. While AGN outflow is relatively confined on a relatively small scale, radiative feedback can influence ISM on a larger scale.

We present the correlation between the photoionization size and the [\OIII] luminosity in Figure \ref{F9_RNLR} (right panel). As previously reported, the photoionization size shows a clear correlation with the [\OIII] luminosity, indicating that when AGN is more luminous, ionizing photons affect the ISM on a larger scale. The best-fit slope is 0.33$\pm$0.04 (black solid line), which is slightly steeper than that of the outflow size--luminosity relation. 

Compared to previous studies, our slope is shallower than 0.4--0.5 from \citet{Bennert02, Husemann14, Bae17} and similar to 0.2--0.3 from \citet{Schmitt03b, Liu13a, Liu14}. However, direct comparison with previous studies is inappropriate due to the diverse definitions of photoionization sizes. For example, the photoionization size was often determined based on either narrow [\OIII] band photometry or spectral analysis of the [\OIII] emission line using a number of criteria such as an S/N-based size \citep{Bennert02, Schmitt03a, Schmitt03b, Fischer18}, a flux-weighted radius \citep{Husemann14, Bae17}, an isophotal radius \citep{Liu13a, Liu14}, and a transiting radius in BPT classification \citep{Bennert06a, Bennert06b, Bennert06c, Deconto-Machado22}. A flux-weighted radius is less affected by observing conditions, yet it can underestimate the photoionization size significantly \citep{Kang18}. On the other hand, as \citet{Bennert06b} pointed out, S/N-based size is highly dependent on the observational depth. Thus, shallower observations may fail to detect more extended NLR structure. The S/N-based size also has the possibility of overetimation due to contamination of star-forming regions. In the case of BPT transiting radius, it requires flux measurements of multiple emission lines. Consequently, this method is more sensitive to observational depth. 
Our method can be classified as an S/N-based method, and it results in 1--10 kpc scales, which is in agreement with previous measurements \citep{Bennert02, Bennert06a, Bennert06b, Bennert06c, Liu13a,  Liu14, Fischer18, Deconto-Machado22}.

For comparing the outflow size ($\mathrm{R}_{\mathrm{out}}$) with the photoionization size ($\mathrm{R}_{\mathrm{NLR}}$), our measurements indicate the ratio, $\mathrm{R}_{\mathrm{out}}/\mathrm{R}_{\mathrm{NLR}}$ is $\sim\;0.72$, which is larger than $\sim$ 0.2--0.3 from \citet{Fischer18, Deconto-Machado22}. However, we emphasize again that the difference of outflow and NLR sizes stems from various 
definitions, sample selection, and observational conditions. More careful comparison and uniform analysis are required to understand the systematic difference.

In conclusion, while better measurements of the photoionization size require deep IFU observations with a relatively large FOV, it is clear that the photoionized region is more extended than the outflow region in general, and correlates with the [\OIII] luminosity. Thus, outflow energetics can be underestimated if photoionization size is used instead of outflow size in calculating outflow energetics.

\subsection{Mass Outflow Rates}\label{subsec:Mdot}

\begin{figure*}[]
    \includegraphics[width=0.49\textwidth]{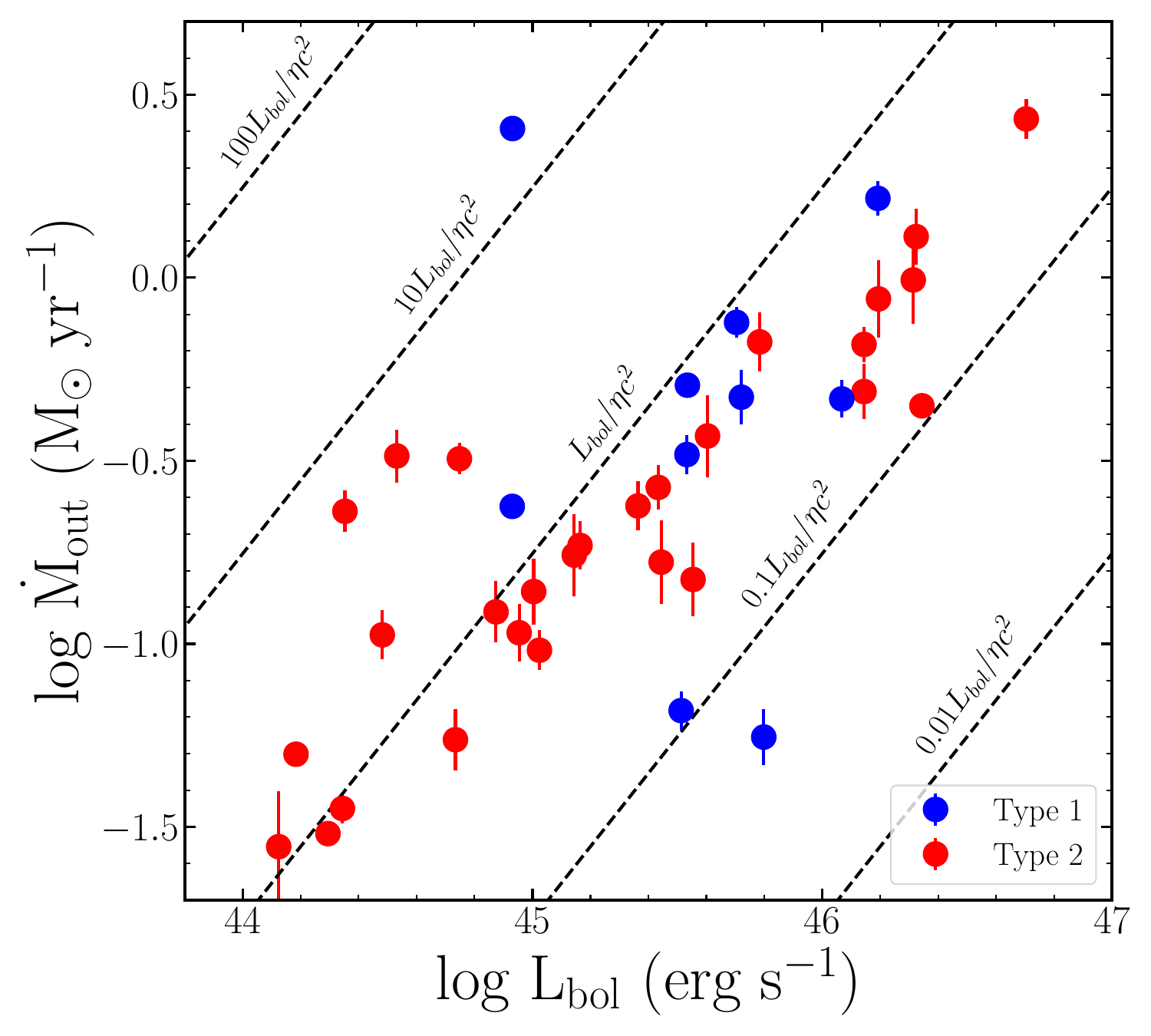}
    \includegraphics[width=0.49\textwidth]{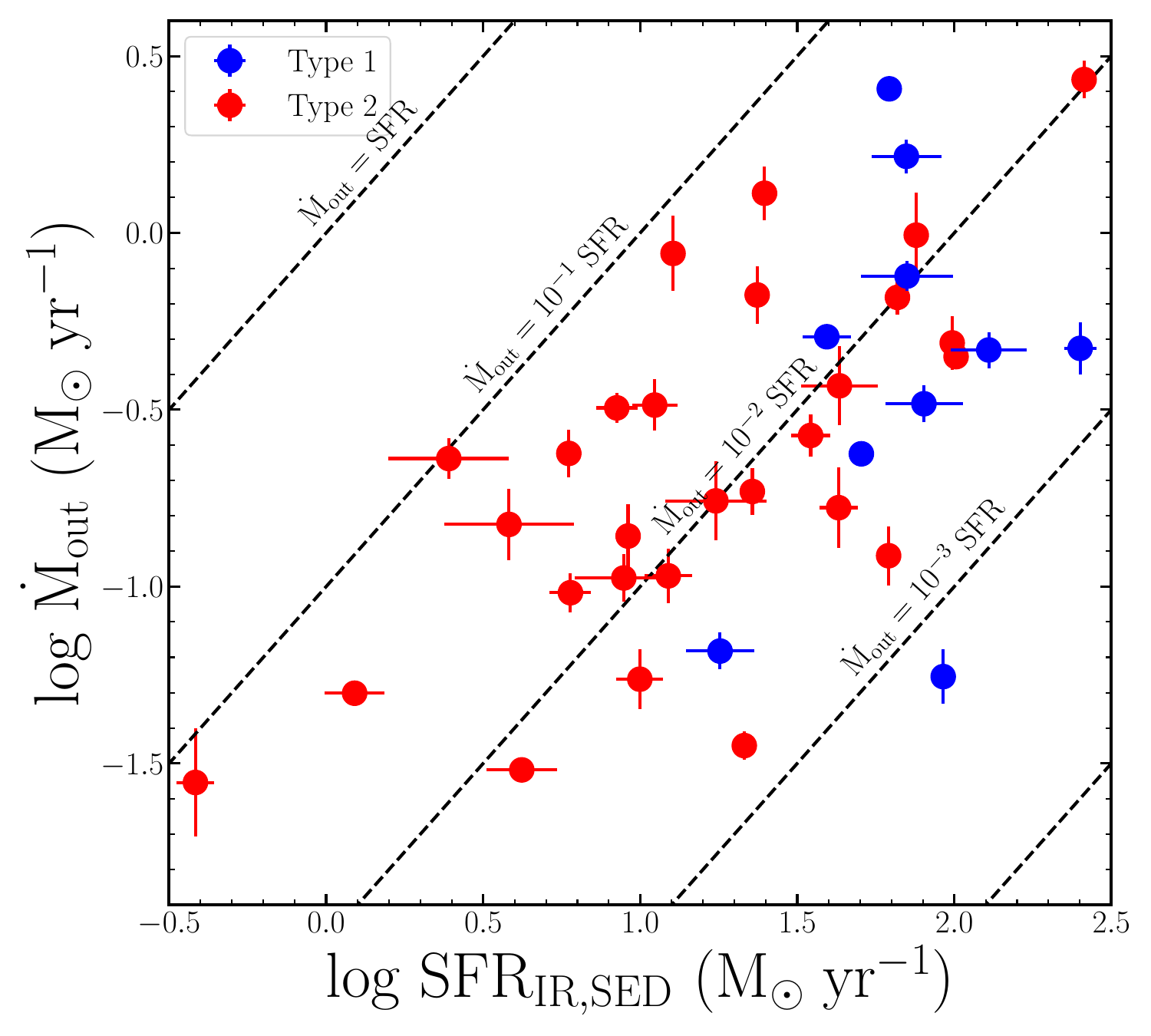}
    \caption{Mass outflow rate ($\dot{\mathrm{M}}_{\mathrm{out}}$) as a function of AGN bolometric luminosity (left) and dust-luminosity-based SFR from CIGALE (right). The dashed lines in each figure indicate where the mass outflow rate becomes a factor of 100, 10, 1, 0.1 and 0.01 of the mass accretion rate (radiative efficiency $\eta$ is assumed to be 0.1) and 1, 0.1, 0.01 and 0.001 of the SFR.}
    \label{F10_Moutdot}
\end{figure*}

\begin{figure*}[]
    \includegraphics[width=0.49\textwidth]{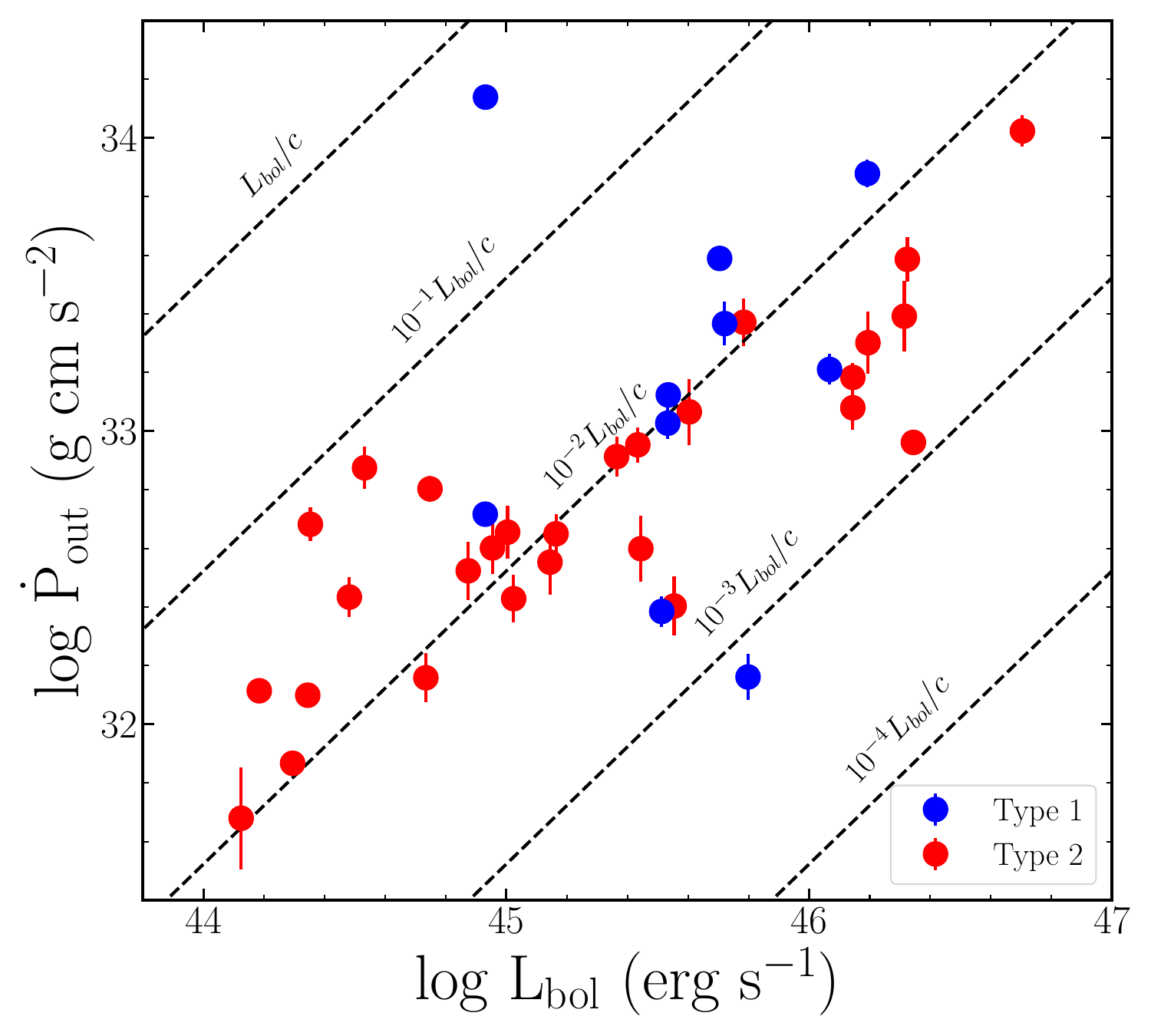}
    \includegraphics[width=0.49\textwidth]{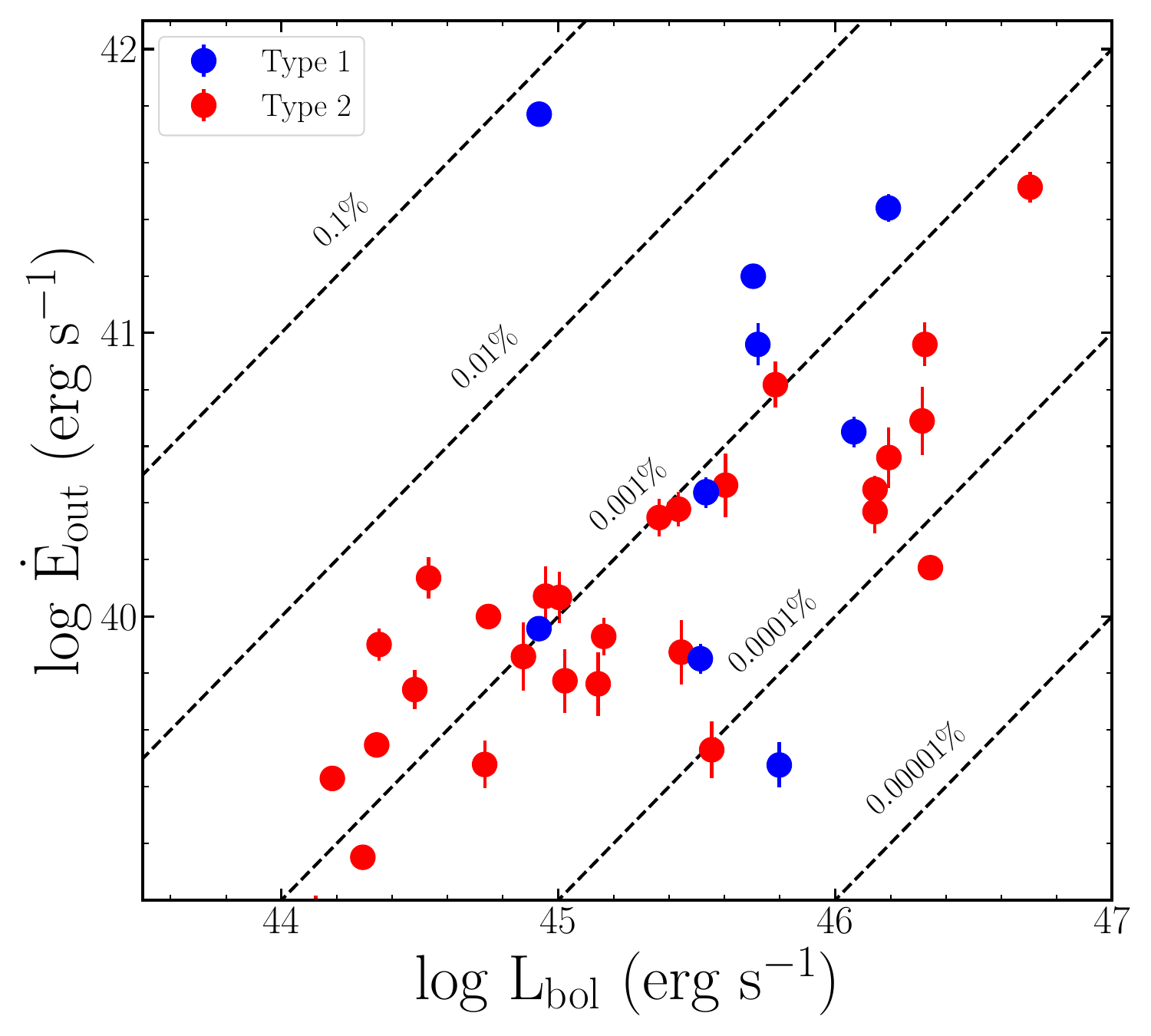}
    \caption{Bolometric luminosity dependency of $\dot{\mathrm{P}}_{\mathrm{out}}$ (left) and $\dot{\mathrm{E}}_{\mathrm{out}}$ (right). Dashed lines show the comparisons with bolometric luminosity.}
    \label{F11_P_KE_outdot}
\end{figure*}

We estimate the mass outflow rates ($\dot{\mathrm{{M}}}_\mathrm{out}$) using the measured outflow size and [\OIII] luminosity to investigate how efficiently outflows can clear gas in the host galaxies. We adopt the [\OIII]-based ionized gas mass equation from \citet{Veilleux20} and calculate the outflow mass within the outflow size.

Electron density is a crucial factor in estimating the ionized gas mass, and often assumed to be uniform with a typical value in the range, $n_e \approx$ 100--1000 cm$^{-3}$ \citep[e.g.,][]{Fiore17, Wylezalek20, Ruschel-Dutra21}, which is calculated based on the [\SII] doublet ratio ($\lambda$6717/$\lambda$6731). We measure the electron density for a half of AGNs in our sample, which show a relatively good [\SII] line profile, using PyNeb package \citep{PyNeb}. The measured $n_e$ typically ranges from 300--900 cm$^{-3}$ with a median of 456 cm$^{-3}$, which is consistent with the $n_e$ measurements reported in the literature. However, recent studies pointed out that the [\SII] method can underestimate the electron density due to the contribution of partially ionized regions \citep{Baron19, Davies20, Revalski22}. In practice, the [\SII] doublet is not covered in the observed spectra or unreliable due to the relatively weak emission line flux for a half of our combined sample. Therefore, we decided to use a representative electron density of 1000 cm$^{-3}$ as suggested by \citet{Revalski22}. 
Considering a large uncertainty of $n_e$ ($\sim$ 1 dex), this value is generally consistent with various $n_e$ at different locations. We adopt 1 dex uncertainty of $n_e$, which propagates to mass outflow rate, momentum rate, and kinetic power.

Note that dust extinction correction based on the Balmer decrement (i.e., H$\alpha$/H$\beta$) cannot be applied to the majority of the sample due to the weak H$\beta$ emission in half of type 1 AGNs and no observation of H$\alpha$ line in type 2 AGNs selected from \citet{Harrison14}. Thus, we used dust-uncorrected [\OIII] luminosity for the mass estimation, which can be conservatively interpreted as a lower limit. For outflow velocity, we used a quadratic sum of flux-weighted means of [\OIII] velocity and velocity dispersion within the outflow size, following \citet{Karouzos16b} and \citet{Rakshit18}. With the determined outflow size ($\mathrm{R}_\mathrm{out}$, see Section \ref{subsec:outflowsize}), mass ($\mathrm{M}_\mathrm{out}$) and velocity (${\mathrm{V}_\mathrm{out}}=\sqrt{v^2_\mathrm{out}+\sigma^2_\mathrm{out}}$), we estimate the mass outflow rate ($\dot{\mathrm{M}}_\mathrm{out}$) of all AGNs in the GMOS-IFU sample, assuming a uniform cone shape of outflows \citep{Fiore17} as:
\begin{equation}\label{eq_Mdot}
    \dot{\mathrm{M}}_\mathrm{out}=3\;\mathrm{M}_\mathrm{out}\frac{\mathrm{V}_\mathrm{out}}{\mathrm{R}_\mathrm{out}}
\end{equation} 

The estimated mass outflow rates range from 0.03 to 3 $\mathrm{M}_{\odot}\;\mathrm{yr^{-1}}$. These values are comparable to the previously reported estimations \citep[e.g., ][]{Baron19, Davies20}, considering the various effect of dust extinction and the range of the bolometric luminosity in different samples. Our sample also generally follows the $\dot{\mathrm{M}}_\mathrm{out}$--$\mathrm{L}_\mathrm{bol}$ relation from \citet{Fiore17} (slope of 1.29), while it is located on the low luminosity regime compared to their sample.

We compare the mass outflow rates with AGN bolometric luminosity and dust luminosity-based SFR, respectively, in Figure \ref{F10_Moutdot}. Bolometric luminosity is calculated as a factor of ten of the monochromatic luminosity at 5100$\mathrm{\AA}$ for type 1, and a factor of 3500 of the uncorrected [\OIII] luminosity for type 2 AGNs \citep{Heckman04}. Dust IR luminosity is obtained from panchromatic SED fitting using CIGALE as described in Section \ref{subsec:SEDfitting}. We find increasing trends of mass outflow rates with both bolometric luminosity and SFR. In the left panel of Figure \ref{F10_Moutdot}, many targets are located close to the diagonal line of mass accretion rate (i.e., $\dot{\mathrm{M}}_\mathrm{out} = \mathrm{L}_{bol}/\eta c^2 = \dot{\mathrm{M}}_\mathrm{acc}$, with $\eta \sim 0.1$), suggesting mass outflow rate is comparable to mass accretion rate. This result is  consistent with that of \citet{Rakshit18} if we match $n_e$ between our and their samples. Interestingly, a PSB, [1]J142230, in our sample shows the highest mass loading factor (i.e., mass outflow rate divided by mass accretion rate), indicating extreme mass outflow activity. As discussed in Section \ref{subsec:kinematics}, this extreme outflow condition might be related to the rapid quenching in this galaxy. Further studies of PSBs with AGN outflow signatures are required to understand the role of outflows in their abrupt quenching history.

In the right panel of Figure \ref{F10_Moutdot}, SFR also shows a positive correlation with mass outflow rate. This might suggest some contribution of starburst activity or supernova (SN) to drive the outflow \citep{Fluetsch19}, but note that SFR also correlates with AGN power \citep{Woo17, Woo20, Zhuang21, Kim22}. \citet{Fiore17} investigated the scaling relations of multiphase outflows and found that energy from star formation is not enough to drive powerful outflows in high luminous AGN hosts. Thus, we interpret this trend as a result of the mass outflow rate--bolometric luminosity correlation instead of a strong contribution of star formation.

Compared to SFRs, mass outflow rates are smaller by 1--3 dex, suggesting that gas consumption by star formation is more efficient than that by outflow \citep{Villar-Martin16}. Thus, gas removal by AGN-driven outflow is not a significant factor of AGN feedback, at least in the ionized gas phase. Further studies of multiphase outflows in these AGNs are necessary to diagnose their ability to clear gas reservoirs in their host galaxies.

We also calculate momentum rates ($\dot{\mathrm{{P}}}_\mathrm{out}=\dot{\mathrm{{M}}}_\mathrm{out}V_\mathrm{out}$) and kinetic power ($\dot{\mathrm{{E}}}_\mathrm{out}=\frac{1}{2}\dot{\mathrm{{M}}}_\mathrm{out}V_\mathrm{out}^2$) and compare them with bolometric luminosity in Figure \ref{F11_P_KE_outdot}. Consistent with previous studies \citep{Fiore17, Fluetsch19, Kakkad22}, similar positive trends are observed. Again, this suggests that AGN radiation is the main driving mechanism of powerful outflows. However, in terms of momentum and energy efficiency (dashed lines), the momentum rate and kinetic power of our sample are driven only by a small portion of AGN power, i.e., $\sim 1\%$ for momentum rate, and $\sim 0.001\%$ for kinetic power. In particular, coupling efficiency between kinetic power and the bolometric luminosity (i.e., $\dot{\mathrm{{E}}}_\mathrm{out}/\mathrm{L}_\mathrm{bol}$) is significantly weaker than some previous observations \citep[e.g., 0.1--10 \% in ][]{Fiore17, Kakkad22}. On the other hand, several studies suggest weak coupling efficiency as consistent with our values \citep[e.g., ][]{Baron19, Davies20, Deconto-Machado22}. Based on \citet{Hopkins10}, $\sim 0.5\%$ of the bolometric luminosity is usually adopted as a threshold of an efficient AGN feedback by outflow. Thus, the low kinetic power compared to the bolometric luminosity suggests that these outflows are unlikely to provide a significant impact on the host galaxy even if we consider the diverse factors of uncertainty such as dust extinction or electron density.

In this section, we estimated several parameters of outflow energetics, i.e., mass outflow, momentum and kinetic energy rates. While a number of observational studies have investigated the relation between mass outflow rates and AGN energetics using spatially resolved spectroscopy data \citep{Fiore17, Davies20, Wylezalek20, Revalski21, Ruschel-Dutra21, Singha22, Kakkad22, Deconto-Machado22},
these values are relatively uncertain and highly inconsistent with each other, even up to several orders of magnitude. This discrepancy originates from various assumptions and definitions adopted in the calculations of these values, i.e., outflow size, electron density, and outflow velocity. For instance, as we discussed in Sec \ref{subsec:outflowsize}, the various definitions of the outflow sizes result in a substantial difference in estimating mass outflow rate \citep[e.g., ][]{Singha22}. In the case of $n_e$, \citet{Baron19, Davies20, Revalski22} claimed that an order of magnitude larger value should be adopted in the NLR gas mass calculation. A different definition of the outflow velocity also changes the mass outflow rates \citep[e.g., $W_{80}$ or $v_{max}=v_{broad}+2\sigma_{broad}$ in ][]{Fiore17}. Dust extinction correction is not applied in some cases due to the limitation of data. Differently assumed outflow geometries can change the coefficient in Equation \ref{eq_Mdot}. In addition, recent studies adopted ``spatially resolved" mass outflow rates, which are calculated based on the velocity and density at each position \citep{Revalski21, Kakkad22}, not a single averaged value. These studies found that spatially resolved mass outflow rates tend to have peak values at certain radii and decrease at the edge. Thus, unifying the definition and assumption used in calculating outflow energetics is essential to compare various observational studies and test AGN feedback by outflows.

In summary, we find that mass outflow rates range from 0.03 to 3 $\mathrm{M}_{\odot}\;\mathrm{yr^{-1}}$, which are too small to clear the entire ISM in the host galaxies. Furthermore, mass outflow rates are significantly smaller than star formation rates, indicating that gas is dominantly consumed by star formation. Positive correlations of mass outflow rates, momentum rates, and kinetic powers with AGN luminosity indicate that AGN radiation is a main driver of the outflow, but there are still a number of uncertainties in outflow rate diagnostics due to the adopted assumptions.


\subsection{Impact on Star Formation}\label{subsec:SFR}
\begin{figure}[]
    \includegraphics[width=\columnwidth]{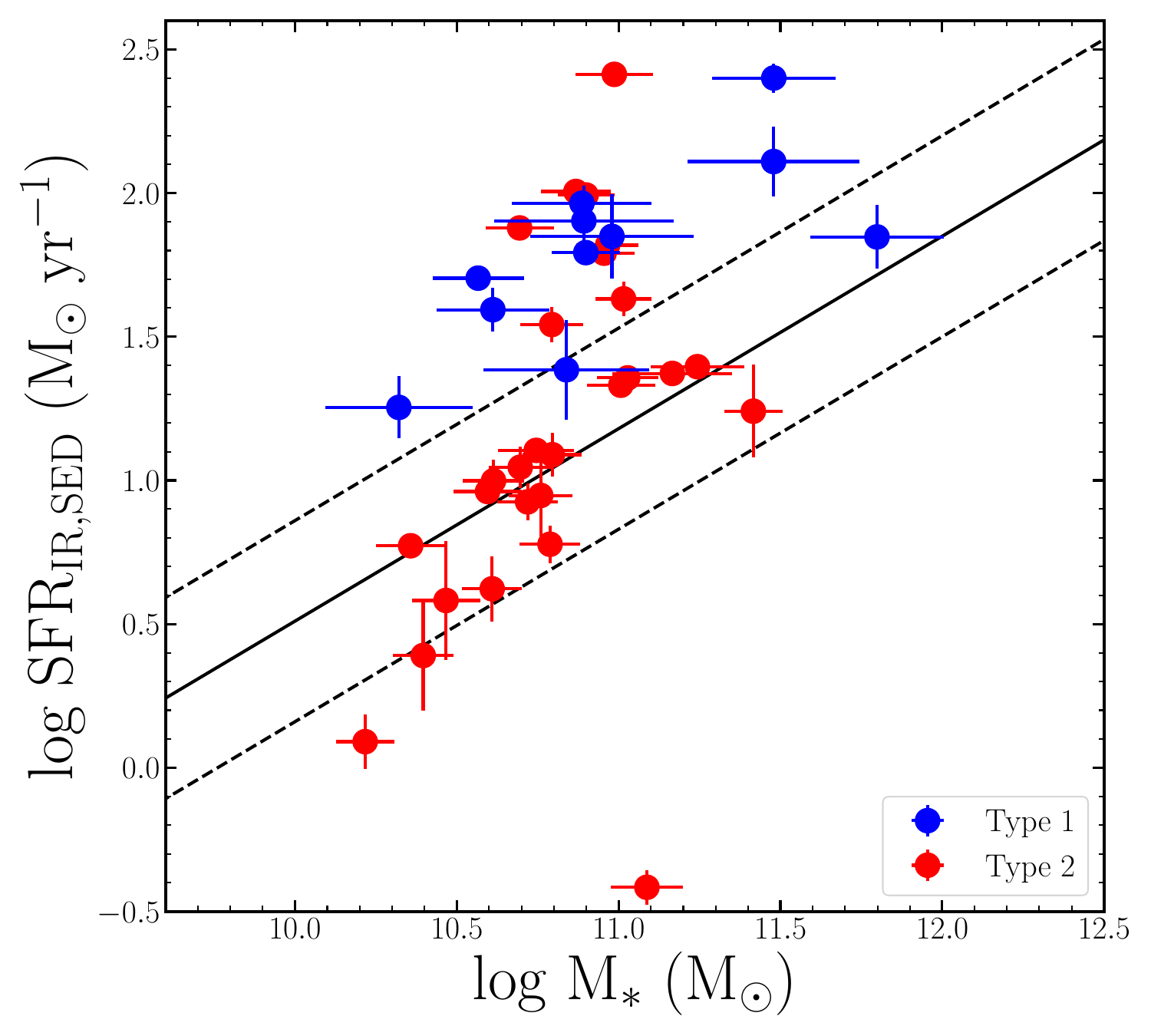}
\caption{Stellar mass--SFR relation. Solid and dashed lines represent the SFMS line and its 1 $\sigma$ distribution from \citet{Noeske07}. Strong outflow AGNs in this study and \citet{Kang18} are located on or above the SFMS line, which implies no evidence of instantaneous quenching.}
    \label{F12_Mstar_SFR}
\end{figure}

In the previous sections, we tested the possibility of global star formation quenching by AGN-driven outflows in terms of outflow size and mass outflow rates. The small values of those quantities indicate a lack of global quenching impact by outflows; rather, they may be confined to the central kpc region. Here, we discuss star formation activity in our GMOS-IFU sample and summarize implications for AGN feedback scenarios.

We plot the stellar mass--SFR relation with the star-forming main sequence (SFMS) distribution from \citet{Noeske07} in Figure \ref{F12_Mstar_SFR}. We find that, consistent with \citet{Woo20} and \citet{Kim22}, strong outflow AGNs are mostly located on or above the SFMS, which indicates that these AGNs are still actively forming stars as many as the SFMS or even starburst galaxies. Therefore, even though strong AGN-driven outflows exist in the central few-kpc region, they have not shut down global star formation in the host galaxies yet. Rather, enhanced star formation activity indicates that the common gas supply may trigger both AGN activity (radiation and outflow) and star formation simultaneously. Note that type 1 AGNs seem to show more actively forming stars than type 2 due to the selection effect since type 1 AGNs are more luminous than type 2 AGNs in our sample, and AGN luminosity correlates with SFR, as reported by \citet{Woo20, Kim22}.

In \citet{Karouzos16b}, we detected circumnuclear star-forming regions in BPT maps of type 2 AGNs with strong outflows near the edge of outflows. We interpreted this as star formation enhancement by contact with outflows or star-forming regions in the disk. We also noted that the presence of nuclear star formation cannot be ruled out even in AGN photoionization regions since relatively strong emission from AGN can cover photoionization by star formation in the central region. The high SFRs of our GMOS-IFU sample are consistent with those scenarios. In other words, we confirm that star formation and AGN activity coexist at the center and that star formation is not rapidly suppressed due to strong outflows.

As suggested in our previous studies, there could be several different scenarios that explain the connection between AGN and star formation \citep{Woo17, Woo20, Kim22}. For example, feed and consumption of gas may control both AGN and star formation. When gas is supplied, both AGN and star formation activity are triggered at the same time. Once the supplied gas is exhausted for black hole accretion and star formation, the AGN activity and star formation become weaker.

However, numerical simulations and semianalytical models can reproduce the coexistence of AGN and star formation activities, while strong AGN feedback plays an important role in quenching star formation \citep[e.g.,][]{Ward22}. Rather, we interpret the correlation between AGN activity and star formation strength with a delayed AGN feedback framework. Once a large amount of gas is supplied to the galaxy, the gas may trigger both star formation and AGN activity (e.g., Eddington ratio, outflow strength), which is the snapshot that we observe in AGN with strong outflow. After a certain time scale, which is required for AGN-driven outflows to terminate/reduce star formation, both AGN and star formation activity decrease.

Note that there were also opposite results from previous IFU studies, claiming that strong outflows can suppress star formation immediately by sweeping up the ISM near the center. For example, \citet{Cano-Diaz12}, \citet{Cresci15}, and \citet{Carniani16} reported the spatial anti-correlation between outflows and star formation, i.e., outflow cavity, and star formation enhancement at the edge of the cavity in four high luminosity quasars at cosmic noon. However, recent studies by \citet{Scholtz20} and \citet{Scholtz21} performed a reanalysis of those quasars along with additional ALMA analysis and found that dust continuum emission by star formation still remains at the center despite the presence of extreme outflows, indicating no evidence of star formation suppression or cavity by outflows. High SFRs based on the dust luminosity of our sample are also in good agreement with later studies, although we were not able to resolve the spatial distribution of star formation.

It is worth noting that [1]J142230 shows a high SFR (SFR$_{\mathrm{IR, SED}} \sim 62\;\mathrm{M}_\odot\;\mathrm{yr^{-1}}$) despite its PSB property. This is another interesting characteristic of this unique target since the optical spectrum and dust luminosity show discrepancies. There could be several interpretations for this target. Obviously, the simplest explanation is that this target had a much higher level of star formation in the past, and its SFR has not yet decreased below the SFMS level. Another possibility is that the dust luminosity-based SFR traces a longer time scale, so starburst in the past can contribute to excess dust emissions. Recently, \citet{Baron22b} found that some PSBs still show a large amount of star formation in FIR, perhaps originating from obscured star formation that is not traced by optical spectra. The specific properties and history of these PSBs are still mysterious and highly important in galaxy evolution. Further studies will investigate them, especially PSBs with strong outflow AGN such as [1]J142230, to elucidate the delayed AGN feedback scenario.

In summary, most of our targets are still forming stars as many as normal SFMS galaxies or even starburst galaxies despite their energetic outflows. Thus, we did not find any sign of instantaneous star formation quenching by strong outflows. The delayed AGN feedback scenario is a possible solution to explain this phenomenon. To investigate this scenario, PSB may be holding a key to revealing galaxy quenching stages in detail.

\section{Summary} \label{sec:summary}
We present spatially resolved characteristics and kinematic outflow sizes of 11 type 1 and 5 type 2 AGN host galaxies with strong outflows at $z<0.3$ using GMOS-IFU data. Combined with the previous IFU sample, we investigated the outflow scaling relations with AGN luminosity and examined whether strong gas outflow has the capability of suppressing global star formation in the host galaxy. In addition, we performed SED analysis with JCMT/SCUBA-2 data to obtain dust luminosity-based SFRs of the sample to search for implications of star formation quenching by AGN-driven outflows. The key findings are as follows.

\begin{itemize}
    \item We find that [\OIII] kinematics clearly show the blueshifted central region indicating the approaching outflow cone in the biconical model, while H$\alpha$ is dominated by rotational motion. The BPT maps of most targets are dominated by AGN photoionization at the center. These results are consistent with our previous IFU studies of type 2 AGNs.

    \item Based on kinematics, we determined the outflow sizes, which are typically several kpc scales. This suggests that the impact of AGN-driven outflows is limited only at the central bulge scale, not the galactic scale.

    \item We expand the previous $\mathrm{R}_{\mathrm{out}}$--$\mathrm{L}_{[\mathrm{OIII}]}$ relation by including new type 1 AGNs and updated the slope as $0.29\pm0.04$ ($0.22\pm0.03$ when including the SNIFS sample). Type 1 AGNs follow the same positive trend, which implies that the opening angle of outflow seems to be a more critical factor than the viewing angle in deciding the outflow size.

    \item The photoionization size is $\sim$ 38\% larger than the outflow size, and clearly correlates with [\OIII] luminosity with the best-fit slope of 0.33$\pm$0.04. This indicates that outflow is less efficiently delivered outward than ionizing photons, presumably due to the influence of the ISM or gravitational force.

    \item Our sample has SFRs similar to or higher than those of SFMS galaxies and low mass outflow rates. This disagrees with the instantaneous star formation quenching scenario by AGN-driven outflows, while the delayed AGN feedback scenario is a possible solution. PSBs with AGN may provide promising clues to determining the AGN feedback mechanisms in the future.
\end{itemize}

\section*{acknowledgements}
We thank the anonymous referee for various comments, which were useful to improve the clarity of the manuscript. This work has been supported by the Basic Science Research Program through the National Research Foundation of Korean Government (NRF-2021R1A2C3008486). This work was supported by the K-GMT Science Program (PID: GN-2016B-Q-30, GS-2016B-Q-30, GN-2019A-Q-201, GN-2019B-Q-137, GN-2019B-Q-236, GN-2019B-327) of the Korea Astronomy and Space Science Institute (KASI). Based on observations obtained at the Gemini Observatory processed using the Gemini IRAF package, which is operated by the Association of Universities for Research in Astronomy, Inc. (AURA) under a cooperative agreement with the NSF on behalf of the Gemini partnership: the US National Science Foundation (NSF), the Canadian National Research Council (NRC), the Chilean Comisión Nacional de Investiga-ción Cientifica y Tecnológica (CONICYT), the Brazilian Ministério da Ciência, the Argentinean Ministerio de Ciencia, Tecnología e Innovación Productiva, Tecnologia e Inovação and the Korea Astronomy and Space Institute (KASI). This research has made use of the NASA/IPAC Extragalactic Database (NED), which is operated by the Jet Propulsion Laboratory, California Institute of Technology, under contract with the National Aeronautics and Space Administration. H.A.N. Le acknowledges support from the National Natural Science Foundation of China (NSFC-12003031) and the “Fundamental Research Funds for the Central Universities".

\software{Astropy \citep{Astropy1, Astropy2}, CIGALE \citep{CIGALE09, CIGALE19}, matplotlib \citep{matplotlib}, numpy \citep{numpy}, ORAC-DR \citep{ORACDR15}, pandas \citep{pandas}, PyNeb \citep{PyNeb}, scipy \citep{scipy}, Starlink \citep{Starlink14}}

\appendix
\section{Kinematics of Core \& Wing components}
\begin{figure*}[]
\centering
\includegraphics[width=\textwidth]{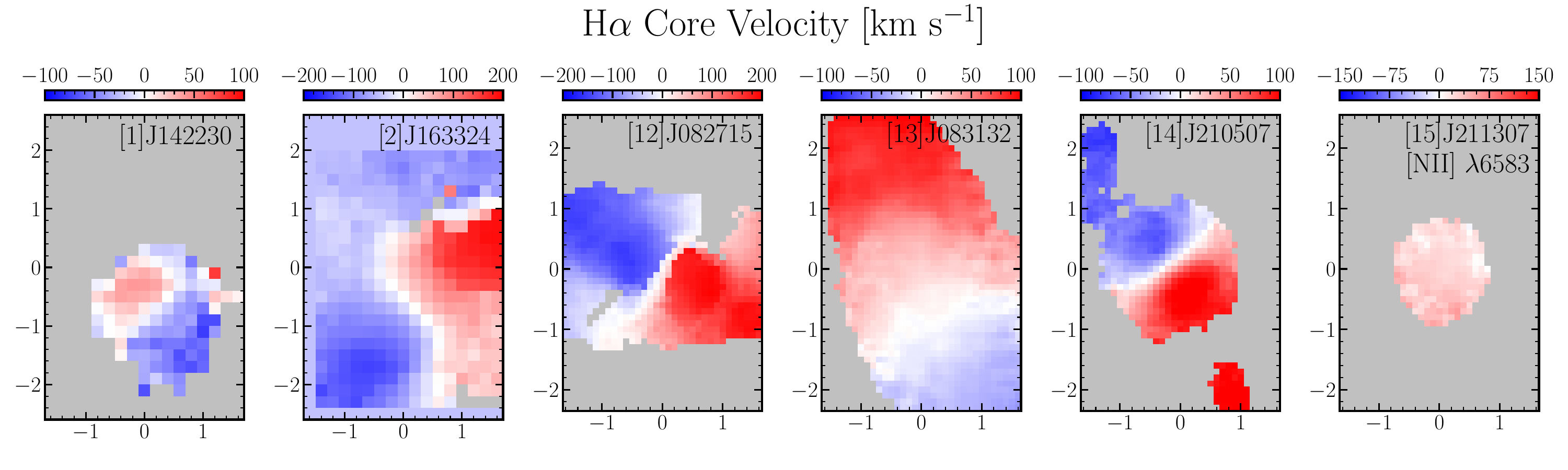}\\
\includegraphics[width=\textwidth]{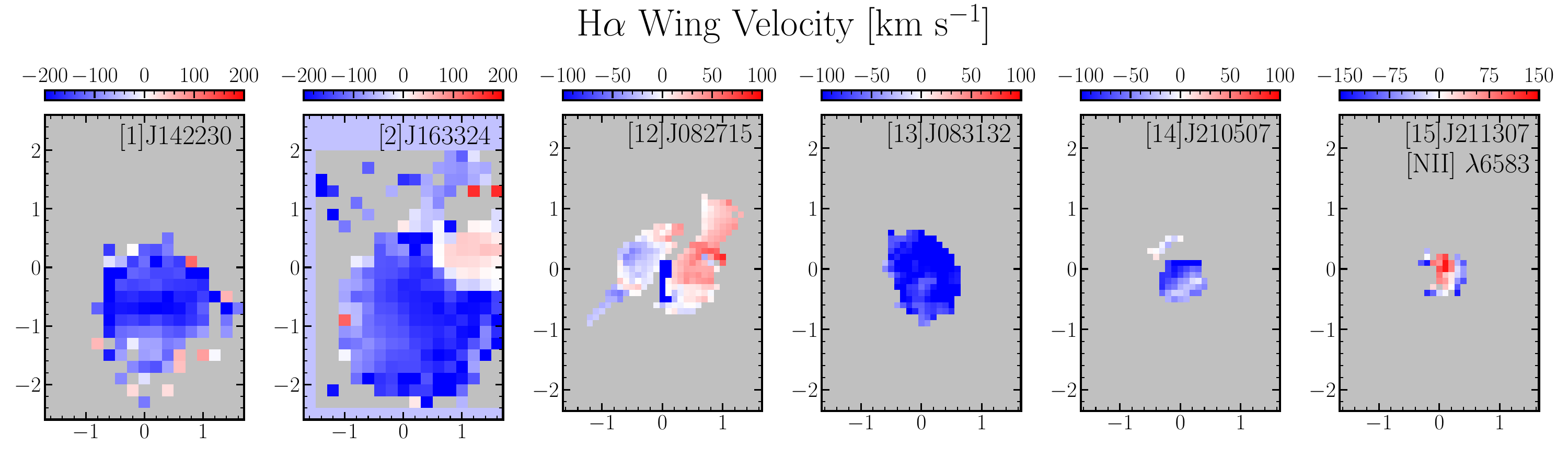}
\caption{Velocity maps of the H$\alpha$ core (top) and wing (bottom) components.}
\label{A1_Ha_Velocitymaps}
\end{figure*}

\begin{figure*}[]
\centering
\includegraphics[width=\textwidth]{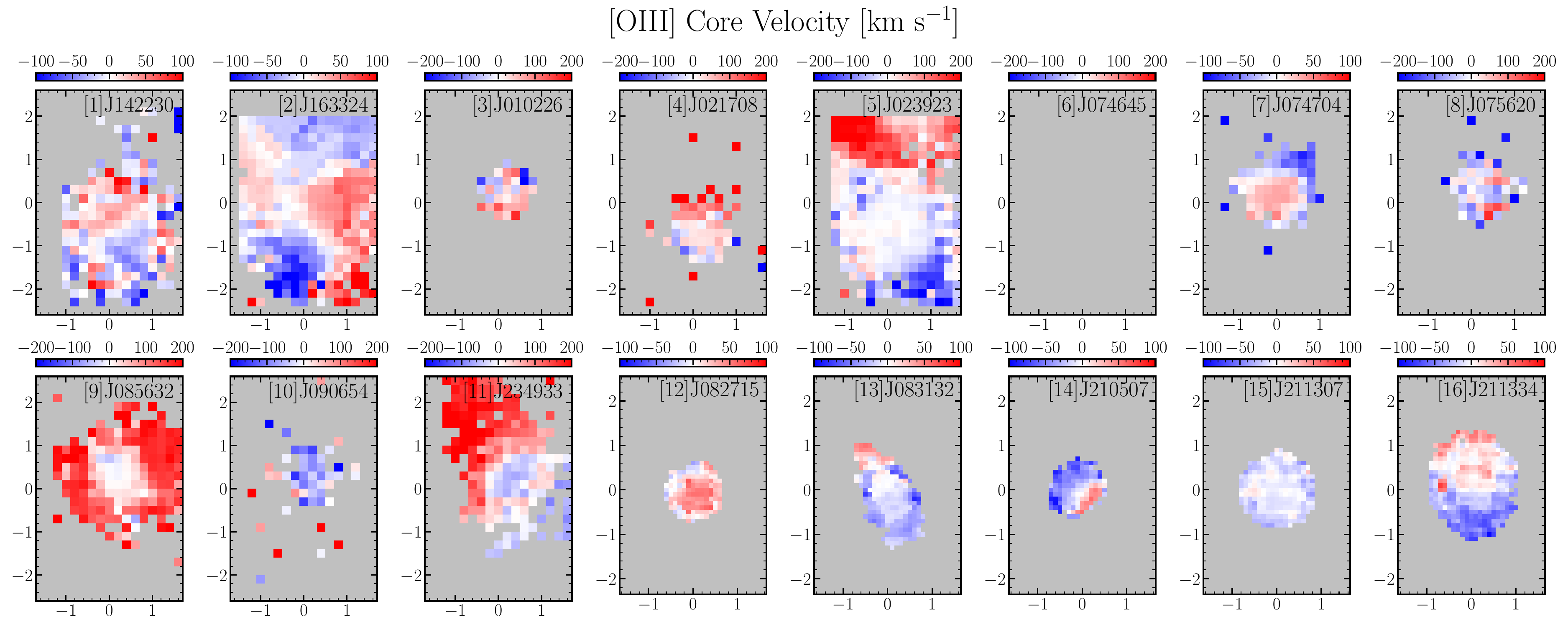}\\
\includegraphics[width=\textwidth]{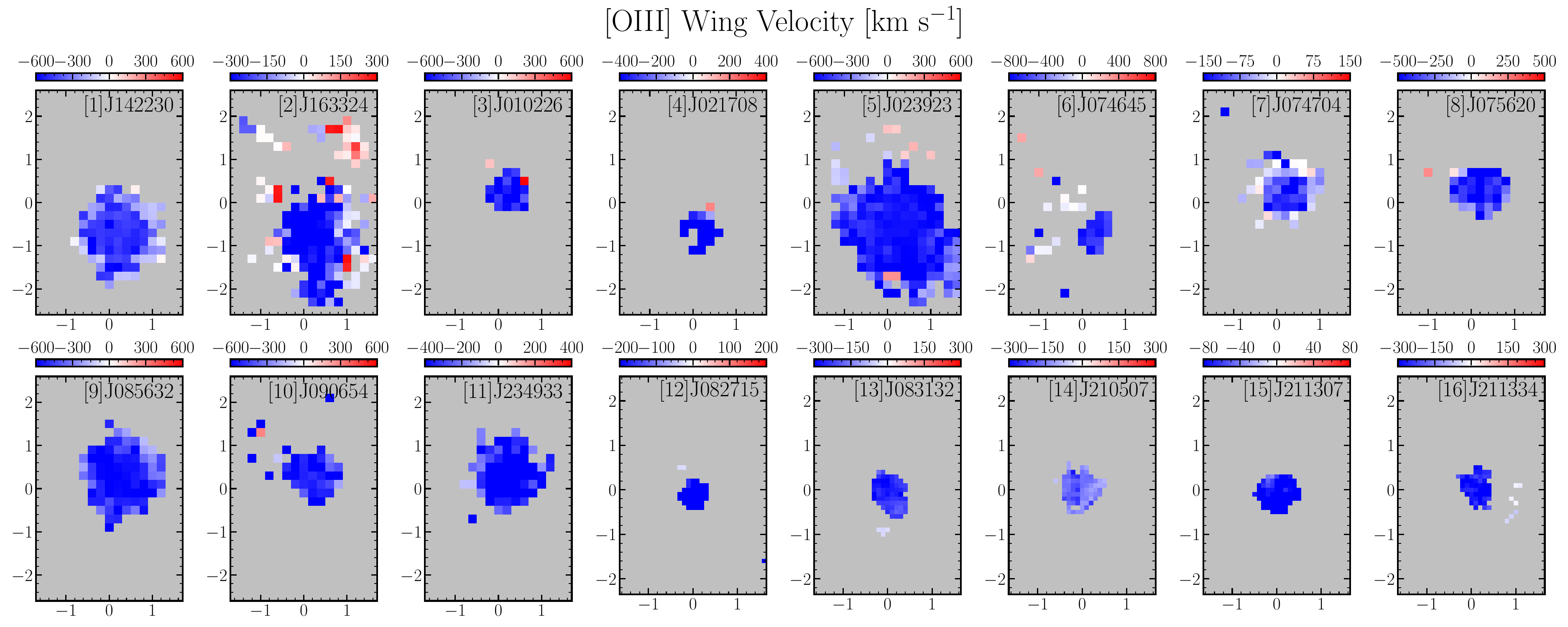}
\caption{Velocity maps of the [\OIII] core (top) and wing (bottom) components.}
\label{A1_OIII_Velocitymaps}
\end{figure*}

\begin{figure*}[]
\centering
\includegraphics[width=\textwidth]{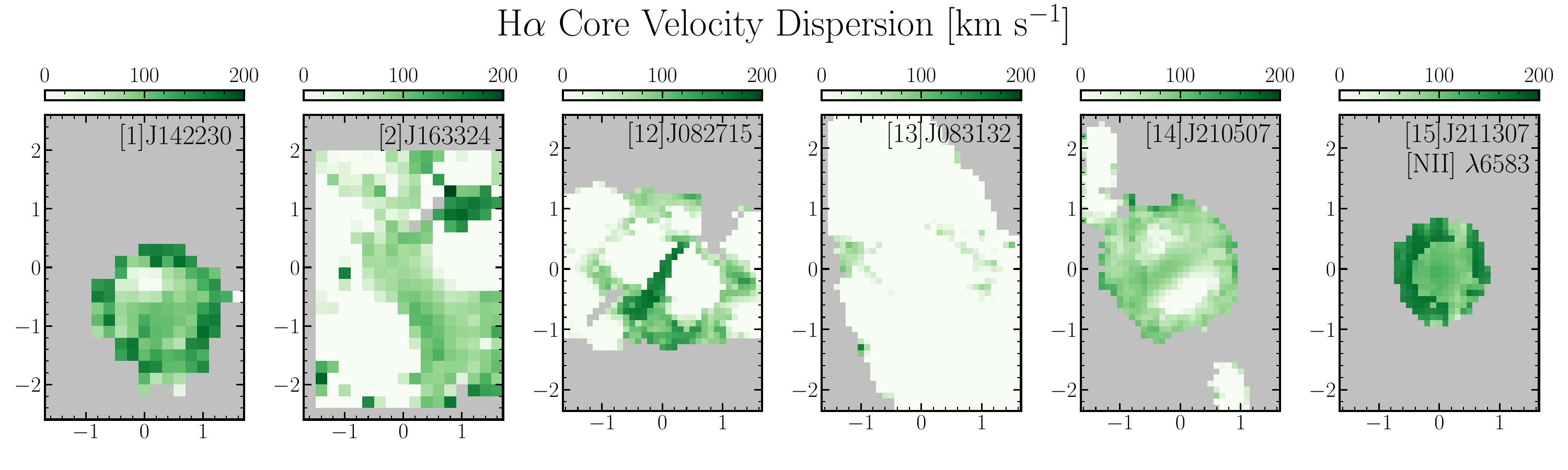}\\
\includegraphics[width=\textwidth]{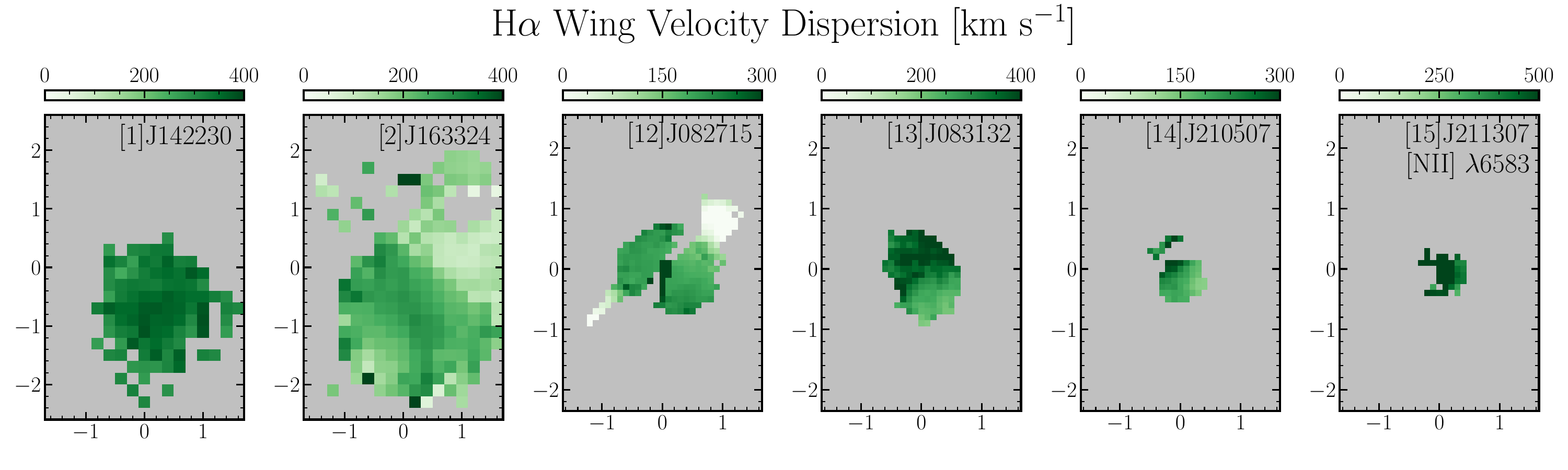}
\caption{Velocity dispersion maps of H$\alpha$ core (top) and wing (bottom) components.}
\label{A2_Ha_Vdispmaps}
\end{figure*}

\begin{figure*}[]
\centering
\includegraphics[width=\textwidth]{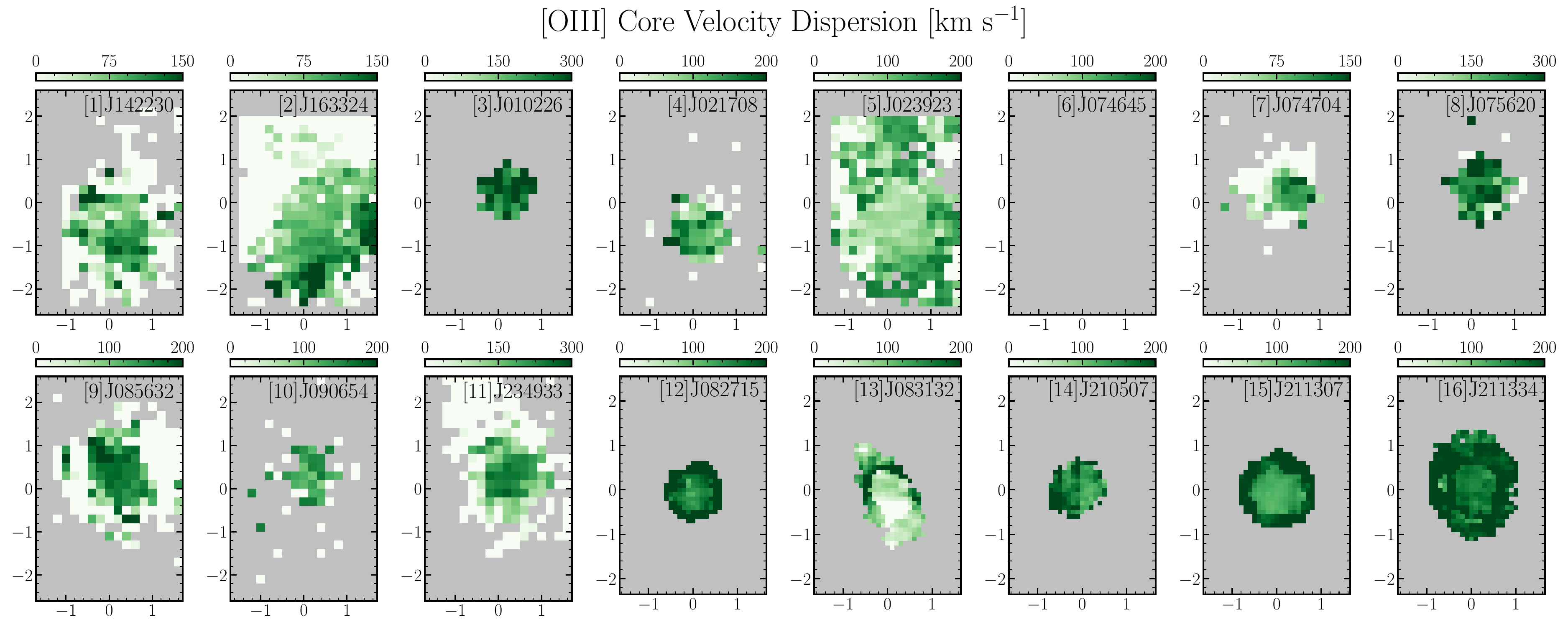}\\
\includegraphics[width=\textwidth]{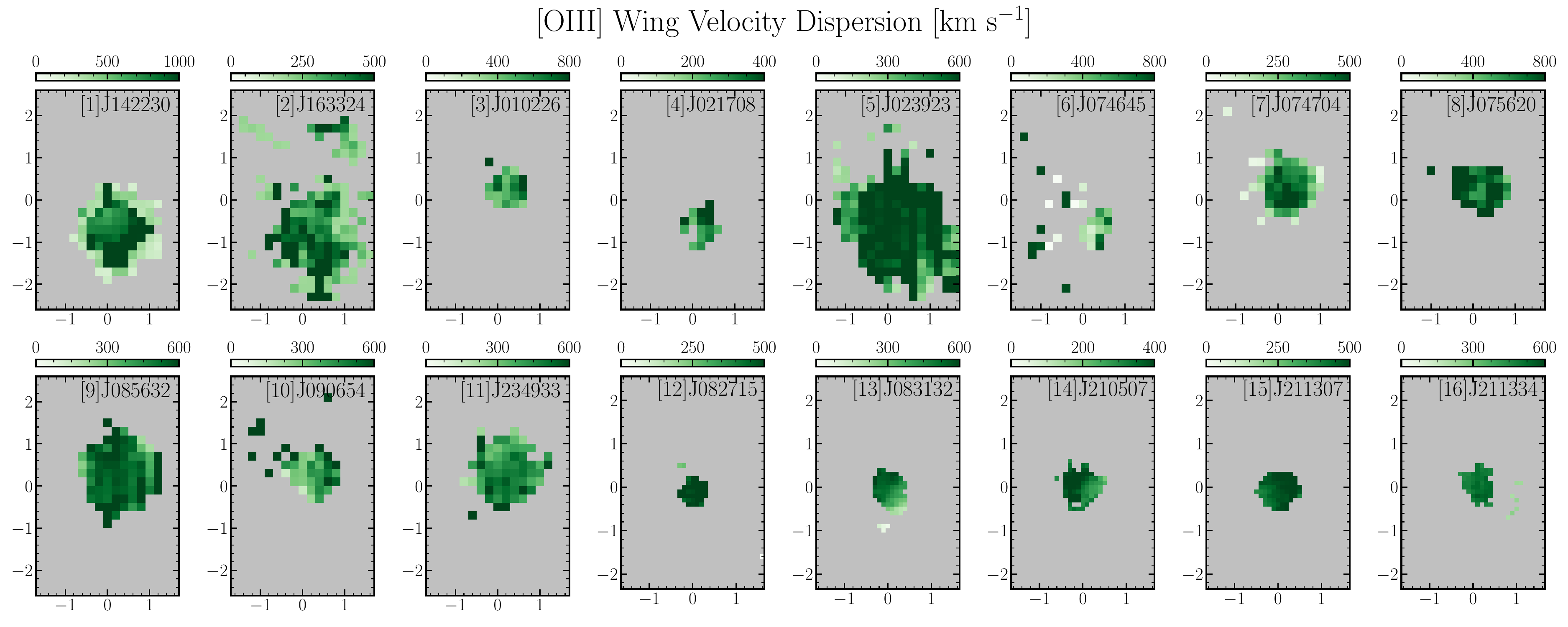}
\caption{Velocity dispersion maps of [\OIII] core (top) and wing (bottom) components.}
\label{A2_OIII_Vdispmaps}
\end{figure*}

We present velocity and velocity dispersion maps of H$\alpha$ and [\OIII], by separating core \& wing components in the emission line profile, respectively, in Figure \ref{A1_Ha_Velocitymaps}, Figure \ref{A1_OIII_Velocitymaps}, Figure \ref{A2_Ha_Vdispmaps}, and Figure \ref{A2_OIII_Vdispmaps}.
Since the [\OIII] of [6]J074645 was fitted with a single Gaussian, we present this target only in the [\OIII] wing map. For H$\alpha$, as six targets were fitted with a double Gaussian, we show these six targets in the H$\alpha$ core and wing maps. 
In general, velocity maps of the [\OIII] core component follow similar rotational patterns compared to the H$\alpha$. The maps of the [\OIII] and H$\alpha$ wing component generally show co-spatial blueshifted region at the center, implying that the two core components trace the same outflow origin, while the [\OIII] wing component shows more extreme kinematics than the H$\alpha$ wing component.

\bibliography{ref}

\begin{thebibliography}{}
\expandafter\ifx\csname natexlab\endcsname\relax\def\natexlab#1{#1}\fi
\providecommand{\url}[1]{\href{#1}{#1}}
\providecommand{\dodoi}[1]{doi:~\href{http://doi.org/#1}{\nolinkurl{#1}}}
\providecommand{\doeprint}[1]{\href{http://ascl.net/#1}{\nolinkurl{http://ascl.net/#1}}}
\providecommand{\doarXiv}[1]{\href{https://arxiv.org/abs/#1}{\nolinkurl{https://arxiv.org/abs/#1}}}

\bibitem[{{Astropy Collaboration} {et~al.}(2013){Astropy Collaboration}, {Robitaille}, {Tollerud}, {Greenfield}, {Droettboom}, {Bray}, {Aldcroft}, {Davis}, {Ginsburg}, {Price-Whelan}, {Kerzendorf}, {Conley}, {Crighton}, {Barbary}, {Muna}, {Ferguson}, {Grollier}, {Parikh}, {Nair}, {Unther}, {Deil}, {Woillez}, {Conseil}, {Kramer}, {Turner}, {Singer}, {Fox}, {Weaver}, {Zabalza}, {Edwards}, {Azalee Bostroem}, {Burke}, {Casey}, {Crawford}, {Dencheva}, {Ely}, {Jenness}, {Labrie}, {Lim}, {Pierfederici}, {Pontzen}, {Ptak}, {Refsdal}, {Servillat}, \& {Streicher}}]{Astropy1}
{Astropy Collaboration}, {Robitaille}, T.~P., {Tollerud}, E.~J., {et~al.} 2013, \aap, 558, A33, \dodoi{10.1051/0004-6361/201322068}

\bibitem[{{Astropy Collaboration} {et~al.}(2018){Astropy Collaboration}, {Price-Whelan}, {Sip{\H{o}}cz}, {G{\"u}nther}, {Lim}, {Crawford}, {Conseil}, {Shupe}, {Craig}, {Dencheva}, {Ginsburg}, {VanderPlas}, {Bradley}, {P{\'e}rez-Su{\'a}rez}, {de Val-Borro}, {Aldcroft}, {Cruz}, {Robitaille}, {Tollerud}, {Ardelean}, {Babej}, {Bach}, {Bachetti}, {Bakanov}, {Bamford}, {Barentsen}, {Barmby}, {Baumbach}, {Berry}, {Biscani}, {Boquien}, {Bostroem}, {Bouma}, {Brammer}, {Bray}, {Breytenbach}, {Buddelmeijer}, {Burke}, {Calderone}, {Cano Rodr{\'\i}guez}, {Cara}, {Cardoso}, {Cheedella}, {Copin}, {Corrales}, {Crichton}, {D'Avella}, {Deil}, {Depagne}, {Dietrich}, {Donath}, {Droettboom}, {Earl}, {Erben}, {Fabbro}, {Ferreira}, {Finethy}, {Fox}, {Garrison}, {Gibbons}, {Goldstein}, {Gommers}, {Greco}, {Greenfield}, {Groener}, {Grollier}, {Hagen}, {Hirst}, {Homeier}, {Horton}, {Hosseinzadeh}, {Hu}, {Hunkeler}, {Ivezi{\'c}}, {Jain}, {Jenness}, {Kanarek}, {Kendrew}, {Kern}, {Kerzendorf}, {Khvalko}, {King}, {Kirkby}, {Kulkarni},
  {Kumar}, {Lee}, {Lenz}, {Littlefair}, {Ma}, {Macleod}, {Mastropietro}, {McCully}, {Montagnac}, {Morris}, {Mueller}, {Mumford}, {Muna}, {Murphy}, {Nelson}, {Nguyen}, {Ninan}, {N{\"o}the}, {Ogaz}, {Oh}, {Parejko}, {Parley}, {Pascual}, {Patil}, {Patil}, {Plunkett}, {Prochaska}, {Rastogi}, {Reddy Janga}, {Sabater}, {Sakurikar}, {Seifert}, {Sherbert}, {Sherwood-Taylor}, {Shih}, {Sick}, {Silbiger}, {Singanamalla}, {Singer}, {Sladen}, {Sooley}, {Sornarajah}, {Streicher}, {Teuben}, {Thomas}, {Tremblay}, {Turner}, {Terr{\'o}n}, {van Kerkwijk}, {de la Vega}, {Watkins}, {Weaver}, {Whitmore}, {Woillez}, {Zabalza}, \& {Astropy Contributors}}]{Astropy2}
{Astropy Collaboration}, {Price-Whelan}, A.~M., {Sip{\H{o}}cz}, B.~M., {et~al.} 2018, \aj, 156, 123, \dodoi{10.3847/1538-3881/aabc4f}

\bibitem[{{Bae} \& {Woo}(2014)}]{Bae14}
{Bae}, H.-J., \& {Woo}, J.-H. 2014, \apj, 795, 30, \dodoi{10.1088/0004-637X/795/1/30}

\bibitem[{{Bae} \& {Woo}(2016)}]{Bae16}
---. 2016, \apj, 828, 97, \dodoi{10.3847/0004-637X/828/2/97}

\bibitem[{{Bae} {et~al.}(2017){Bae}, {Woo}, {Karouzos}, {Gallo}, {Flohic}, {Shen}, \& {Yoon}}]{Bae17}
{Bae}, H.-J., {Woo}, J.-H., {Karouzos}, M., {et~al.} 2017, \apj, 837, 91, \dodoi{10.3847/1538-4357/aa5f5c}

\bibitem[{{Baron} \& {Netzer}(2019)}]{Baron19}
{Baron}, D., \& {Netzer}, H. 2019, \mnras, 486, 4290, \dodoi{10.1093/mnras/stz1070}

\bibitem[{{Baron} {et~al.}(2022){Baron}, {Netzer}, {French}, {Lutz}, {Davies}, \& {Prochaska}}]{Baron22b}
{Baron}, D., {Netzer}, H., {French}, K.~D., {et~al.} 2022, arXiv e-prints, arXiv:2204.11881.
\newblock \doarXiv{2204.11881}

\bibitem[{{Baron} {et~al.}(2018){Baron}, {Netzer}, {Prochaska}, {Cai}, {Cantalupo}, {Martin}, {Matuszewski}, {Moore}, {Morrissey}, \& {Neill}}]{Baron18}
{Baron}, D., {Netzer}, H., {Prochaska}, J.~X., {et~al.} 2018, \mnras, 480, 3993, \dodoi{10.1093/mnras/sty2113}

\bibitem[{{Bennert} {et~al.}(2002){Bennert}, {Falcke}, {Schulz}, {Wilson}, \& {Wills}}]{Bennert02}
{Bennert}, N., {Falcke}, H., {Schulz}, H., {Wilson}, A.~S., \& {Wills}, B.~J. 2002, \apjl, 574, L105, \dodoi{10.1086/342420}

\bibitem[{{Bennert} {et~al.}(2006{\natexlab{a}}){Bennert}, {Jungwiert}, {Komossa}, {Haas}, \& {Chini}}]{Bennert06a}
{Bennert}, N., {Jungwiert}, B., {Komossa}, S., {Haas}, M., \& {Chini}, R. 2006{\natexlab{a}}, \aap, 446, 919, \dodoi{10.1051/0004-6361:20053571}

\bibitem[{{Bennert} {et~al.}(2006{\natexlab{b}}){Bennert}, {Jungwiert}, {Komossa}, {Haas}, \& {Chini}}]{Bennert06b}
---. 2006{\natexlab{b}}, \aap, 456, 953, \dodoi{10.1051/0004-6361:20065319}

\bibitem[{{Bennert} {et~al.}(2006{\natexlab{c}}){Bennert}, {Jungwiert}, {Komossa}, {Haas}, \& {Chini}}]{Bennert06c}
---. 2006{\natexlab{c}}, \aap, 459, 55, \dodoi{10.1051/0004-6361:20065477}

\bibitem[{{Bessiere} \& {Ramos Almeida}(2022)}]{Bessiere22}
{Bessiere}, P.~S., \& {Ramos Almeida}, C. 2022, \mnras, 512, L54, \dodoi{10.1093/mnrasl/slac016}

\bibitem[{{Bianchin} {et~al.}(2022){Bianchin}, {Riffel}, {Storchi-Bergmann}, {Riffel}, {Ruschel-Dutra}, {Harrison}, {Dahmer-Hahn}, {Mainieri}, {Sch{\"o}nell}, \& {Dametto}}]{Bianchin22}
{Bianchin}, M., {Riffel}, R.~A., {Storchi-Bergmann}, T., {et~al.} 2022, \mnras, 510, 639, \dodoi{10.1093/mnras/stab3468}

\bibitem[{{Boquien} {et~al.}(2019){Boquien}, {Burgarella}, {Roehlly}, {Buat}, {Ciesla}, {Corre}, {Inoue}, \& {Salas}}]{CIGALE19}
{Boquien}, M., {Burgarella}, D., {Roehlly}, Y., {et~al.} 2019, \aap, 622, A103, \dodoi{10.1051/0004-6361/201834156}

\bibitem[{{Boroson}(2005)}]{Boroson05}
{Boroson}, T. 2005, \aj, 130, 381, \dodoi{10.1086/431722}

\bibitem[{{Boroson} \& {Green}(1992)}]{BG92_FeII}
{Boroson}, T.~A., \& {Green}, R.~F. 1992, \apjs, 80, 109, \dodoi{10.1086/191661}

\bibitem[{{Cano-D{\'\i}az} {et~al.}(2012){Cano-D{\'\i}az}, {Maiolino}, {Marconi}, {Netzer}, {Shemmer}, \& {Cresci}}]{Cano-Diaz12}
{Cano-D{\'\i}az}, M., {Maiolino}, R., {Marconi}, A., {et~al.} 2012, \aap, 537, L8, \dodoi{10.1051/0004-6361/201118358}

\bibitem[{{Cappellari}(2017)}]{ppxf17}
{Cappellari}, M. 2017, \mnras, 466, 798, \dodoi{10.1093/mnras/stw3020}

\bibitem[{{Carniani} {et~al.}(2015){Carniani}, {Marconi}, {Maiolino}, {Balmaverde}, {Brusa}, {Cano-D{\'\i}az}, {Cicone}, {Comastri}, {Cresci}, {Fiore}, {Feruglio}, {La Franca}, {Mainieri}, {Mannucci}, {Nagao}, {Netzer}, {Piconcelli}, {Risaliti}, {Schneider}, \& {Shemmer}}]{Carniani15}
{Carniani}, S., {Marconi}, A., {Maiolino}, R., {et~al.} 2015, \aap, 580, A102, \dodoi{10.1051/0004-6361/201526557}

\bibitem[{{Carniani} {et~al.}(2016){Carniani}, {Marconi}, {Maiolino}, {Balmaverde}, {Brusa}, {Cano-D{\'\i}az}, {Cicone}, {Comastri}, {Cresci}, {Fiore}, {Feruglio}, {La Franca}, {Mainieri}, {Mannucci}, {Nagao}, {Netzer}, {Piconcelli}, {Risaliti}, {Schneider}, \& {Shemmer}}]{Carniani16}
---. 2016, \aap, 591, A28, \dodoi{10.1051/0004-6361/201528037}

\bibitem[{{Castell{\'o}-Mor} {et~al.}(2012){Castell{\'o}-Mor}, {Barcons}, {Ballo}, {Carrera}, {Ward}, \& {Jin}}]{Castello-Mor12}
{Castell{\'o}-Mor}, N., {Barcons}, X., {Ballo}, L., {et~al.} 2012, \aap, 544, A48, \dodoi{10.1051/0004-6361/201118301}

\bibitem[{{Choi} {et~al.}(2015){Choi}, {Ostriker}, {Naab}, {Oser}, \& {Moster}}]{Choi15}
{Choi}, E., {Ostriker}, J.~P., {Naab}, T., {Oser}, L., \& {Moster}, B.~P. 2015, \mnras, 449, 4105, \dodoi{10.1093/mnras/stv575}

\bibitem[{{Cicone} {et~al.}(2014){Cicone}, {Maiolino}, {Sturm}, {Graci{\'a}-Carpio}, {Feruglio}, {Neri}, {Aalto}, {Davies}, {Fiore}, {Fischer}, {Garc{\'\i}a-Burillo}, {Gonz{\'a}lez-Alfonso}, {Hailey-Dunsheath}, {Piconcelli}, \& {Veilleux}}]{Cicone14}
{Cicone}, C., {Maiolino}, R., {Sturm}, E., {et~al.} 2014, \aap, 562, A21, \dodoi{10.1051/0004-6361/201322464}

\bibitem[{{Cid Fernandes} {et~al.}(2011){Cid Fernandes}, {Stasi{\'n}ska}, {Mateus}, \& {Vale Asari}}]{CidFernandes11}
{Cid Fernandes}, R., {Stasi{\'n}ska}, G., {Mateus}, A., \& {Vale Asari}, N. 2011, \mnras, 413, 1687, \dodoi{10.1111/j.1365-2966.2011.18244.x}

\bibitem[{{Cid Fernandes} {et~al.}(2010){Cid Fernandes}, {Stasi{\'n}ska}, {Schlickmann}, {Mateus}, {Vale Asari}, {Schoenell}, \& {Sodr{\'e}}}]{CidFernandes10}
{Cid Fernandes}, R., {Stasi{\'n}ska}, G., {Schlickmann}, M.~S., {et~al.} 2010, \mnras, 403, 1036, \dodoi{10.1111/j.1365-2966.2009.16185.x}

\bibitem[{{Condon} {et~al.}(1998){Condon}, {Cotton}, {Greisen}, {Yin}, {Perley}, {Taylor}, \& {Broderick}}]{VLASS98}
{Condon}, J.~J., {Cotton}, W.~D., {Greisen}, E.~W., {et~al.} 1998, \aj, 115, 1693, \dodoi{10.1086/300337}

\bibitem[{{Cracco} {et~al.}(2016){Cracco}, {Ciroi}, {Berton}, {Di Mille}, {Foschini}, {La Mura}, \& {Rafanelli}}]{Cracco16}
{Cracco}, V., {Ciroi}, S., {Berton}, M., {et~al.} 2016, \mnras, 462, 1256, \dodoi{10.1093/mnras/stw1689}

\bibitem[{{Cresci} {et~al.}(2015){Cresci}, {Mainieri}, {Brusa}, {Marconi}, {Perna}, {Mannucci}, {Piconcelli}, {Maiolino}, {Feruglio}, {Fiore}, {Bongiorno}, {Lanzuisi}, {Merloni}, {Schramm}, {Silverman}, \& {Civano}}]{Cresci15}
{Cresci}, G., {Mainieri}, V., {Brusa}, M., {et~al.} 2015, \apj, 799, 82, \dodoi{10.1088/0004-637X/799/1/82}

\bibitem[{{Cresci} {et~al.}(2023){Cresci}, {Tozzi}, {Perna}, {Brusa}, {Marconcini}, {Marconi}, {Carniani}, {Brienza}, {Giroletti}, {Belfiore}, {Ginolfi}, {Mannucci}, {Ulivi}, {Scholtz}, {Venturi}, {Arribas}, {{\"U}bler}, {D'Eugenio}, {Mingozzi}, {Balmaverde}, {Capetti}, {Parlanti}, \& {Zana}}]{Cresci23}
{Cresci}, G., {Tozzi}, G., {Perna}, M., {et~al.} 2023, \aap, 672, A128, \dodoi{10.1051/0004-6361/202346001}

\bibitem[{{Currie} {et~al.}(2014){Currie}, {Berry}, {Jenness}, {Gibb}, {Bell}, \& {Draper}}]{Starlink14}
{Currie}, M.~J., {Berry}, D.~S., {Jenness}, T., {et~al.} 2014, in Astronomical Society of the Pacific Conference Series, Vol. 485, Astronomical Data Analysis Software and Systems XXIII, ed. N.~{Manset} \& P.~{Forshay}, 391

\bibitem[{{Dav{\'e}} {et~al.}(2019){Dav{\'e}}, {Angl{\'e}s-Alc{\'a}zar}, {Narayanan}, {Li}, {Rafieferantsoa}, \& {Appleby}}]{Dave19}
{Dav{\'e}}, R., {Angl{\'e}s-Alc{\'a}zar}, D., {Narayanan}, D., {et~al.} 2019, \mnras, 486, 2827, \dodoi{10.1093/mnras/stz937}

\bibitem[{{Davies} {et~al.}(2020){Davies}, {Baron}, {Shimizu}, {Netzer}, {Burtscher}, {de Zeeuw}, {Genzel}, {Hicks}, {Koss}, {Lin}, {Lutz}, {Maciejewski}, {M{\"u}ller-S{\'a}nchez}, {Orban de Xivry}, {Ricci}, {Riffel}, {Riffel}, {Rosario}, {Schartmann}, {Schnorr-M{\"u}ller}, {Shangguan}, {Sternberg}, {Sturm}, {Storchi-Bergmann}, {Tacconi}, \& {Veilleux}}]{Davies20}
{Davies}, R., {Baron}, D., {Shimizu}, T., {et~al.} 2020, \mnras, 498, 4150, \dodoi{10.1093/mnras/staa2413}

\bibitem[{{Deconto-Machado} {et~al.}(2022){Deconto-Machado}, {Riffel}, {Ilha}, {Rembold}, {Storchi-Bergmann}, {Riffel}, {Schimoia}, {Schneider}, {Bizyaev}, {Feng}, {Wylezalek}, {da Costa}, {do Nascimento}, \& {Maia}}]{Deconto-Machado22}
{Deconto-Machado}, A., {Riffel}, R.~A., {Ilha}, G.~S., {et~al.} 2022, \aap, 659, A131, \dodoi{10.1051/0004-6361/202140613}

\bibitem[{{Ferrarese} \& {Merritt}(2000)}]{Ferrarese00}
{Ferrarese}, L., \& {Merritt}, D. 2000, \apjl, 539, L9, \dodoi{10.1086/312838}

\bibitem[{{Fiore} {et~al.}(2017){Fiore}, {Feruglio}, {Shankar}, {Bischetti}, {Bongiorno}, {Brusa}, {Carniani}, {Cicone}, {Duras}, {Lamastra}, {Mainieri}, {Marconi}, {Menci}, {Maiolino}, {Piconcelli}, {Vietri}, \& {Zappacosta}}]{Fiore17}
{Fiore}, F., {Feruglio}, C., {Shankar}, F., {et~al.} 2017, \aap, 601, A143, \dodoi{10.1051/0004-6361/201629478}

\bibitem[{{Fischer} {et~al.}(2018){Fischer}, {Kraemer}, {Schmitt}, {Longo Micchi}, {Crenshaw}, {Revalski}, {Vestergaard}, {Elvis}, {Gaskell}, {Hamann}, {Ho}, {Hutchings}, {Mushotzky}, {Netzer}, {Storchi-Bergmann}, {Straughn}, {Turner}, \& {Ward}}]{Fischer18}
{Fischer}, T.~C., {Kraemer}, S.~B., {Schmitt}, H.~R., {et~al.} 2018, \apj, 856, 102, \dodoi{10.3847/1538-4357/aab03e}

\bibitem[{{Fluetsch} {et~al.}(2019){Fluetsch}, {Maiolino}, {Carniani}, {Marconi}, {Cicone}, {Bourne}, {Costa}, {Fabian}, {Ishibashi}, \& {Venturi}}]{Fluetsch19}
{Fluetsch}, A., {Maiolino}, R., {Carniani}, S., {et~al.} 2019, \mnras, 483, 4586, \dodoi{10.1093/mnras/sty3449}

\bibitem[{{Foreman-Mackey} {et~al.}(2013){Foreman-Mackey}, {Hogg}, {Lang}, \& {Goodman}}]{emcee}
{Foreman-Mackey}, D., {Hogg}, D.~W., {Lang}, D., \& {Goodman}, J. 2013, \pasp, 125, 306, \dodoi{10.1086/670067}

\bibitem[{{Harris} {et~al.}(2020){Harris}, {Millman}, {van der Walt}, {Gommers}, {Virtanen}, {Cournapeau}, {Wieser}, {Taylor}, {Berg}, {Smith}, {Kern}, {Picus}, {Hoyer}, {van Kerkwijk}, {Brett}, {Haldane}, {del R{\'\i}o}, {Wiebe}, {Peterson}, {G{\'e}rard-Marchant}, {Sheppard}, {Reddy}, {Weckesser}, {Abbasi}, {Gohlke}, \& {Oliphant}}]{numpy}
{Harris}, C.~R., {Millman}, K.~J., {van der Walt}, S.~J., {et~al.} 2020, \nat, 585, 357, \dodoi{10.1038/s41586-020-2649-2}

\bibitem[{{Harrison}(2017)}]{Harrison17}
{Harrison}, C.~M. 2017, Nature Astronomy, 1, 0165, \dodoi{10.1038/s41550-017-0165}

\bibitem[{{Harrison} {et~al.}(2014){Harrison}, {Alexander}, {Mullaney}, \& {Swinbank}}]{Harrison14}
{Harrison}, C.~M., {Alexander}, D.~M., {Mullaney}, J.~R., \& {Swinbank}, A.~M. 2014, \mnras, 441, 3306, \dodoi{10.1093/mnras/stu515}

\bibitem[{{Heckman} {et~al.}(2004){Heckman}, {Kauffmann}, {Brinchmann}, {Charlot}, {Tremonti}, \& {White}}]{Heckman04}
{Heckman}, T.~M., {Kauffmann}, G., {Brinchmann}, J., {et~al.} 2004, \apj, 613, 109, \dodoi{10.1086/422872}

\bibitem[{{Hopkins} \& {Elvis}(2010)}]{Hopkins10}
{Hopkins}, P.~F., \& {Elvis}, M. 2010, \mnras, 401, 7, \dodoi{10.1111/j.1365-2966.2009.15643.x}

\bibitem[{Hunter(2007)}]{matplotlib}
Hunter, J.~D. 2007, Computing in Science \& Engineering, 9, 90, \dodoi{10.1109/MCSE.2007.55}

\bibitem[{{Husemann} {et~al.}(2014){Husemann}, {Jahnke}, {S{\'a}nchez}, {Wisotzki}, {Nugroho}, {Kupko}, \& {Schramm}}]{Husemann14}
{Husemann}, B., {Jahnke}, K., {S{\'a}nchez}, S.~F., {et~al.} 2014, \mnras, 443, 755, \dodoi{10.1093/mnras/stu1167}

\bibitem[{{Husemann} {et~al.}(2012){Husemann}, {Kamann}, {Sandin}, {S{\'a}nchez}, {Garc{\'\i}a-Benito}, \& {Mast}}]{Pycosmic}
{Husemann}, B., {Kamann}, S., {Sandin}, C., {et~al.} 2012, \aap, 545, A137, \dodoi{10.1051/0004-6361/201220102}

\bibitem[{{Husemann} {et~al.}(2016){Husemann}, {Scharw{\"a}chter}, {Bennert}, {Mainieri}, {Woo}, \& {Kakkad}}]{Husemann16}
{Husemann}, B., {Scharw{\"a}chter}, J., {Bennert}, V.~N., {et~al.} 2016, \aap, 594, A44, \dodoi{10.1051/0004-6361/201527992}

\bibitem[{{Husemann} {et~al.}(2013){Husemann}, {Jahnke}, {S{\'a}nchez}, {Barrado}, {Bekerait{\.{e}}}, {Bomans}, {Castillo-Morales}, {Catal{\'a}n-Torrecilla}, {Cid Fernandes}, {Falc{\'o}n-Barroso}, {Garc{\'\i}a-Benito}, {Gonz{\'a}lez Delgado}, {Iglesias-P{\'a}ramo}, {Johnson}, {Kupko}, {L{\'o}pez-Fernandez}, {Lyubenova}, {Marino}, {Mast}, {Miskolczi}, {Monreal-Ibero}, {Gil de Paz}, {P{\'e}rez}, {P{\'e}rez}, {Rosales-Ortega}, {Ruiz-Lara}, {Schilling}, {van de Ven}, {Walcher}, {Alves}, {de Amorim}, {Backsmann}, {Barrera-Ballesteros}, {Bland-Hawthorn}, {Cortijo}, {Dettmar}, {Demleitner}, {D{\'\i}az}, {Enke}, {Florido}, {Flores}, {Galbany}, {Gallazzi}, {Garc{\'\i}a-Lorenzo}, {Gomes}, {Gruel}, {Haines}, {Holmes}, {Jungwiert}, {Kalinova}, {Kehrig}, {Kennicutt}, {Klar}, {Lehnert}, {L{\'o}pez-S{\'a}nchez}, {de Lorenzo-C{\'a}ceres}, {M{\'a}rmol-Queralt{\'o}}, {M{\'a}rquez}, {Mendez-Abreu}, {Moll{\'a}}, {del Olmo}, {Meidt}, {Papaderos}, {Puschnig}, {Quirrenbach}, {Roth}, {S{\'a}nchez-Bl{\'a}zquez}, {Spekkens}, {Singh},
  {Stanishev}, {Trager}, {Vilchez}, {Wild}, {Wisotzki}, {Zibetti}, \& {Ziegler}}]{Husemann13}
{Husemann}, B., {Jahnke}, K., {S{\'a}nchez}, S.~F., {et~al.} 2013, \aap, 549, A87, \dodoi{10.1051/0004-6361/201220582}

\bibitem[{{Husemann} {et~al.}(2019){Husemann}, {Scharw{\"a}chter}, {Davis}, {P{\'e}rez-Torres}, {Smirnova-Pinchukova}, {Tremblay}, {Krumpe}, {Combes}, {Baum}, {Busch}, {Connor}, {Croom}, {Gaspari}, {Kraft}, {O'Dea}, {Powell}, {Singha}, \& {Urrutia}}]{Husemann19}
{Husemann}, B., {Scharw{\"a}chter}, J., {Davis}, T.~A., {et~al.} 2019, \aap, 627, A53, \dodoi{10.1051/0004-6361/201935283}

\bibitem[{{Jenness} \& {Economou}(2015)}]{ORACDR15}
{Jenness}, T., \& {Economou}, F. 2015, Astronomy and Computing, 9, 40, \dodoi{10.1016/j.ascom.2014.10.005}

\bibitem[{{Juneau} {et~al.}(2022){Juneau}, {Goulding}, {Banfield}, {Bianchi}, {Duc}, {Ho}, {Dopita}, {Scharw{\"a}chter}, {Bauer}, {Groves}, {Alexander}, {Davies}, {Elbaz}, {Freeland}, {Hampton}, {Kewley}, {Nikutta}, {Shastri}, {Shu}, {Vogt}, {Wang}, {Wong}, \& {Woo}}]{Juneau22}
{Juneau}, S., {Goulding}, A.~D., {Banfield}, J., {et~al.} 2022, \apj, 925, 203, \dodoi{10.3847/1538-4357/ac425f}

\bibitem[{{Kakkad} {et~al.}(2020){Kakkad}, {Mainieri}, {Vietri}, {Carniani}, {Harrison}, {Perna}, {Scholtz}, {Circosta}, {Cresci}, {Husemann}, {Bischetti}, {Feruglio}, {Fiore}, {Marconi}, {Padovani}, {Brusa}, {Cicone}, {Comastri}, {Lanzuisi}, {Mannucci}, {Menci}, {Netzer}, {Piconcelli}, {Puglisi}, {Salvato}, {Schramm}, {Silverman}, {Vignali}, {Zamorani}, \& {Zappacosta}}]{Kakkad20}
{Kakkad}, D., {Mainieri}, V., {Vietri}, G., {et~al.} 2020, \aap, 642, A147, \dodoi{10.1051/0004-6361/202038551}

\bibitem[{{Kakkad} {et~al.}(2022){Kakkad}, {Sani}, {Rojas}, {Mallmann}, {Veilleux}, {Bauer}, {Ricci}, {Mushotzky}, {Koss}, {Ricci}, {Treister}, {Privon}, {Nguyen}, {B{\"a}r}, {Harrison}, {Oh}, {Powell}, {Riffel}, {Stern}, {Trakhtenbrot}, \& {Urry}}]{Kakkad22}
{Kakkad}, D., {Sani}, E., {Rojas}, A.~F., {et~al.} 2022, \mnras, 511, 2105, \dodoi{10.1093/mnras/stac103}

\bibitem[{{Kakkad} {et~al.}(2023){Kakkad}, {Mainieri}, {Vietri}, {Lamperti}, {Carniani}, {Cresci}, {Harrison}, {Marconi}, {Bischetti}, {Cicone}, {Circosta}, {Husemann}, {Man}, {Mannucci}, {Netzer}, {Padovani}, {Perna}, {Puglisi}, {Scholtz}, {Tozzi}, {Vignali}, \& {Zappacosta}}]{Kakkad23}
{Kakkad}, D., {Mainieri}, V., {Vietri}, G., {et~al.} 2023, \mnras, 520, 5783, \dodoi{10.1093/mnras/stad439}

\bibitem[{{Kang} \& {Woo}(2018)}]{Kang18}
{Kang}, D., \& {Woo}, J.-H. 2018, \apj, 864, 124, \dodoi{10.3847/1538-4357/aad561}

\bibitem[{{Karouzos} {et~al.}(2016{\natexlab{a}}){Karouzos}, {Woo}, \& {Bae}}]{Karouzos16a}
{Karouzos}, M., {Woo}, J.-H., \& {Bae}, H.-J. 2016{\natexlab{a}}, \apj, 819, 148, \dodoi{10.3847/0004-637X/819/2/148}

\bibitem[{{Karouzos} {et~al.}(2016{\natexlab{b}}){Karouzos}, {Woo}, \& {Bae}}]{Karouzos16b}
---. 2016{\natexlab{b}}, \apj, 833, 171, \dodoi{10.3847/1538-4357/833/2/171}

\bibitem[{{Kauffmann} {et~al.}(2003){Kauffmann}, {Heckman}, {Tremonti}, {Brinchmann}, {Charlot}, {White}, {Ridgway}, {Brinkmann}, {Fukugita}, {Hall}, {Ivezi{\'c}}, {Richards}, \& {Schneider}}]{Kauffmann03}
{Kauffmann}, G., {Heckman}, T.~M., {Tremonti}, C., {et~al.} 2003, \mnras, 346, 1055, \dodoi{10.1111/j.1365-2966.2003.07154.x}

\bibitem[{{Kennicutt}(1998)}]{Kennicutt98}
{Kennicutt}, Robert~C., J. 1998, \araa, 36, 189, \dodoi{10.1146/annurev.astro.36.1.189}

\bibitem[{{Kewley} {et~al.}(2001){Kewley}, {Dopita}, {Sutherland}, {Heisler}, \& {Trevena}}]{Kewley01}
{Kewley}, L.~J., {Dopita}, M.~A., {Sutherland}, R.~S., {Heisler}, C.~A., \& {Trevena}, J. 2001, \apj, 556, 121, \dodoi{10.1086/321545}

\bibitem[{{Kim} {et~al.}(2022){Kim}, {Woo}, {Jadhav}, {Chung}, {Baek}, {Lee}, {Shin}, {Hwang}, {Luo}, {Son}, {Kim}, \& {Woo}}]{Kim22}
{Kim}, C., {Woo}, J.-H., {Jadhav}, Y., {et~al.} 2022, \apj, 928, 73, \dodoi{10.3847/1538-4357/ac5407}

\bibitem[{{Komossa}(2008)}]{Komossa08}
{Komossa}, S. 2008, in Revista Mexicana de Astronomia y Astrofisica Conference Series, Vol.~32, Revista Mexicana de Astronomia y Astrofisica Conference Series, 86--92.
\newblock \doarXiv{0710.3326}

\bibitem[{{Kormendy} \& {Ho}(2013)}]{Kormendy13}
{Kormendy}, J., \& {Ho}, L.~C. 2013, \araa, 51, 511, \dodoi{10.1146/annurev-astro-082708-101811}

\bibitem[{{Lamperti} {et~al.}(2022){Lamperti}, {Pereira-Santaella}, {Perna}, {Colina}, {Arribas}, {Garc{\'\i}a-Burillo}, {Gonz{\'a}lez-Alfonso}, {Aalto}, {Alonso-Herrero}, {Combes}, {Labiano}, {Piqueras-L{\'o}pez}, {Rigopoulou}, \& {van der Werf}}]{Lamperti22}
{Lamperti}, I., {Pereira-Santaella}, M., {Perna}, M., {et~al.} 2022, \aap, 668, A45, \dodoi{10.1051/0004-6361/202244054}

\bibitem[{{Liu} {et~al.}(2014){Liu}, {Zakamska}, \& {Greene}}]{Liu14}
{Liu}, G., {Zakamska}, N.~L., \& {Greene}, J.~E. 2014, \mnras, 442, 1303, \dodoi{10.1093/mnras/stu974}

\bibitem[{{Liu} {et~al.}(2013){Liu}, {Zakamska}, {Greene}, {Nesvadba}, \& {Liu}}]{Liu13a}
{Liu}, G., {Zakamska}, N.~L., {Greene}, J.~E., {Nesvadba}, N. P.~H., \& {Liu}, X. 2013, \mnras, 430, 2327, \dodoi{10.1093/mnras/stt051}

\bibitem[{{Luo} {et~al.}(2019){Luo}, {Woo}, {Shin}, {Kang}, {Bae}, \& {Karouzos}}]{Luo19}
{Luo}, R., {Woo}, J.-H., {Shin}, J., {et~al.} 2019, \apj, 874, 99, \dodoi{10.3847/1538-4357/ab08e6}

\bibitem[{{Luo} {et~al.}(2021){Luo}, {Woo}, {Karouzos}, {Bae}, {Shin}, {McConnell}, {Shih}, {Kim}, \& {Park}}]{Luo21}
{Luo}, R., {Woo}, J.-H., {Karouzos}, M., {et~al.} 2021, \apj, 908, 221, \dodoi{10.3847/1538-4357/abd5ac}

\bibitem[{{Luridiana} {et~al.}(2015){Luridiana}, {Morisset}, \& {Shaw}}]{PyNeb}
{Luridiana}, V., {Morisset}, C., \& {Shaw}, R.~A. 2015, \aap, 573, A42, \dodoi{10.1051/0004-6361/201323152}

\bibitem[{{Maiolino} {et~al.}(2017){Maiolino}, {Russell}, {Fabian}, {Carniani}, {Gallagher}, {Cazzoli}, {Arribas}, {Belfiore}, {Bellocchi}, {Colina}, {Cresci}, {Ishibashi}, {Marconi}, {Mannucci}, {Oliva}, \& {Sturm}}]{Maiolino17}
{Maiolino}, R., {Russell}, H.~R., {Fabian}, A.~C., {et~al.} 2017, \nat, 544, 202, \dodoi{10.1038/nature21677}

\bibitem[{{Martinsson} {et~al.}(2013){Martinsson}, {Verheijen}, {Westfall}, {Bershady}, {Schechtman-Rook}, {Andersen}, \& {Swaters}}]{Martinsson13}
{Martinsson}, T. P.~K., {Verheijen}, M. A.~W., {Westfall}, K.~B., {et~al.} 2013, \aap, 557, A130, \dodoi{10.1051/0004-6361/201220515}

\bibitem[{{Marton} {et~al.}(2017){Marton}, {Calzoletti}, {Perez Garcia}, {Kiss}, {Paladini}, {Altieri}, {Sanchez Portal}, {Kidger}, \& {the Herschel Point Source Catalogue Working Group}}]{Herschel-PACS17}
{Marton}, G., {Calzoletti}, L., {Perez Garcia}, A.~M., {et~al.} 2017, arXiv e-prints, arXiv:1705.05693.
\newblock \doarXiv{1705.05693}

\bibitem[{{Mingozzi} {et~al.}(2019){Mingozzi}, {Cresci}, {Venturi}, {Marconi}, {Mannucci}, {Perna}, {Belfiore}, {Carniani}, {Balmaverde}, {Brusa}, {Cicone}, {Feruglio}, {Gallazzi}, {Mainieri}, {Maiolino}, {Nagao}, {Nardini}, {Sani}, {Tozzi}, \& {Zibetti}}]{Mingozzi19}
{Mingozzi}, M., {Cresci}, G., {Venturi}, G., {et~al.} 2019, \aap, 622, A146, \dodoi{10.1051/0004-6361/201834372}

\bibitem[{{Molina} {et~al.}(2022){Molina}, {Ho}, {Wang}, {Shangguan}, {Bauer}, {Treister}, {Zhuang}, {Ricci}, \& {Bian}}]{Molina22}
{Molina}, J., {Ho}, L.~C., {Wang}, R., {et~al.} 2022, \apj, 935, 72, \dodoi{10.3847/1538-4357/ac7d4d}

\bibitem[{{Mullaney} {et~al.}(2013){Mullaney}, {Alexander}, {Fine}, {Goulding}, {Harrison}, \& {Hickox}}]{Mullaney13}
{Mullaney}, J.~R., {Alexander}, D.~M., {Fine}, S., {et~al.} 2013, \mnras, 433, 622, \dodoi{10.1093/mnras/stt751}

\bibitem[{{Netzer}(2009)}]{Netzer09}
{Netzer}, H. 2009, \mnras, 399, 1907, \dodoi{10.1111/j.1365-2966.2009.15434.x}

\bibitem[{{Noeske} {et~al.}(2007){Noeske}, {Weiner}, {Faber}, {Papovich}, {Koo}, {Somerville}, {Bundy}, {Conselice}, {Newman}, {Schiminovich}, {Le Floc'h}, {Coil}, {Rieke}, {Lotz}, {Primack}, {Barmby}, {Cooper}, {Davis}, {Ellis}, {Fazio}, {Guhathakurta}, {Huang}, {Kassin}, {Martin}, {Phillips}, {Rich}, {Small}, {Willmer}, \& {Wilson}}]{Noeske07}
{Noeske}, K.~G., {Weiner}, B.~J., {Faber}, S.~M., {et~al.} 2007, \apjl, 660, L43, \dodoi{10.1086/517926}

\bibitem[{{Noll} {et~al.}(2009){Noll}, {Burgarella}, {Giovannoli}, {Buat}, {Marcillac}, \& {Mu{\~n}oz-Mateos}}]{CIGALE09}
{Noll}, S., {Burgarella}, D., {Giovannoli}, E., {et~al.} 2009, \aap, 507, 1793, \dodoi{10.1051/0004-6361/200912497}

\bibitem[{{Oh} {et~al.}(2022){Oh}, {Colless}, {D'Eugenio}, {Croom}, {Cortese}, {Groves}, {Kewley}, {van de Sande}, {Zovaro}, {Varidel}, {Barsanti}, {Bland-Hawthorn}, {Brough}, {Bryant}, {Casura}, {Lawrence}, {Lorente}, {Medling}, {Owers}, \& {Yi}}]{Oh22}
{Oh}, S., {Colless}, M., {D'Eugenio}, F., {et~al.} 2022, \mnras, 512, 1765, \dodoi{10.1093/mnras/stac509}

\bibitem[{{Rakshit} {et~al.}(2017){Rakshit}, {Stalin}, {Chand}, \& {Zhang}}]{Rakshit17}
{Rakshit}, S., {Stalin}, C.~S., {Chand}, H., \& {Zhang}, X.-G. 2017, \apjs, 229, 39, \dodoi{10.3847/1538-4365/aa6971}

\bibitem[{{Rakshit} \& {Woo}(2018)}]{Rakshit18}
{Rakshit}, S., \& {Woo}, J.-H. 2018, \apj, 865, 5, \dodoi{10.3847/1538-4357/aad9f8}

\bibitem[{{Ramos Almeida} {et~al.}(2019){Ramos Almeida}, {Acosta-Pulido}, {Tadhunter}, {Gonz{\'a}lez-Fern{\'a}ndez}, {Cicone}, \& {Fern{\'a}ndez-Torreiro}}]{Ramos-Almeida19}
{Ramos Almeida}, C., {Acosta-Pulido}, J.~A., {Tadhunter}, C.~N., {et~al.} 2019, \mnras, 487, L18, \dodoi{10.1093/mnrasl/slz072}

\bibitem[{{Revalski} {et~al.}(2021){Revalski}, {Meena}, {Martinez}, {Polack}, {Crenshaw}, {Kraemer}, {Collins}, {Fischer}, {Schmitt}, {Schmidt}, {Maksym}, \& {Rafelski}}]{Revalski21}
{Revalski}, M., {Meena}, B., {Martinez}, F., {et~al.} 2021, \apj, 910, 139, \dodoi{10.3847/1538-4357/abdcad}

\bibitem[{{Revalski} {et~al.}(2022){Revalski}, {Crenshaw}, {Rafelski}, {Kraemer}, {Polack}, {Falc{\~a}o}, {Fischer}, {Meena}, {Martinez}, {Schmitt}, {Collins}, \& {Falcone}}]{Revalski22}
{Revalski}, M., {Crenshaw}, D.~M., {Rafelski}, M., {et~al.} 2022, \apj, 930, 14, \dodoi{10.3847/1538-4357/ac5f3d}

\bibitem[{{Riffel} {et~al.}(2023){Riffel}, {Storchi-Bergmann}, {Riffel}, {Bianchin}, {Zakamska}, {Ruschel-Dutra}, {Bentz}, {Burtscher}, {Crenshaw}, {Dahmer-Hahn}, {Dametto}, {Davies}, {Diniz}, {Fischer}, {Harrison}, {Mainieri}, {Revalski}, {Rodriguez-Ardila}, {Rosario}, \& {Sch{\"o}nell}}]{Riffel23}
{Riffel}, R.~A., {Storchi-Bergmann}, T., {Riffel}, R., {et~al.} 2023, \mnras, 521, 1832, \dodoi{10.1093/mnras/stad599}

\bibitem[{{Rupke} {et~al.}(2017){Rupke}, {G{\"u}ltekin}, \& {Veilleux}}]{Rupke17}
{Rupke}, D. S.~N., {G{\"u}ltekin}, K., \& {Veilleux}, S. 2017, \apj, 850, 40, \dodoi{10.3847/1538-4357/aa94d1}

\bibitem[{{Ruschel-Dutra} {et~al.}(2021){Ruschel-Dutra}, {Storchi-Bergmann}, {Schnorr-M{\"u}ller}, {Riffel}, {Dall'Agnol de Oliveira}, {Lena}, {Robinson}, {Nagar}, \& {Elvis}}]{Ruschel-Dutra21}
{Ruschel-Dutra}, D., {Storchi-Bergmann}, T., {Schnorr-M{\"u}ller}, A., {et~al.} 2021, \mnras, 507, 74, \dodoi{10.1093/mnras/stab2058}

\bibitem[{{Scharw{\"a}chter} {et~al.}(2017){Scharw{\"a}chter}, {Husemann}, {Busch}, {Komossa}, \& {Dopita}}]{Scharwachter17}
{Scharw{\"a}chter}, J., {Husemann}, B., {Busch}, G., {Komossa}, S., \& {Dopita}, M.~A. 2017, \apj, 848, 35, \dodoi{10.3847/1538-4357/aa8ad8}

\bibitem[{{Schmidt} {et~al.}(2018){Schmidt}, {Oio}, {Ferreiro}, {Vega}, \& {Weidmann}}]{Schmidt18}
{Schmidt}, E.~O., {Oio}, G.~A., {Ferreiro}, D., {Vega}, L., \& {Weidmann}, W. 2018, \aap, 615, A13, \dodoi{10.1051/0004-6361/201731557}

\bibitem[{{Schmitt} {et~al.}(2003{\natexlab{a}}){Schmitt}, {Donley}, {Antonucci}, {Hutchings}, \& {Kinney}}]{Schmitt03a}
{Schmitt}, H.~R., {Donley}, J.~L., {Antonucci}, R.~R.~J., {Hutchings}, J.~B., \& {Kinney}, A.~L. 2003{\natexlab{a}}, \apjs, 148, 327, \dodoi{10.1086/377440}

\bibitem[{{Schmitt} {et~al.}(2003{\natexlab{b}}){Schmitt}, {Donley}, {Antonucci}, {Hutchings}, {Kinney}, \& {Pringle}}]{Schmitt03b}
{Schmitt}, H.~R., {Donley}, J.~L., {Antonucci}, R.~R.~J., {et~al.} 2003{\natexlab{b}}, \apj, 597, 768, \dodoi{10.1086/381224}

\bibitem[{{Scholtz} {et~al.}(2020){Scholtz}, {Harrison}, {Rosario}, {Alexander}, {Chen}, {Kakkad}, {Mainieri}, {Tiley}, {Turner}, {Cirasuolo}, {Sharples}, \& {Stach}}]{Scholtz20}
{Scholtz}, J., {Harrison}, C.~M., {Rosario}, D.~J., {et~al.} 2020, \mnras, 492, 3194, \dodoi{10.1093/mnras/staa030}

\bibitem[{{Scholtz} {et~al.}(2021){Scholtz}, {Harrison}, {Rosario}, {Alexander}, {Knudsen}, {Stanley}, {Chen}, {Kakkad}, {Mainieri}, \& {Mullaney}}]{Scholtz21}
---. 2021, \mnras, 505, 5469, \dodoi{10.1093/mnras/stab1631}

\bibitem[{{Schulz} {et~al.}(2017){Schulz}, {Marton}, {Valtchanov}, {P{\'e}rez Garc{\'\i}a}, {Pint{\'e}r}, {Appleton}, {Kiss}, {Lim}, {Lu}, {Papageorgiou}, {Pearson}, {Rector}, {S{\'a}nchez Portal}, {Shupe}, {T{\'o}th}, {Van Dyk}, {Varga-Vereb{\'e}lyi}, \& {Xu}}]{Herschel-SPIRE17}
{Schulz}, B., {Marton}, G., {Valtchanov}, I., {et~al.} 2017, arXiv e-prints, arXiv:1706.00448.
\newblock \doarXiv{1706.00448}

\bibitem[{{Shimizu} {et~al.}(2015){Shimizu}, {Mushotzky}, {Mel{\'e}ndez}, {Koss}, \& {Rosario}}]{Shimizu15}
{Shimizu}, T.~T., {Mushotzky}, R.~F., {Mel{\'e}ndez}, M., {Koss}, M., \& {Rosario}, D.~J. 2015, \mnras, 452, 1841, \dodoi{10.1093/mnras/stv1407}

\bibitem[{{Shin} {et~al.}(2019){Shin}, {Woo}, {Chung}, {Baek}, {Cho}, {Kang}, \& {Bae}}]{Shin19}
{Shin}, J., {Woo}, J.-H., {Chung}, A., {et~al.} 2019, \apj, 881, 147, \dodoi{10.3847/1538-4357/ab2e72}

\bibitem[{{Shin} {et~al.}(2021){Shin}, {Woo}, {Kim}, \& {Wang}}]{Shin21}
{Shin}, J., {Woo}, J.-H., {Kim}, M., \& {Wang}, J. 2021, \apj, 908, 81, \dodoi{10.3847/1538-4357/abd779}

\bibitem[{{Silk}(2013)}]{Silk13}
{Silk}, J. 2013, \apj, 772, 112, \dodoi{10.1088/0004-637X/772/2/112}

\bibitem[{{Silk} \& {Rees}(1998)}]{Silk98}
{Silk}, J., \& {Rees}, M.~J. 1998, \aap, 331, L1.
\newblock \doarXiv{astro-ph/9801013}

\bibitem[{{Singha} {et~al.}(2022){Singha}, {Husemann}, {Urrutia}, {O'Dea}, {Scharw{\"a}chter}, {Gaspari}, {Combes}, {Nevin}, {Terrazas}, {P{\'e}rez-Torres}, {Rose}, {Davis}, {Tremblay}, {Neumann}, {Smirnova-Pinchukova}, \& {Baum}}]{Singha22}
{Singha}, M., {Husemann}, B., {Urrutia}, T., {et~al.} 2022, \aap, 659, A123, \dodoi{10.1051/0004-6361/202040122}

\bibitem[{{Smirnova-Pinchukova} {et~al.}(2022){Smirnova-Pinchukova}, {Husemann}, {Davis}, {Smith}, {Singha}, {Tremblay}, {Klessen}, {Powell}, {Connor}, {Baum}, {Combes}, {Croom}, {Gaspari}, {Neumann}, {O'Dea}, {P{\'e}rez-Torres}, {Rosario}, {Rose}, {Scharw{\"a}chter}, \& {Winkel}}]{Smirnova-Pinchukova22}
{Smirnova-Pinchukova}, I., {Husemann}, B., {Davis}, T.~A., {et~al.} 2022, \aap, 659, A125, \dodoi{10.1051/0004-6361/202142011}

\bibitem[{{Vayner} {et~al.}(2021){Vayner}, {Zakamska}, {Riffel}, {Alexandroff}, {Cosens}, {Hamann}, {Perrotta}, {Rupke}, {Bergmann}, {Veilleux}, {Walth}, {Wright}, \& {Wylezalek}}]{Vayner21a}
{Vayner}, A., {Zakamska}, N.~L., {Riffel}, R.~A., {et~al.} 2021, \mnras, 504, 4445, \dodoi{10.1093/mnras/stab1176}

\bibitem[{{Vayner} {et~al.}(2023){Vayner}, {Zakamska}, {Ishikawa}, {Sankar}, {Wylezalek}, {Rupke}, {Veilleux}, {Bertemes}, {Barrera-Ballesteros}, {Chen}, {Diachenko}, {Goulding}, {Greene}, {Hainline}, {Hamann}, {Heckman}, {Johnson}, {Lim}, {Liu}, {Lutz}, {Lutzgendorf}, {Mainieri}, {McCrory}, {Murphree}, {Nesvadba}, {Ogle}, {Sturm}, \& {Whitesell}}]{Vayner23}
{Vayner}, A., {Zakamska}, N.~L., {Ishikawa}, Y., {et~al.} 2023, arXiv e-prints, arXiv:2307.13751, \dodoi{10.48550/arXiv.2307.13751}

\bibitem[{{Vazdekis} {et~al.}(2016){Vazdekis}, {Koleva}, {Ricciardelli}, {R{\"o}ck}, \& {Falc{\'o}n-Barroso}}]{EMILES16}
{Vazdekis}, A., {Koleva}, M., {Ricciardelli}, E., {R{\"o}ck}, B., \& {Falc{\'o}n-Barroso}, J. 2016, \mnras, 463, 3409, \dodoi{10.1093/mnras/stw2231}

\bibitem[{{Veilleux} {et~al.}(2020){Veilleux}, {Maiolino}, {Bolatto}, \& {Aalto}}]{Veilleux20}
{Veilleux}, S., {Maiolino}, R., {Bolatto}, A.~D., \& {Aalto}, S. 2020, \aapr, 28, 2, \dodoi{10.1007/s00159-019-0121-9}

\bibitem[{{Veilleux} {et~al.}(2023){Veilleux}, {Liu}, {Vayner}, {Wylezalek}, {Rupke}, {Zakamska}, {Ishikawa}, {Bertemes}, {Barrera-Ballesteros}, {Chen}, {Diachenko}, {Goulding}, {Greene}, {Hainline}, {Hamann}, {Heckman}, {Johnson}, {Grace Lim}, {Lutz}, {L{\"u}tzgendorf}, {Mainieri}, {Maiolino}, {McCrory}, {Murphree}, {Nesvadba}, {Ogle}, {Sankar}, {Sturm}, \& {Whitesell}}]{Veilleux23}
{Veilleux}, S., {Liu}, W., {Vayner}, A., {et~al.} 2023, \apj, 953, 56, \dodoi{10.3847/1538-4357/ace10f}

\bibitem[{{Villar-Mart{\'\i}n} {et~al.}(2016){Villar-Mart{\'\i}n}, {Arribas}, {Emonts}, {Humphrey}, {Tadhunter}, {Bessiere}, {Cabrera Lavers}, \& {Ramos Almeida}}]{Villar-Martin16}
{Villar-Mart{\'\i}n}, M., {Arribas}, S., {Emonts}, B., {et~al.} 2016, \mnras, 460, 130, \dodoi{10.1093/mnras/stw901}

\bibitem[{Virtanen {et~al.}(2020)Virtanen, Gommers, Oliphant, Haberland, Reddy, Cournapeau, Burovski, Peterson, Weckesser, Bright, {van der Walt}, Brett, Wilson, Millman, Mayorov, Nelson, Jones, Kern, Larson, Carey, Polat, Feng, Moore, {VanderPlas}, Laxalde, Perktold, Cimrman, Henriksen, Quintero, Harris, Archibald, Ribeiro, Pedregosa, {van Mulbregt}, \& {SciPy 1.0 Contributors}}]{scipy}
Virtanen, P., Gommers, R., Oliphant, T.~E., {et~al.} 2020, Nature Methods, 17, 261, \dodoi{10.1038/s41592-019-0686-2}

\bibitem[{{Ward} {et~al.}(2022){Ward}, {Harrison}, {Costa}, \& {Mainieri}}]{Ward22}
{Ward}, S.~R., {Harrison}, C.~M., {Costa}, T., \& {Mainieri}, V. 2022, \mnras, 514, 2936, \dodoi{10.1093/mnras/stac1219}

\bibitem[{{Wei} {et~al.}(2018){Wei}, {Gu}, {Brotherton}, {Shi}, \& {Chen}}]{Wei18}
{Wei}, P., {Gu}, Y., {Brotherton}, M.~S., {Shi}, Y., \& {Chen}, Y. 2018, \apj, 857, 27, \dodoi{10.3847/1538-4357/aab499}

\bibitem[{{Weinberger} {et~al.}(2017){Weinberger}, {Springel}, {Hernquist}, {Pillepich}, {Marinacci}, {Pakmor}, {Nelson}, {Genel}, {Vogelsberger}, {Naiman}, \& {Torrey}}]{Weinberger17}
{Weinberger}, R., {Springel}, V., {Hernquist}, L., {et~al.} 2017, \mnras, 465, 3291, \dodoi{10.1093/mnras/stw2944}

\bibitem[{{Wellons} {et~al.}(2022){Wellons}, {Faucher-Gigu{\`e}re}, {Hopkins}, {Quataert}, {Angl{\'e}s-Alc{\'a}zar}, {Feldmann}, {Hayward}, {Kere{\v{s}}}, {Su}, \& {Wetzel}}]{Wellons22}
{Wellons}, S., {Faucher-Gigu{\`e}re}, C.-A., {Hopkins}, P.~F., {et~al.} 2022, arXiv e-prints, arXiv:2203.06201.
\newblock \doarXiv{2203.06201}

\bibitem[{{W}es {M}c{K}inney(2010)}]{pandas}
{W}es {M}c{K}inney. 2010, in {P}roceedings of the 9th {P}ython in {S}cience {C}onference, ed. {S}t\'efan van~der {W}alt \& {J}arrod {M}illman, 56 -- 61, \dodoi{10.25080/Majora-92bf1922-00a}

\bibitem[{{Winkel} {et~al.}(2022){Winkel}, {Husemann}, {Davis}, {Smirnova-Pinchukova}, {Bennert}, {Combes}, {Gaspari}, {Jahnke}, {Neumann}, {O'Dea}, {P{\'e}rez-Torres}, {Singha}, {Tremblay}, \& {Rix}}]{Winkel22}
{Winkel}, N., {Husemann}, B., {Davis}, T.~A., {et~al.} 2022, \aap, 663, A104, \dodoi{10.1051/0004-6361/202243697}

\bibitem[{{Woo} {et~al.}(2016){Woo}, {Bae}, {Son}, \& {Karouzos}}]{Woo16}
{Woo}, J.-H., {Bae}, H.-J., {Son}, D., \& {Karouzos}, M. 2016, \apj, 817, 108, \dodoi{10.3847/0004-637X/817/2/108}

\bibitem[{{Woo} {et~al.}(2017){Woo}, {Son}, \& {Bae}}]{Woo17}
{Woo}, J.-H., {Son}, D., \& {Bae}, H.-J. 2017, \apj, 839, 120, \dodoi{10.3847/1538-4357/aa6894}

\bibitem[{{Woo} {et~al.}(2020){Woo}, {Son}, \& {Rakshit}}]{Woo20}
{Woo}, J.-H., {Son}, D., \& {Rakshit}, S. 2020, \apj, 901, 66, \dodoi{10.3847/1538-4357/abad97}

\bibitem[{{Woo} {et~al.}(2015){Woo}, {Yoon}, {Park}, {Park}, \& {Kim}}]{Woo15}
{Woo}, J.-H., {Yoon}, Y., {Park}, S., {Park}, D., \& {Kim}, S.~C. 2015, \apj, 801, 38, \dodoi{10.1088/0004-637X/801/1/38}

\bibitem[{{Wylezalek} {et~al.}(2020){Wylezalek}, {Flores}, {Zakamska}, {Greene}, \& {Riffel}}]{Wylezalek20}
{Wylezalek}, D., {Flores}, A.~M., {Zakamska}, N.~L., {Greene}, J.~E., \& {Riffel}, R.~A. 2020, \mnras, 492, 4680, \dodoi{10.1093/mnras/staa062}

\bibitem[{{Wylezalek} \& {Zakamska}(2016)}]{Wylezalek16}
{Wylezalek}, D., \& {Zakamska}, N.~L. 2016, \mnras, 461, 3724, \dodoi{10.1093/mnras/stw1557}

\bibitem[{{Wylezalek} {et~al.}(2022){Wylezalek}, {Vayner}, {Rupke}, {Zakamska}, {Veilleux}, {Ishikawa}, {Bertemes}, {Liu}, {Barrera-Ballesteros}, {Chen}, {Goulding}, {Greene}, {Hainline}, {Hamann}, {Heckman}, {Johnson}, {Lutz}, {L{\"u}tzgendorf}, {Mainieri}, {Maiolino}, {Nesvadba}, {Ogle}, \& {Sturm}}]{Wylezalek22}
{Wylezalek}, D., {Vayner}, A., {Rupke}, D. S.~N., {et~al.} 2022, \apjl, 940, L7, \dodoi{10.3847/2041-8213/ac98c3}

\bibitem[{{Yamamura} {et~al.}(2010){Yamamura}, {Makiuti}, {Ikeda}, {Fukuda}, {Oyabu}, {Koga}, \& {White}}]{AKARI10}
{Yamamura}, I., {Makiuti}, S., {Ikeda}, N., {et~al.} 2010, VizieR Online Data Catalog, II/298

\bibitem[{{Zhuang} \& {Ho}(2020)}]{Zhuang20}
{Zhuang}, M.-Y., \& {Ho}, L.~C. 2020, \apj, 896, 108, \dodoi{10.3847/1538-4357/ab8f2e}

\bibitem[{{Zhuang} {et~al.}(2021){Zhuang}, {Ho}, \& {Shangguan}}]{Zhuang21}
{Zhuang}, M.-Y., {Ho}, L.~C., \& {Shangguan}, J. 2021, \apj, 906, 38, \dodoi{10.3847/1538-4357/abc94d}

\bibitem[{{Zubovas} \& {King}(2012)}]{Zubovas12}
{Zubovas}, K., \& {King}, A. 2012, \apjl, 745, L34, \dodoi{10.1088/2041-8205/745/2/L34}

\end{thebibliography}
\end{document}